\documentclass[a4paper,11pt]{article}

\usepackage{jheppub}
\pdfoutput=1
\usepackage{amsmath}
\usepackage{graphicx}
\usepackage{hyperref}
\usepackage{xspace}
\usepackage{xcolor}
\usepackage{placeins}

\graphicspath{ {./figs/} }

\title{Euclidean mirrors: enhanced vacuum decay from reflected instantons}

\author[1]{Ibrahim Akal,}
\author[2]{Gudrid Moortgat-Pick,}

\affiliation[1]{Theory Group, Deutsches Elektronen-Synchrotron DESY, Hamburg, Germany}
\affiliation[2]{University Hamburg, II. Institute for Theoretical Physics, Hamburg, Germany}

\emailAdd{ibrahim.akal@desy.de}
\emailAdd{gudrid.moortgat-pick@desy.de}

\abstract{We study the tunneling of virtual matter-antimatter pairs from the quantum vacuum in the presence of a spatially uniform, time-dependent electric background composed of a strong, slow field superimposed with a weak, rapid field.
After analytic continuation to Euclidean spacetime, we obtain from the instanton equations two critical points. While one of them is the closing point of the instanton path, the other serves as an Euclidean mirror which reflects and squeezes the instanton.
It is this reflection and shrinking which is responsible for an enormous enhancement of the vacuum pair production rate. We discuss how important features of two different mechanisms can be analysed and understood via such a rotation in the complex plane.
a) Consistent with previous studies,
we first discuss the standard assisted mechanism with a static strong field and certain weak fields
with a distinct pole structure in order to show that the reflection takes place exactly at the poles. 
We also discuss the effect of possible sub-cycle structures.
We extend this reflection picture then to weak fields which have no poles present and illustrate the effective reflections with explicit examples.
An additional field strength dependence for the rate occurs in such cases.
We analytically compute the characteristic threshold for the assisted mechanism given by the critical combined Keldysh parameter. We discuss significant differences between these two types of fields.
For various backgrounds, we present the contributing instantons and perform analytical computations for the corresponding rates treating both fields nonperturbatively.
b) In addition, we also study the case with a nonstatic strong field which gives rise to the assisted dynamical mechanism.
For different strong field profiles we investigate the impact on the critical combined Keldysh parameter.
As an explicit example, we analytically compute the rate by employing the exact reflection points.
The validity of the predictions for both mechanisms is confirmed by numerical computations.
}

\preprint{DESY 17-074}

\begin{document}
\maketitle
\flushbottom

\section{Introduction}
The decay of the quantum vacuum against the production of matter-antimatter pairs in the presence of a spatially uniform, static electric background has been predicted long time ago by Schwinger \cite{Schwinger:1951nm}. The rate in the weak field and weak coupling limit is given by
\begin{align}
\mathcal{R} \simeq \frac{(eE)^2}{(2 \pi)^3} e^{-\mathcal{W}_0},
\label{eq:static-R}
\end{align}
with $e$ being the particle charge and $E$ the electric field strength, respectively.
The prefactor counts for the quantum fluctuation
and the characteristic exponential signals the quantum mechanical tunneling behaviour of this purely nonperturbative\footnote{The perturbative treatment in $E,e$ fails. For a comprehensive overview see \cite{Dunne:2004nc}.} absorptive process.
The damping factor
has the following dependence on the field strength,
\begin{align}
\mathcal{W}_0 = \pi \frac{E_\mathrm{S}}{E},
\label{eq:W0-n=1}
\end{align}
where
$E_\mathrm{S} := m^2/e$, with $m$ being the particle mass, denotes the Schwinger electric field strength beyond which the vacuum decay is expected to become significant.
Despite its elegant and relatively simple derivation in Quantum Electrodynamics (QED), vacuum pair production still could not yet be realised in the laboratory, mainly due to $E_\mathrm{S} = 1.3 \times 10^{18}$ V/m being extremely large.

However, significant efforts, on experimental as well as theoretical side, have been made in recent years that may bring vacuum pair production within experimental reach, see \cite{Dunne:2008kc,Gelis:2015kya} and references therein.
Strong field facilities with field strengths approaching $E \sim 10^{-3} E_\mathrm{S}$ (ELI) \cite{Gies:2008wv}, or even higher, will be available within the next couple of years. Besides, there have been predicted highly promising scenarios in order to enhance the decay rate with time-dependent, inhomogeneous electric fields. One of the most prominent is the so-called assisted mechanism \cite{Dunne:2009gi,Schutzhold:2008pz,Nuriman:2012hn,Akal:2014eua,Otto:2014ssa,Otto:2015gla,Linder:2015vta}.
Overlapping multiple pulsed fields \cite{Bulanov:2010ei} and optimisation of the field profile \cite{Kohlfurst:2012rb,Gonoskov:2013ada,Hebenstreit:2014lra,Hebenstreit:2015jaa,Fillion-Gourdeau:2017uss} can drive the rate higher as well. Besides, there appear remarkable effects if
magnetic field components interact \cite{Tanji:2008ku,Dumlu:2015paa,Copinger:2016llk}.
There are also substantial effects if
thermalisation is involved \cite{ELMFORS1995141,Gies:1999vb,Kim:2008em,Dittrich:2000zu,Gould:2017fve,Akal-therm:2017}. Meanwhile, investigations of analogue condensed matter systems like graphene layers \cite{Katsnelson:2012cz,Zubkov:2012ht,Akal:2016stu,Akal:2017vem,Fillion-Gourdeau:2016izx,Fillion-Gourdeau:2015dga}, semiconductors \cite{PhysRevLett.95.137601,Linder:2015fba} and semimetals
\cite{Abramchuk:2016afc} have become
an active area.

The Schwinger effect is not necessarily restricted to QED. More general, it applies to quantum field theories (QFTs) containing a $U(1)$ gauge field. There may be also generalisations to non-Abelian gauge fields as for instance in Quantum Chromodynamics (QCD) \cite{Casher:1978wy}. Furthermore, in recent years there has been interesting progress on investigations from a much more fundamental point of view within holographic models to study, for instance, the effect of catastrophic instability \cite{Semenoff:2011ng,Bolognesi:2012gr,Sato:2013iua,Hashimoto:2014dza,Fadafan:2015iwa}, effects in confining gauge theories \cite{Kawai:2013xya,Sato:2013pxa,Sato:2013hyw,Sato:2013dwa,Dietrich:2014ala,Hashimoto:2014yya,Ghodrati:2015rta,Akal-holo:2017}, and making even the connection to the recently proposed ER = EPR conjecture \cite{Sonner:2013mba,Fischler:2014ama}.

In this paper we will focus on the enormous enhancement of the decay rate via two mechanisms, the usual assisted mechanism \cite{Schutzhold:2008pz} and the assisted dynamical mechanism. 
Both basically apply if the strong field is superimposed with a much faster varying weak field with the difference that in the 
standard assisted mechanism the former one is assumed to be (locally) static.
A fully analytical treatment for such combined electric backgrounds is in general extremely challenging. We utilise the semiclassical instanton approach to employ the reflection\footnote{It is important to note that the effect of assistance requires in general the nonperturbative treatment of the weak rapid field, see e.g. \cite{Dunne:2009gi}. Such a treatment is realised in the reflection picture.} picture, cf. \cite{Schutzhold:2008pz}.
According to this description, the enhancement can be simply understood as instanton reflections in Euclidean spacetime. We find that many properties and characteristic features
can be explained by means of such reflections even when poles in the instanton plane do not exist or the strong field is inhomogeneous as well. Particularly, we focus on the role of the assisting weak rapid field that, depending on its profile, has remarkable affects on the vacuum decay rate \cite{Nuriman:2012hn,Linder:2015vta,Aleksandrov:2017owa,Torgrimsson:2017pzs}. 
In the assisted dynamical mechanism, we demonstrate how the effect of the weak field alters in the presence of a nonstatic strong field. 
Based on geometric considerations, we explain the origin for significant differences.

The remaining part of the paper is structured as follows:
in Sec.~\ref{sec:instantons} we give a brief discussion of vacuum pair production in the worldline formalism and introduce the stationary instanton equations.
Afterwards, we set up the general form of time-dependent electric backgrounds we will consider in this paper.
In Sec.~\ref{sec:reflection} we work out in detail the instanton reflections.
We first discuss in Sec.~\ref{sec:pole-fields} weak fields leading to a distinct pole structure in the instanton plane.
Sec.~\ref{sec:fields-without-poles} is the first main part of this work dealing with the standard assisted mechanism where we extend the reflection picture to weak fields without poles. We analytically compute the effective reflection points and critical Keldysh parameters for several weak field profiles. Using these results, we then compute the tunneling rates and compare our analytical predictions with numerical computations. 
In Sec.~\ref{sec:critical-inhomogeneity} 
we present the second main part of our work.
We first discuss the impact of the strong field on the critical Keldysh parameter where the former is assumed to be nonstatic giving rise to the assisted dynamical mechanism.
For such a background, we analytically compute the tunneling rate and show the perfect agreement with numerical results.
In Sec.~\ref{sec:conclusion} we summarise and conclude.
In App.~\ref{sec:inhomo-effects} we discuss the effects of temporal background inhomogeneities in more detail.
\section{Instantons}
\label{sec:instantons}
We begin with a brief introduction within the string-inspired \cite{Bern:1991aq} worldline formalism in quantum field theory \cite{Strassler:1992zr,Schubert:2001he}.
Although this technique is completely different from Schwinger's original work, it leads in an elegant way to the correct rate even beyond the weak coupling regime \cite{Affleck:1981bma}.
According to \cite{Schwinger:1951nm}, the probability for vacuum decay against the production of matter-antimatter pairs is given by
\begin{align}
\mathcal{P} = 1 -  e^{ - 2 \mathcal{R} }.
\end{align}
The rate $\mathcal{R}$ is determined by the imaginary part of the one-loop Euler-Heisenberg (EH) effective action \cite{Schwinger:1951nm,Dittrich:1985yb}.
One can analytically continue to Euclidean spacetime and integrate out the high energy degrees of freedom. Such an effective description will then be restricted to $\omega \ll m$ where $\omega$ denotes the background frequency. In addition, the former step will imply
that the decay rate will be determined by the real part of the corresponding effective action. In other words, we would resign the (Lorentzian) time direction via a rotation in the complex plane to obtain a real valued action describing the physics of the quantum vacuum.

In the present work we focus on the weak coupling regime and neglect the contribution coming from the dynamical part of the vector field that is split into $A_\mu + \mathcal{A}_\mu$. The latter stands for the background (vector) potential whose shape will be introduced further below. Introducing the proper time $s$ \cite{Nambu:1950rs} and applying the quantum mechanical path integral representation \cite{Feynman:1950ir,Feynman:1951gn} for the trace using position eigenstates \cite{Schubert:2001he} (paths in spacetime, so-called worldlines), one arrives at the following worldline representation
\begin{align}
\mathcal{R} \simeq  \int_0^\infty \frac{ds}{s}\ e^{ -m^2 s } \oint \mathcal{D}x(\tau)\ e^{ -\int_0^s d\tau \left(\frac{\dot x^2(\tau)}{4} + i e \mathcal{A} \cdot \dot x(\tau)\right) }
\label{eq:rate-weak}
\end{align}
where $\mathcal{D}x$ denotes the path integration measure over the closed worldlines. Here, we may consider for simplifying reasons scalar QED, since the fermion spin does not play a decisive role \cite{Dunne:2004nc}.
Note that $x_\mu(\tau)$ satisfies the boundary condition $x_\mu(0)= x_\mu(s)$.
Rescaling the worldline under $\tau = us$ and performing a saddle-point analysis in $s$, one obtains over all closed paths, now specified by
\begin{align}
  \mathfrak{p} = \{ x_\mu | x_\mu(0) = x_\mu(1) \},
  \label{eq:rescaled-boundary}
\end{align}
the rate
\begin{align}
\mathcal{R} \simeq  \oint \mathcal{D}x(u)\ \sqrt{\frac{2 \pi}{m^2 s_0}} \exp \left( - \mathcal{W} \right).
\end{align}
The stationary point is
\begin{align}
s_0 = \frac{1}{m} \sqrt{ \int_0^1 du\ \dot x^2(u)}.
\label{eq:s_0}
\end{align}
This method is applicable to the fermionic case as well \cite{Dunne:2005sx}.
The latter result is valid in the so-called large mass approximation (LMA),
\begin{align}
  m^2 s_0 \gg 1.
  \label{eq:lma}
\end{align}
The worldline action $\mathcal{W}$ consists of a kinetic part $\mathcal{W}_{\mathrm{kin}}$ and an external part $\mathcal{W}_{\mathrm{ext}}$,
\begin{align}
\mathcal{W} \stackrel{\mathrm{weak}}\simeq \mathcal{W}_{\mathrm{kin}} + \mathcal{W}_{\mathrm{ext}} = m^2 s_0 + i e \int_0^1 du\ \mathcal{A} \cdot \dot x(u).
\label{eq:wl-action-weak}
\end{align}
Note that performing the integrations in the opposite order, i.e. first the integral for $x_\mu(u)$ and then the $s$ integration, will lead to the EH effective action.

The $x_\mu(u)$ integration can be done using the method of steepest descents \cite{Affleck:1981bma}.
Expanding the worldline action in the fluctuations over the (worldline) instanton, i.e. $x_\mu (u) \rightarrow x_\mu(u) + \delta x_\mu(u)$, this leads to
\begin{align}
\mathcal{R} = \frac{1}{\sqrt{\text{Det} M}} e^{ - \mathcal{W}_0},
\label{eq:R-tot}
\end{align}
where
\begin{align}
  \mathcal{W}_0 \equiv \mathcal{W} \left[\mathrm{instanton}\right].
\end{align}
Note that $M_{\mu \nu}$ is the second order variation operator \cite{Affleck:1981bma}.
It is the expression in front of the exponential in \eqref{eq:R-tot} that leads to the given quantum fluctuation prefactor in \eqref{eq:static-R}.
The general form of the prefactor for temporally or spatially inhomogeneous electric fields as a function of one spacetime coordinate has been derived in \cite{Dunne:2006st} evaluating the different determinants from the corresponding integrations, three in total, separately.
Using the Gutzwiller trace formula, it has been shown that all prefactors are encoded in a single determinant that is determined by the monodromy matrix \cite{Dietrich:2007vw}.
Recently, it has been shown in detailed numerical computations that important features in the assisted mechanism are mainly unaffected by the prefactor \cite{Schneider:2016vrl}.
For the present discussions we will only focus on the exponential $e^{-\mathcal{W}_0}$. Detailed studies regarding the corresponding fluctuation prefactors will be presented elsewhere \cite{Akal-pre:2017}.
Using the anti-symmetry of the field tensor $\mathcal{F}_{\mu \nu} = \partial_\mu \mathcal{A}_\nu - \partial_\nu \mathcal{A}_\mu$
one can show that
\begin{align}
  \dot x^2 = \text{constant} \equiv a^2.
  \label{eq:kin-invariant}
\end{align}
Thus, the closed instanton path is determined by the following set of equations \cite{Affleck:1981bma}
\begin{align}
m \ddot x_\mu = i a e \mathcal{F}_{\mu \nu} \dot x_\nu.
\label{eq:eom}
\end{align}
The complete solution including the fluctuation prefactor is exact \cite{Affleck:1981bma,Strobel:2013vza,Gordon:2014aba} for a quadratic action in $x_\mu(u)$ and approximate for actions with higher order dependence.

In the following we consider time dependent backgrounds oriented in $\hat x_3$ direction. The instanton equations \eqref{eq:eom} from above then read as
\begin{equation}
    \begin{split}
    \ddot x_4 &= + \frac{i e a}{m} \partial_4 \mathcal{A}_3 (x_4) \dot x_3,\\
    \ddot x_3 &= - \frac{i e a}{m} \partial_4 \mathcal{A}_3 (x_4) \dot x_4.
    \label{eq:gen-instanton-eqs}
  \end{split}
\end{equation}
Furthermore, to allow only real instanton solutions, we will consider electric backgrounds which will be described by analytic even functions in Minkowskian time $t$.
Note that in general those can be complex as well, e.g. cf. \cite{Dumlu:2011cc}.
The Euclidean vector potential ($x_4 = i t$) can be written in the form
\begin{align}
   \mathcal{A}_3(x_4) = -i E F(x_4),
\end{align}
where $F$ is assumed to be an odd real function. The complex $i$ in front guarantees that the instanton equations \eqref{eq:gen-instanton-eqs} will have real, closed solutions $x_\mu$.

The effect of temporal inhomogeneities in vacuum pair production has been studied extensively in the literature, e.g. cf.
\cite{Popov:1971iga,PhysRevD.2.1191,Marinov:1977gq,Dietrich:2003qf,Dunne:2005sx,Gies:2005bz,Hebenstreit:2008ae,Schutzhold:2008pz}.
It has been shown that those
can enhance the production rate even with field
strengths far below $E_\mathrm{S}$.
The intuitive physical picture is that the vacuum energy gap, which has to be overcome by the virtual pair, can effectively be lowered by the additional inhomogeneity in spacetime. This is in close analogy to atomic ionisation processes, e.g. cf. \cite{Fedorov2016}. A more detailed discussion can be found in App.~\ref{sec:inhomo-effects}.

We consider a linearly combined electric background of the form
\begin{align}
  \pmb{E}(t) = \left( E f(t) + \tilde E g(t) \right) \hat x_3.
  \label{eq:gen-E-field}
\end{align}
The weak rapid field, $\propto \tilde E$ with frequency $\tilde \omega$, is described by an analytic function $g(t)$, whereas the
the strong slow field, $\propto E$ with frequency $\omega$, is characterised by a function $f(t)$. The fields shall satisfy $E_\mathrm{S} \gg E \gg \tilde E$ and $m \gg \tilde \omega \gg \omega$. Both functions $f,g$ are even in Minkowski time $t$. The corresponding gauge potential
after analytic continuation to Euclidean spacetime reads
 \begin{align}
    \mathcal{A}_3(x_4) = -i E F(x_4)  - i \tilde E G(x_4).
    \label{eq:gen-A-potential}
 \end{align}
Here $F(x_4)$ and $G(x_4)$ denote the corresponding odd functions obtained after the integration of $f(t)$ and $g(t)$, respectively. Inserting the derivative of the gauge potential
\begin{align}
 \partial_4 \mathcal{A}_3(x_4) = -i \left( E F^\prime(x_4) + \tilde E G^\prime(x_4) \right)
 \end{align}
into the instanton equations, we find the following nonlinearly coupled system of differential equations
 \begin{equation}
    \begin{split}
     \ddot x_4 &= + \frac{e a E}{m}  \left( F^\prime(x_4) + \epsilon G^\prime(x_4) \right) \dot x_3,\\
     \ddot x_3 &= - \frac{e a E}{m}  \left( F^\prime(x_4) + \epsilon G^\prime(x_4) \right) \dot x_4,
     \label{eq:x4-x3ddot}
   \end{split}
 \end{equation}
 where $\epsilon := \tilde E / E$ has been introduced as a dimensionless parameter.

\section{Reflections}
\label{sec:reflection}
For the seek of convenience we first introduce the following dimensionless Keldysh parameters,
\begin{align}
  \gamma = \frac{m \omega}{e E},\qquad
  \tilde \gamma = \frac{m \tilde \omega}{e E}.
  \label{eq:Keldyshs}
\end{align}
The strong field parameter $\gamma$ interpolates between the adiabatic, nonperturbative tunneling regime ($\gamma \ll 1$) and the anti-adiabatic, perturbative multi-photon regime ($\gamma \gg 1$) \cite{Ringwald:2001ib}.
A background composed of only one inhomogeneous field with $\gamma > 0$ gives rise to the dynamical (Schwinger) mechanism without any assistance. Such a strong field in the presence of an additional assisting weak field, the assisted dynamical mechanism, 
will be discussed in detail in Sec.~\ref{sec:critical-inhomogeneity}.
The second parameter $\tilde \gamma$ in \eqref{eq:Keldyshs} is usually called the combined Keldysh parameter that involves the strong field amplitude $E$ and the weak field frequency $\tilde \omega$.
For $\tilde \gamma \ll 1$ the usual Schwinger mechanism is approached. Here, $\tilde \gamma \gg 1$ does not correspond to a pure perturbative multi-photon weak field.
It involves both multi-photons with the weak field frequency $\tilde\omega$ as well
as a dependence on the non-perturbative strong field strength $E$, see e.g. \cite{Schutzhold:2008pz}.
We will see that in cases where the weak field $\propto \tilde E$ possesses a distinct pole structure, this parameter\footnote{We will discuss in detail that weak fields without a distinct pole structure lead in general to an additional $\epsilon$ dependence of the vacuum pair production rate even $\epsilon$ is taken to be very small, cf. Sec.~\ref{sec:fields-without-poles}.} becomes the main quantity in the standard assisted mechanism.
Now, the second equation in \eqref{eq:x4-x3ddot} can be integrated as
\begin{align}
    \dot x_3 &= - \frac{e a E}{m}  \left( F(x_4) + \epsilon G(x_4) \right).
    \label{eq:x3dot}
\end{align}
Using the kinematic invariant $a = \sqrt{\dot x_3^2 + \dot x_4^2}$, we may write the equation for $\dot x_4$. However, the integral is generally difficult to solve analytically. An effective reflection picture \cite{Schutzhold:2008pz} will provide a simplified way to tackle it, cf. e.g. \cite{Schneider:2014mla}.

Namely, since we are interested in the limit $\epsilon \ll 1$, we may omit the second term in \eqref{eq:x3dot}. However, going back to the original instanton equations \eqref{eq:x4-x3ddot}, this is allowed as long as $G^\prime(x_4)$ is sufficiently small. As soon as it becomes very large, which happens for sure at some pole, determined by
\begin{align}
  1/g(x_4^\mathrm{p}) = 0,
\end{align}
one expects a substantial contribution from this term.
For the moment, let us assume that the weak field has a distinct pole structure.
Note that due to symmetry reasons, which apply for the specific field configurations considered here, it is sufficient to do the present analysis with respect to the pole on the positive Euclidean time axis, i.e. $+x_4^\mathrm{p}$. Thus, the weak field pole acts as an infinite wall where the instanton will be reflected with a non-vanishing velocity $\dot x_4$.
Away from these Euclidean mirrors, as we call such reflection points,
we can neglect the second terms in the brackets and integrate the instanton equations approximately as
\begin{align}
  \begin{split}
  \dot x_3 &\approx -  a \frac{\omega}{\gamma} F(x_4),\\
  \dot x_4 &\approx \pm a \sqrt{ 1 - \left( \frac{\omega}{\gamma} F(x_4) \right)^2 }.
  \label{eq:x3dot-x4dot}
  \end{split}
\end{align}
Since in the reflection points we will find $\dot x_4 \neq 0$, the invariant $a$ is expected to be modified.
For $\epsilon \ll 1$ we then write the external part in $\mathcal{W}$ as
\begin{align}
  \mathcal{W}_\mathrm{ext} = i e \int_0^1 du\ \dot x \cdot \mathcal{A}
  \approx e E \int_0^1 du\ \dot x_3(u) F(x_4(u)).
\end{align}
Due to the instanton symmetry
\begin{align}
  \begin{split}
  x_3 \rightarrow - x_3,\quad
  x_4 \rightarrow - x_4.
\end{split}
\end{align}
we may use the relation
\begin{align}
  \int_0^{1/4} du\ \dot x_3 = \int_0^{x_4^\mathrm{c}} dx_4\ \frac{\dot x_3}{\dot x_4}
\end{align}
where $x_4^\mathrm{c}$ is the closing point at the intersection between the first and second quarter (at $x_3 = 0$),
since the derivatives $F^\prime,G^\prime$ are even functions.
So, we proceed with the expression for the external part of the stationary worldline action,
\begin{align}
  \mathcal{W}_{0,\mathrm{ext}} = 4 e E \int_0^{x_4^\mathrm{c}} dx_4\ \frac{  - \frac{\omega}{\gamma} F(x_4) F(x_4) }{ \sqrt{ 1 - \left( \frac{\omega}{\gamma} F(x_4) \right)^2 } },
  \label{eq:gen-ext}
\end{align}
where $x_4^\mathrm{c}$ is the closing point in which the instanton  has to be closed.
Now, here comes the question
about the value for the closing point.
One may think about the critical point
\begin{align}
  \frac{\omega}{\gamma} F(x_4^*) = 1,
  \label{eq:x4star}
\end{align}
that we can read off directly from the denominator.
This point, however, corresponds to $\dot x_4 = 0$ which cannot be allowed in the reflection picture with poles in the instanton plane present.
But we know, if the instanton is reflected at the weak field's pole, the path has to be closed there as well.
\begin{figure}[h]
  \centering
\includegraphics[width=.98\textwidth]{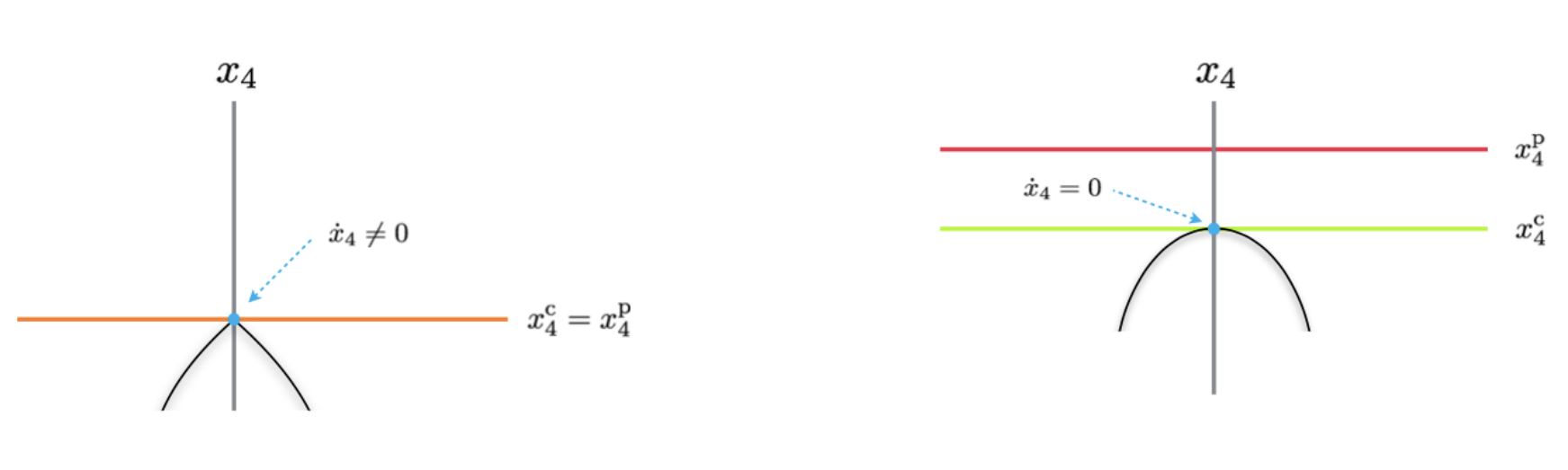}
\caption{Comparison of assisted (left) and standard Schwinger mechanism (right) in temporally inhomogeneous electric background.
The former is characterised by the additional instanton reflection due to the superposition with a weak rapid field.}
\label{fig:reflection-closing}
\end{figure}
So, in case of reflection, means if the latter field assists the vacuum decay,
we have to set
\begin{align}
x_4^\mathrm{c} \overset{!}{=} x_4^\mathrm{p}.
\label{eq:ref-cond}
\end{align}
Otherwise, if
\begin{align}
x_4^\mathrm{c} = x_4^*,
\end{align}
the instanton path will be closed much earlier, means no resizing due to reflections. This behaviour is schematically illustrated in Fig.~\ref{fig:reflection-closing}. In the left panel we have the reflection picture where $x_4^\mathrm{c} = x_4^\mathrm{p}$. However, in the right panel the instanton is closed before reaching the pole, i.e. $x_4^\mathrm{c} < x_4^\mathrm{p}$.
The reflection condition \eqref{eq:ref-cond} shows that the profile of the weak high-frequent field, which basically determines the position of the pole $x_4^\mathrm{p}$, is crucial for the assisted vacuum decay.
Moreover, the profile of the strong slow field, $f$, determines the form of the integrand in \eqref{eq:gen-ext}.
Hence, the interplay between both contributions will be relevant.
Now, we still need to find $a$ in order to compute the kinetic term
\begin{align}
\mathcal{W}_{0,\mathrm{kin}} = m a,
\end{align}
see \eqref{eq:s_0} and \eqref{eq:kin-invariant}. In fact, we will see that demanding the closing point $x_4^\mathrm{c}$ to be placed in $x_4^\mathrm{p}$, cf. \eqref{eq:ref-cond}, will lead to substantial
modifications of the invariant $a$.
We first rewrite the integration measure and set $x_4^\mathrm{c} = x_4^\mathrm{p}$ in order to find
\begin{align}
  \frac{1}{4} = \int_0^{1/4} du = \int_0^{x_4^\mathrm{p}} \frac{dx_4}{\dot x_4}.
\end{align}
From this equality we can determine the kinematic invariant after inserting $\dot x_4$ from \eqref{eq:x3dot-x4dot}, that gives us the remaining relation
\begin{align}
  a = 4 \int_0^{x_4^\mathrm{p}} dx_4\ \frac{1}{ \sqrt{ 1 - \left( \frac{\omega}{\gamma} F(x_4) \right)^2 } }.
  \label{eq:gen-a}
\end{align}
Altogether, combining \eqref{eq:gen-ext} and \eqref{eq:gen-a}, we find the following expression
\begin{align}
  \mathcal{W}_0
  = 4 m \int_0^{x_4^\mathrm{p}} dx_4\ \sqrt{ 1 - \left( \frac{\omega}{\gamma} F(x_4) \right)^2 }.
  \label{eq:gen-W}
\end{align}

\section{Assisted mechanism: fields with poles}
\label{sec:pole-fields}
Let us first begin by illustrating the previous modifications for known examples in the literature.
We first assume the strong field, $\propto E$, to be static, i.e.
\begin{align}
 f(t) = 1,\qquad
 F(x_4) = x_4.
\end{align}
A background of this type will result in the usual assisted mechanism \cite{Schutzhold:2008pz}.
Note that for $\gamma \gg 1$
the effect of such an assistance vanishes if $\tilde\gamma$ remains relatively small. This is due to a threshold value for the latter which is characteristic for this mechanism \cite{Schutzhold:2008pz,Linder:2015vta}. In such a case the strong field alone will be sufficient to induce the enhancement which corresponds to the usual dynamical mechanism in temporally inhomogeneous electric fields. In Sec.~\ref{subsec:effects-strong-mode} we will discuss the generalisation with an additional assisting field (assisted dynamical mechanism).
However, let us first focus on the static strong field case for which the integral in \eqref{eq:gen-a} can now be solved analytically as
\begin{align}
  a = 4 \int_0^{x_4^\mathrm{p}} dx_4\ \frac{1}{ \sqrt{ 1 - \left( \frac{\omega}{\gamma} x_4 \right)^2 } }
  = 4 \frac{\gamma}{\omega} \mathrm{arcsin} \left( \frac{\omega}{\gamma} x_4  \right) \bigg|_{x_4=0}^{x_4^\mathrm{p}}
\end{align}
leading to the following kinematic invariant
\begin{align}
  a = 4 \frac{\gamma}{\omega} \mathrm{arcsin} \left( \frac{\omega}{\gamma} x_4^\mathrm{p} \right)
  \label{eq:mod-a}
\end{align}
that depends on the pole $x_4^\mathrm{p}$ we have not specified yet.
The latter expression already signals the appearance of the mentioned critical value for $\tilde \gamma$ depending on $x_4^\mathrm{p}$, since $\omega/\gamma = \tilde \omega/\tilde \gamma \leq 1/x_4^\mathrm{p}$.
So, the instanton path in the right half plane, i.e. $u \in [-1/4,1/4]$, see \eqref{eq:rescaled-boundary}, is simply an arch-like curve. With this the LMA condition \eqref{eq:lma} becomes
\begin{align}
  \frac{m}{\omega} 4 \gamma \mathrm{arcsin} \left( \frac{\omega}{\gamma} x_4^\mathrm{p} \right) \gg 1.
\end{align}
After the integration of \eqref{eq:gen-instanton-eqs} and inserting the modified invariant \eqref{eq:mod-a}, we obtain
\begin{align}
  \begin{split}
  x_4(u) &= \frac{m}{e E}  \sin \left(4 u \mathrm{arcsin}\left(\frac{\omega  x_4^\mathrm{p}}{\gamma }\right)\right),\\
  x_3(u) &= \frac{m}{e E}  \cos \left(4 u \mathrm{arcsin}\left(\frac{\omega  x_4^\mathrm{p}}{\gamma }\right)\right) - \mathcal{C}.
  \label{eq:computed-instanton}
  \end{split}
\end{align}
The closed path is realised by the integration constant
\begin{align}
  \mathcal{C} = x_3(u=\pm 1/4) =  \frac{m}{e E} \cos \left( \mathrm{arcsin}\left(\frac{\omega  x_4^\mathrm{p}}{\gamma }\right)\right).
\end{align}
It shifts the instanton along the $\hat x_3$ axis to guarantee the condition $x_3(u=\pm 1/4) = 0$. Evaluating the stationary worldline action with \eqref{eq:gen-W} gives
\begin{align}
  \begin{split}
  \mathcal{W}_0
  &= 4 m \int_0^{x_4^\mathrm{p}} dx_4\ \sqrt{ 1 - \left( \omega x_4 / \gamma \right)^2 }\\
  &=
  \frac{2m}{\omega} \left( x_4^\mathrm{p} \omega  \sqrt{1-\left(\frac{x_4^{\mathrm{p}} \omega}{\gamma}\right)^2}+\gamma  \mathrm{arcsin}\left(\frac{x_4^\mathrm{p} \omega }{\gamma }\right) \right).
\end{split}
  \label{eq:computed-W}
\end{align}
\subsection{Examples}
In the following we demonstrate the reflections for two different backgrounds as a superposition of a strong static field
and a weak rapid field which has poles in the instanton plane.
\subsubsection{Weak Sauter}
\begin{figure}[h!]
  \centering
\includegraphics[width=.5\textwidth]{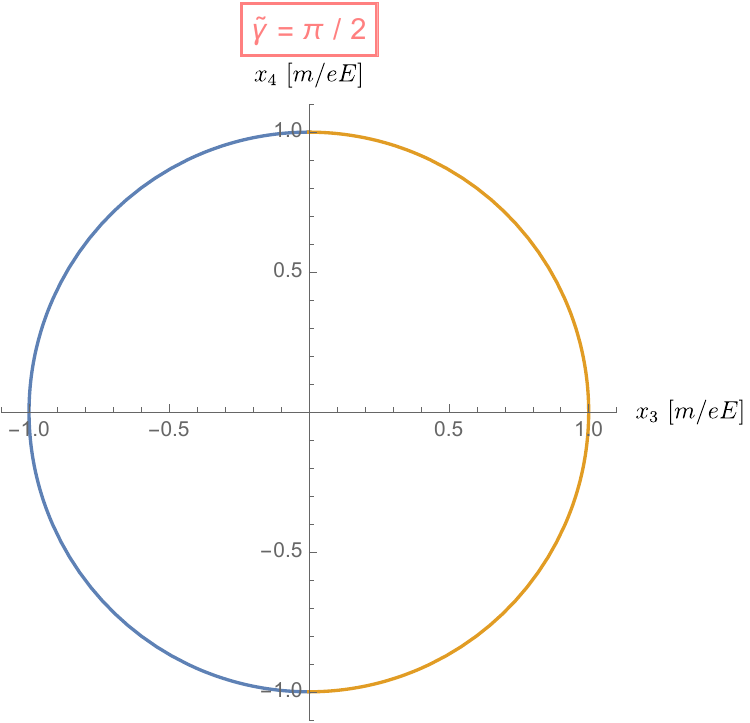}\\
\includegraphics[width=.325\textwidth]{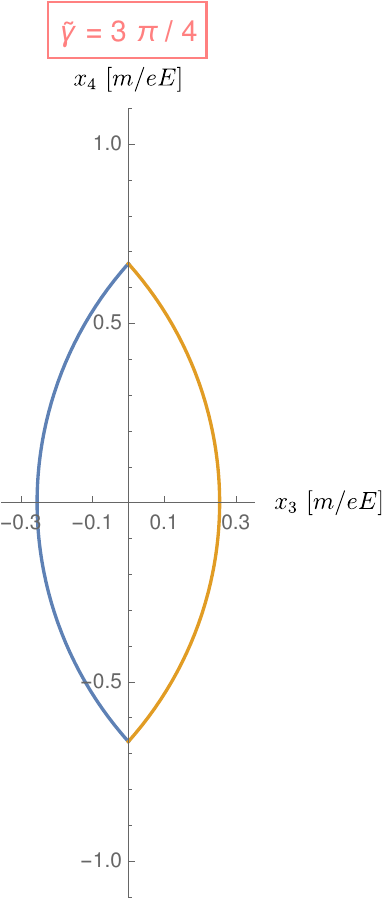}
\includegraphics[width=.26\textwidth]{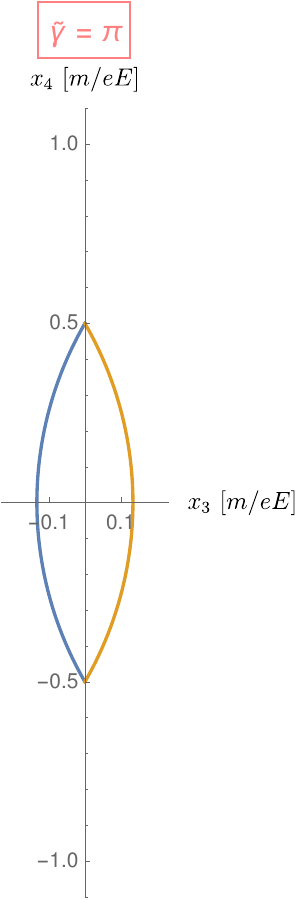}
\caption{Instanton path for an electric background as superposition of a strong static and weak Sauter field for different $\tilde\gamma$ given in the plot labels.}
\label{fig:instantons-sauter}
\end{figure}
We begin with a weak Sauter field, i.e.
\begin{align}
  g(t) = \mathrm{sech}^2(\tilde \omega t),\qquad
  G(x_4) = \frac{\tan(\tilde \omega x_4)}{\tilde \omega}.
  \label{eq:sauter}
\end{align}
The pole structure for this field is of multitype. However, the (first) relevant pole for $g(x_4) = \sec^2(\tilde \omega x_4)$ is located in
\begin{align}
  x_4^\mathrm{p} = \frac{\pi}{2 \tilde \omega},
  \label{eq:sauter-pole}
\end{align}
which leads to the following invariant
\begin{align}
  a = 4 \frac{m}{e E} \mathrm{arcsin} \left( \frac{\pi}{2 \tilde \gamma} \right).
\end{align}
The above result indicates that the combined Keldysh parameter has to satisfy $\tilde \gamma > \pi/2$, otherwise we would find $a \in \mathbb{C}$, means no closed instanton path. Below this critical value,
\begin{align}
  {\tilde \gamma}^\mathrm{crit} = \frac{\pi}{2},
  \label{eq:sauter-gamma-crit}
\end{align}
there will be no  effect of the weak field and we are left with the strong static field contribution, see \cite{Schutzhold:2008pz}.
Note that even the weak mode does not contribute, there will be a non-zero, but small decay rate due to the non-perturbative strong mode.
Here, the strong field can be assumed as locally static as it is seen by the weak but rapid field.
It is clear that this can only be expected for the case of a large frequency difference. As soon as $\gamma$ approaches larger values, it will have substantial effects below the critical value ${\tilde \gamma}^\mathrm{crit}$, see Sec.~\ref{sec:critical-inhomogeneity}.

Coming back to the present example,
inserting the pole $x_4^\mathrm{p}$ into the modified solutions \eqref{eq:computed-instanton}, we plot the instanton path for different frequencies $\tilde \omega$ as shown in Fig.~\ref{fig:instantons-sauter}. The size of the instanton decreases with larger $\tilde \gamma$. This shrinking will then increase the rate
$\mathcal{R}$, since $\mathcal{W}_0$ decreases as soon as the size of the instanton is reduced. Such lens shaped instantons also apply if the strong field is a spatially inhomogeneous Sauter field \cite{Schneider:2014mla}.
After inserting the pole into the solution \eqref{eq:computed-W} we find \cite{Schutzhold:2008pz}
\begin{align}
  \mathcal{W}_0
  = \frac{m^2}{e E} \left( \frac{\pi}{2 {\tilde \gamma}^2} \sqrt{4 {\tilde \gamma}^2 - \pi^2}  + 2 \mathrm{arcsin}\left( \frac{\pi}{2 \tilde \gamma} \right) \right).
\end{align}
Alternatively, one would get this result by plugging the instanton solution from above into \eqref{eq:gen-W} and integrating over $u \in [0,1/4]$.

\subsubsection{Weak Lorentzian}
\begin{figure}[h!]
  \centering
\includegraphics[width=.5\textwidth]{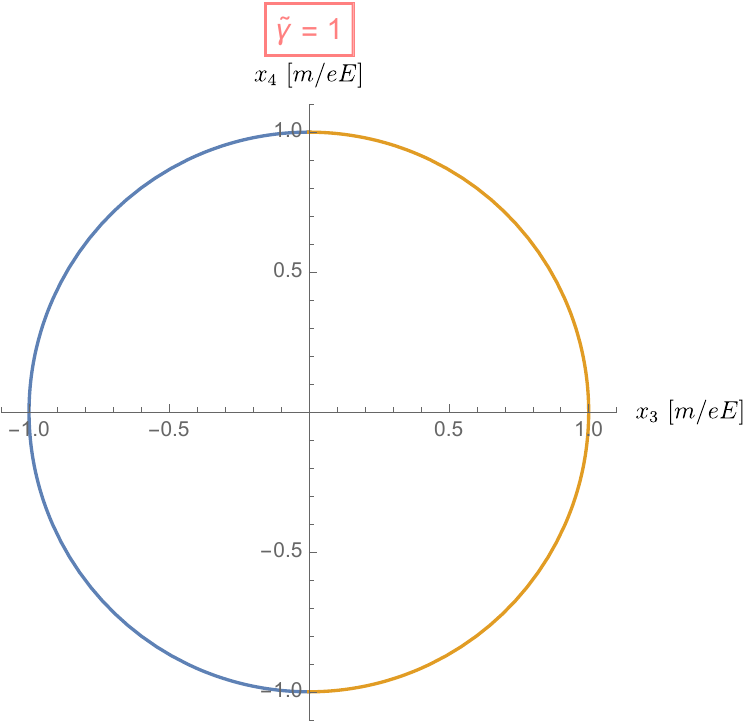}\\
\includegraphics[width=.35\textwidth]{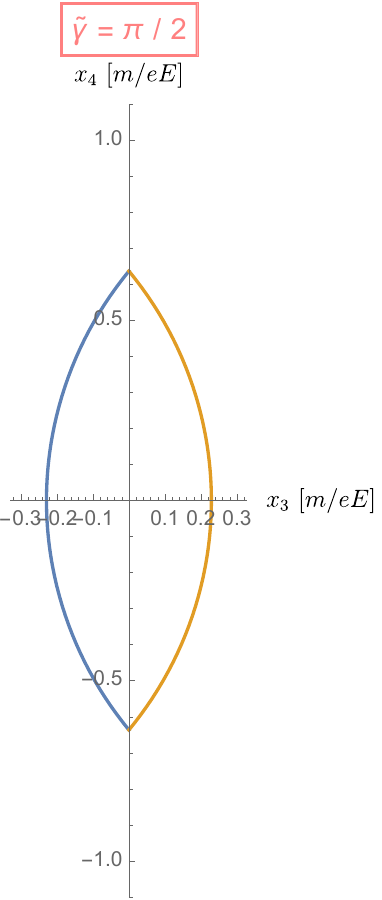}
\includegraphics[width=.27\textwidth]{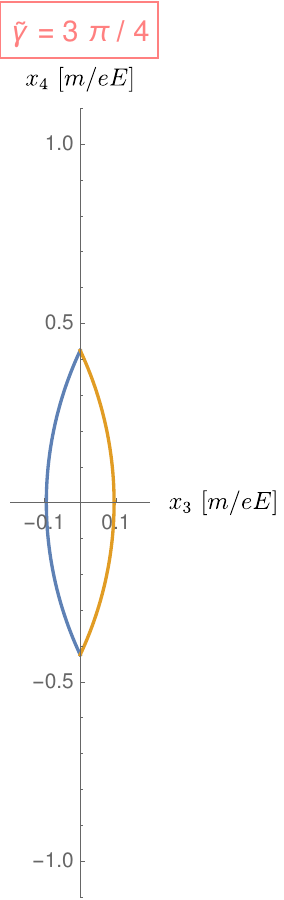}
\caption{Instanton path for an electric background as superposition of a strong static and weak Lorentzian field for different $\tilde\gamma$ given in the plot labels.}
\label{fig:instantons-lorentzian}
\end{figure}
\begin{figure}[h!]
  \centering
\includegraphics[width=.98\textwidth]{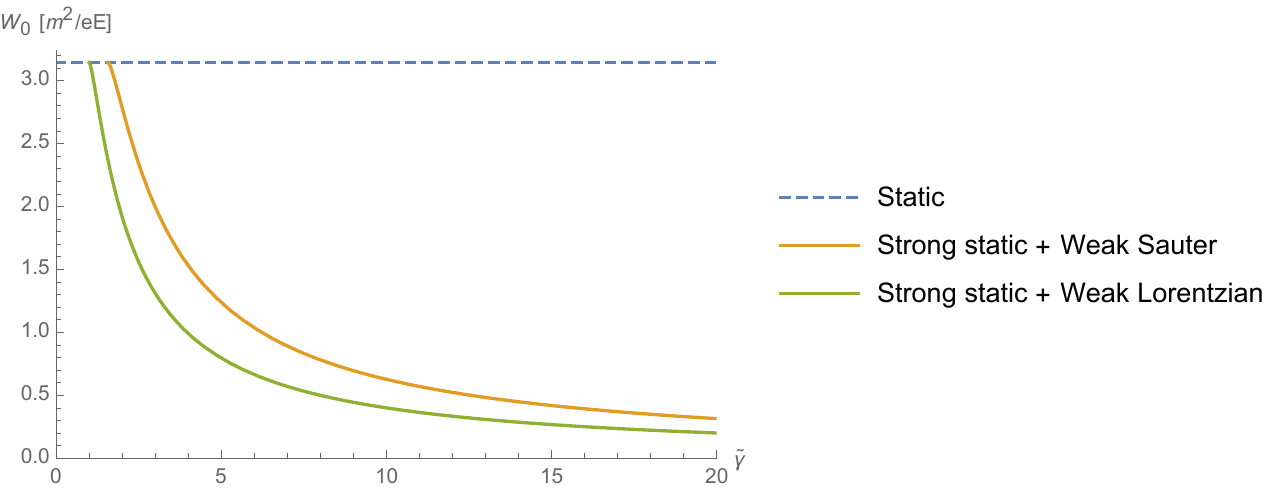}
\caption{Stationary worldline action $\mathcal{W}_0\ [\frac{m^2}{eE}]$ for the electric background as superposition of a strong static and weak Sauter/Lorentzian (yellow/green curve) field. The horizontal blue line corresponds to the static field case with $\mathcal{W}_0\ [\frac{m^2}{eE}] = \pi$.}
\label{fig:W0-sauter-lorentzian}
\end{figure}

Another example we can compute is a weak Lorentzian field described by
\begin{align}
  g(t) = \frac{1}{\left( 1 + (\tilde \omega t)^{2} \right)^{3/2}},\qquad
  G(x_4) = \frac{1}{\tilde \omega}\frac{\tilde \omega x_4}{\sqrt{1 - (\tilde \omega x_4)^2}}.
  \label{eq:lorentzian}
\end{align}
The function $g(x_4) = 1/\left( 1 - (\tilde \omega x_4)^{2} \right)^{3/2}$ leads to
\begin{align}
  x_4^\mathrm{p} = \frac{1}{\tilde \omega}.
  \label{eq:lorentzian-pole}
\end{align}
Except the factor $\pi/2$, it resembles the case before. Therefore, similar results are expected. However, one should remark that not the visually indistinguishable bell-shaped profile is responsible for this similarity\footnote{It is the distinct pole structure of the field which is responsible for such a similarity.}, cf. e.g. \cite{Linder:2015vta}. This will be discussed in detail in Sec.~\ref{sec:fields-without-poles}. The modified invariant reads
\begin{align}
  a = 4 \frac{m}{e E} \mathrm{arcsin} \left( \frac{1}{\tilde \gamma} \right)
\end{align}
leading to
\begin{align}
  {\tilde \gamma}^\mathrm{crit} = 1,
  \label{eq:lorentzian-gamma-crit}
\end{align}
cf. e.g. \cite{Schneider:2016vrl}.
Consequently, the weak Lorentzian field will start to contribute much earlier compared to the previous case. This behaviour is illustrated in Fig.~\ref{fig:instantons-lorentzian}.
Inserting the pole into \eqref{eq:computed-W}, we find the stationary worldline action \cite{Schneider:2016vrl}
\begin{align}
  \mathcal{W}_0
  = \frac{m^2}{e E} \left( \frac{2}{{\tilde \gamma}^2} \sqrt{{\tilde \gamma}^2 - 1}  + 2 \mathrm{arcsin}\left( \frac{1}{\tilde \gamma} \right) \right).
\end{align}
The comparison of $\mathcal{W}_0$ for both fields is shown in Fig.~\ref{fig:W0-sauter-lorentzian}. The difference with respect to the critical threshold is clearly observable.
Despite the relative difference between the curves, we observe that both behave similarly.
Identical results will apply, if we increase the frequency of the Sauter field by a factor $\frac{\pi}{2}$. This we can already observe directly in Fig.~\ref{fig:W0-sauter-lorentzian}. For instance, in case of the Lorentzian field the value $\approx 1.8$ is reached at $\tilde \gamma \approx 2$. The same result applies for the Sauter field at $\tilde \gamma \approx 3.2$, the mentioned factor $\frac{\pi}{2}$.
Now, suppose we consider a more general Lorentzian described by
\begin{align}
  g(t) = \frac{1}{\left( 1 + (\tilde \omega t)^{2} \right)^{d/2}},
\end{align}
where $d \in \mathbb{N}$. Apparently, we will obtain the same pole as before, i.e the inverse of $\tilde \omega$.
It turns out that in the relevant regime $\epsilon \ll 1$ it is sufficient to have only the pole. Namely, after rotation in the complex plane the variation of $d$ will have negligible effects on the rate.
It is the reflection at the pole \eqref{eq:lorentzian-pole} that predominantly determines the strength of the enhancement.

\subsection{Effects of sub-cycle structure}
\begin{figure}[h!]
  \centering
\includegraphics[width=.4\textwidth]{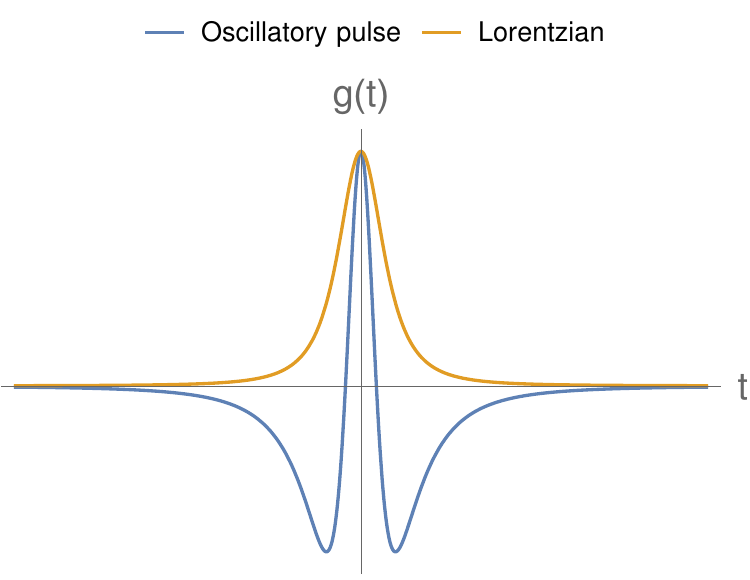}
\hfill%
\includegraphics[width=.55\textwidth]{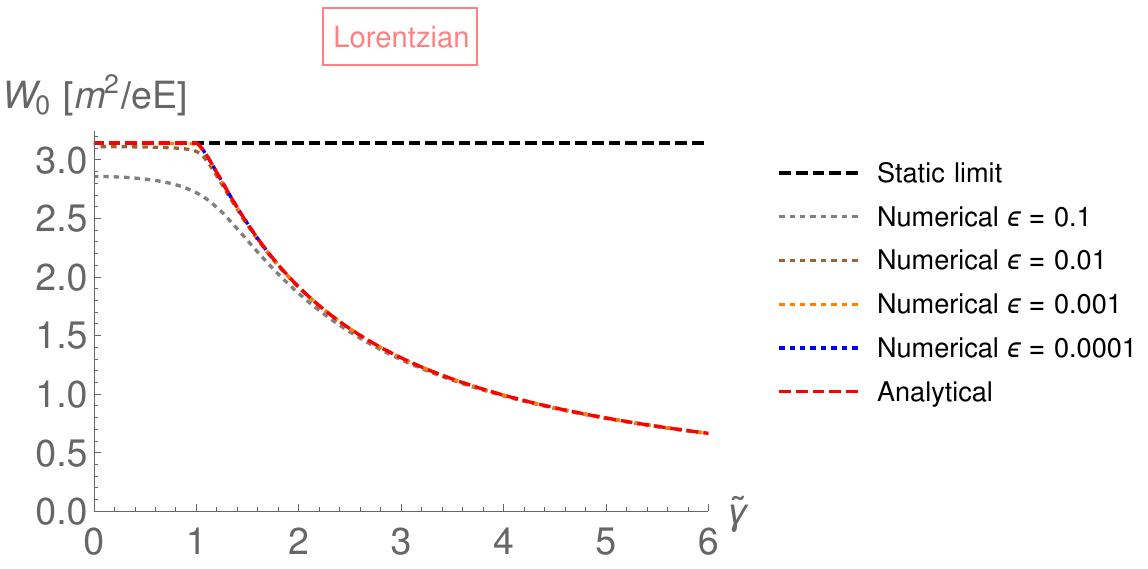}
\hfill%
\includegraphics[width=.36\textwidth]{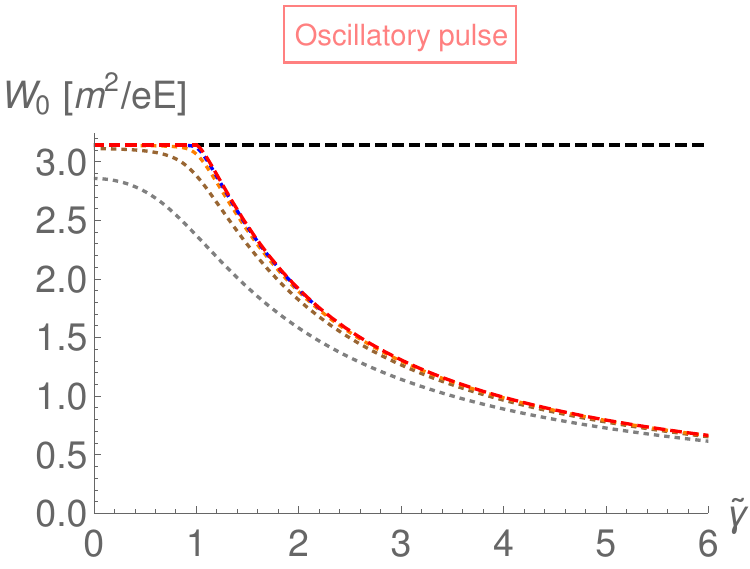}
\hspace*{2cm}
\includegraphics[width=.32\textwidth]{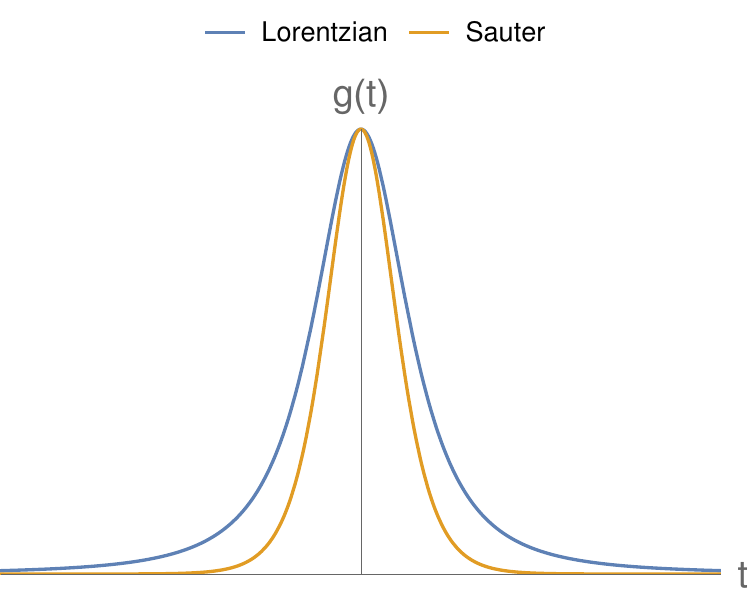}
\caption{Analytically and numerically computed stationary worldline action for the oscillatory pulse \eqref{eq:oscpulse} with $N=1$ (bottom-left) and Lorentzian \eqref{eq:lorentzian} (top-right) profile. The values for $\epsilon$ are given in the plot legend. The function $g(t)$ for the oscillatory pulse (blue) and Lorentzian field (yellow) is plotted in the top-left panel. The comparison between the Sauter and Lorentzian field is depicted in the bottom-right panel where the frequency for the former one is multiplied by $\pi/2$. This leads to the same reflection point
$x_4^\mathrm{p}$, cf. \eqref{eq:sauter-pole} and \eqref{eq:lorentzian-pole}.
Even the profiles look different, the corresponding stationary actions $\mathcal{W}_0$ are identical for $\epsilon \lesssim 10^{-3}$.}
\label{fig:oscpulse-lorentzian}
\end{figure}

In the following we discuss the possible impact of an additional oscillatory sub-structure. This situation is reflected in laser setups where the field consists of a very short wave packet\footnote{Despite the fact that those are electromagnetic pulses, a pure electric field of this type is still a good approximation. It can be realised, for instance, through a collision of two counter-propagating pulses equal in their (linear) polarisation and intensity.}. Hence, the question is, how will the rate be influenced? Let us assume a simple oscillatory pulse described by
\begin{align}
  g(t) = \frac{1 - 3 (\tilde \omega t)^2 - 2 (N\tilde \omega t)^2}{(1 + (N \tilde\omega t)^2)^{5/2}},\qquad
  G(x_4) = \frac{1}{\tilde\omega}\frac{\tilde \omega x_4  + (\tilde \omega x_4)^3}{(1 - (N \tilde \omega x_4)^2)^{3/2}},\quad N \geq 1.
  \label{eq:oscpulse}
\end{align}
Its comparison with the Lorentzian profile \eqref{eq:lorentzian} is depicted in Fig.~\ref{fig:oscpulse-lorentzian} (top-left) for $N=1$. For this we find the generalised Lorentzian pole
\begin{align}
  x_4^\mathrm{p} = \frac{1}{N \tilde \omega}.
\end{align}
For sufficiently small $\epsilon$ the sub-cycle structure is expected to be irrelevant, in some sense in analogy to the considerations before. It is the pole structure of the bell-shaped Lorentzian that regulates the strength of the enhancement. 

Such a behaviour is indeed confirmed in Fig.~\ref{fig:oscpulse-lorentzian}.
In the top-right panel we have plotted again $\mathcal{W}_0$ for the previous Lorentzian field, but now comparing it with numerical computations for different $\epsilon$. For $\epsilon < 0.01$ the analytical prediction and the numerical result are almost identical. Only for $\epsilon = 0.01$ there appears a notable difference. Doing the same computation for the oscillatory pulse, \eqref{eq:oscpulse}, we identify a similar picture. Despite the fact, that for larger
$\epsilon$ one observes a stronger deviation which is completely plausible, since the total effective field strength becomes larger,
we observe that the prediction agrees very well with the numerical results\footnote{To compute $\mathcal{W}_0$ numerically we transform the worldline action as in \cite{Linder:2015vta} making use of the underlying structure of the instanton equations \eqref{eq:gen-instanton-eqs}. The integration is done with the standard $\mathtt{Mathematica}$ routine.}. Hence, having obtained the exact reflection point turns out to be sufficient to predict $\mathcal{W}_0$ if $\epsilon \ll 1$. Note, that for $\tilde \gamma \gg 1$, both curves converge, independent of $\epsilon$.

The previous observations are interesting, since pulsed fields are described by an appropriate oscillatory function multiplied with some bell-shaped function.
Usually such envelope functions have poles which are closer to the origin than the effective reflection points for (infinitely extended) oscillating fields, as the sinusoidal field, see Sec.~\ref{sec:fields-without-poles}. Thus, it is exactly the latter pole originating from the envelope function which will predominantly determine the reflection of the instanton. Note that an envelope function which can model such a pulsed field may also exist without a distinct pole structure. An example is the Gaussian field studied in Sec.~\ref{sec:fields-without-poles}.

Hence, at least for $\epsilon \ll 1$, we expect that the assistance is mainly determined by the pole structure of the envelope function and not by the encased oscillatory structure.
However, this strictly applies for the total tunneling rate, cf. e.g. \cite{Aleksandrov:2017owa}. 
Differences in the momentum spectrum due to the inner sub-cycle structure would still be visible.
Namely, the latter can be very decisive,
basically in form of quantum interference effects, cf. e.g. \cite{Hebenstreit:2009km,Akkermans:2011yn,Dumlu:2011rr,Orthaber:2011cm,Akal:2014eua}. Moreover, note that those features will substantially change for fields which do not fulfil the symmetry properties we have assumed for our studies.

The critical pole for the Lorentzian field is reached if we multiply the frequency of the Sauter field by a factor $\frac{\pi}{2}$. For a sufficiently small $\epsilon$, say $\epsilon = 10^{-3}$, the rate is identical for both cases, which is also confirmed by numerical computations.
However, looking on the field profiles with this frequency ratio, one observes a clear difference, cf. Fig.~\ref{fig:oscpulse-lorentzian} (bottom-right).
This is similar to the previous situation with the oscillatory pulse. Despite the visual differences in Minkowski spacetime, we find identical results due to same critical points in the instanton plane.

\section{Assisted mechanism: fields without poles}
\label{sec:fields-without-poles}
So far we have illustrated the reflection picture for known setups with a distinct pole structure in the instanton plane.
The assisted mechanism operates more general, i.e. also for inhomogeneous weak fields with a completely different profile \cite{Nuriman:2012hn,Akal:2014eua,Otto:2014ssa,Hebenstreit:2014lra,Li:2014psw,Otto:2015gla,Linder:2015vta,Torgrimsson:2017pzs}.
A well considered candidate is the sinusoidal field.
It is extended to infinity and does not have poles as in the previous examples.
However, as we have seen, instanton reflections turned out to be the main mechanism behind assistance.
This brings us to the first main goal of this paper.
In the following, after a brief motivation and some basic observations, we will try to find out analogue effective reflectors in order to employ the above picture even in the absence of poles.
\subsection{Motivation}
\begin{figure}[h!]
  \centering
\includegraphics[width=.32\textwidth]{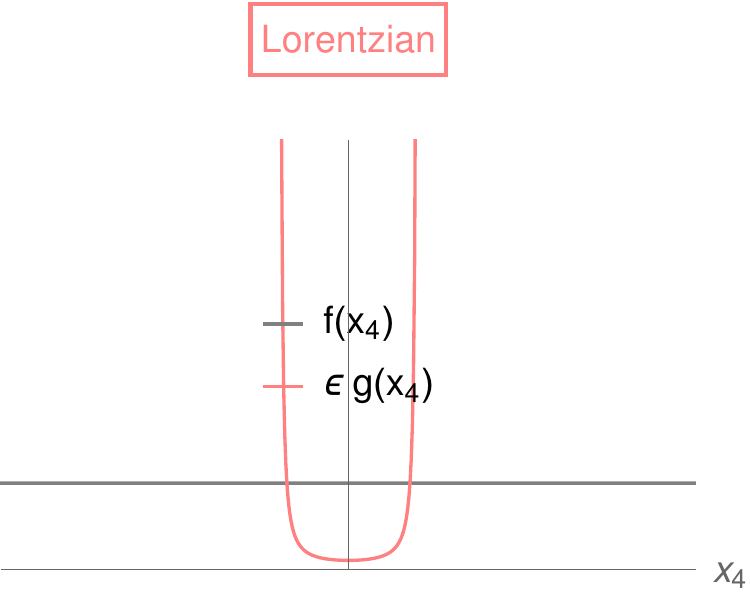}
\hfill%
\includegraphics[width=.32\textwidth]{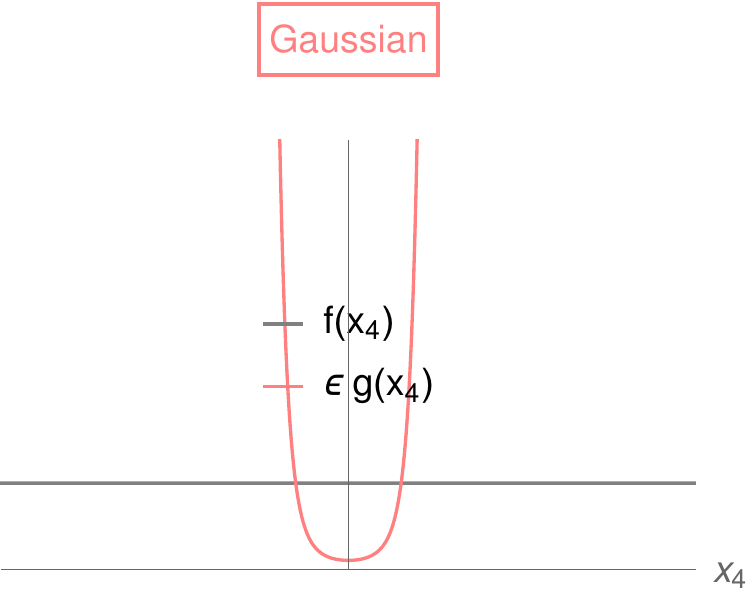}
\hfill%
\includegraphics[width=.32\textwidth]{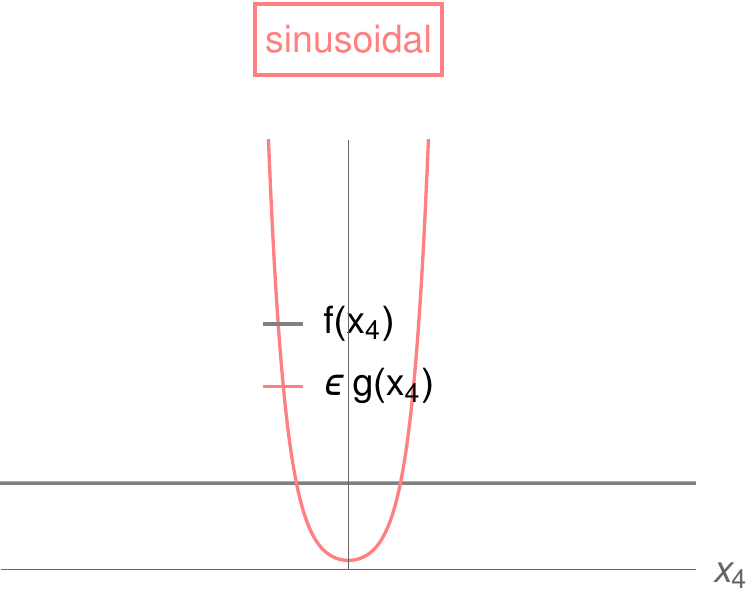}
\hfill%
\caption{Intersection points between the strong and weak field as reflecting mirrors: The functions $f(x_4)$ (gray line) and $\epsilon g(x_4)$ (pink curves) are plotted schematically versus $x_4$. After rotation in the complex plane the strong static field becomes again static (gray line). The inhomogeneous weak field, however, becomes a positive monotonic function (pink curves).
The weak field is assumed to be Lorentzian (left), Gaussian (center) and sinusoidal (right). Note that, in contrast to the Lorentzian field, the last two have not a distinct pole structure.}
\label{fig:intersection1}
\end{figure}
Let us begin with the following observation.
In the limit $\tilde \omega \gg \omega$, the function $g(x_4)$ will be curving much stronger than the slower varying function $f(x_4)$. For sufficiently large frequencies $\tilde \omega$, such a bending results in a potential-wall-like structure confining a considerable region of the strong static curve, $f$, as sketched in Fig.~\ref{fig:intersection1}.
The left panel shows the case with a weak Lorentzian field. The infinite wall is formed at the poles $x_4^\mathrm{p}$. In the remaining two other cases, the Gaussian (center) and sinusoidal (right) field, there appear similar structures. Such fields seem to result in some effective reflectors located around the intersection points between $f$ and $\epsilon g$. Therefore, we expect, at least for a sufficiently large frequency
$\tilde \omega$ that such intersection points, denoted in the following by $x_4^\mathrm{i}$, will play a similar role as the poles for the Lorentzian field.
Of course this is a very rough picture. Indeed, we will see later that improving the location of such effective reflectors will be necessary.
However, according to the described analogy, let us set as a first attempt
\begin{align}
    x_4^\mathrm{p} = x_4^\mathrm{i}.
    \label{eq:intersection-cond}
\end{align}
Taking into account $f \sim F^\prime$ and $g \sim G^\prime$, except the prefactor, we have to solve the following conditional equation
\begin{align}
  F^\prime(x_4^\mathrm{i}) \overset{!}{=} \epsilon G^\prime(x_4^\mathrm{i})
  \label{eq:intersection-eq}
\end{align}
which can be also derived from the original instanton equations \eqref{eq:x4-x3ddot} in order to obtain the point at which both terms contribute equally.
Note that it was expression \eqref{eq:x4-x3ddot} where we have neglected the term $\propto \epsilon$ away from the poles.
So, it is natural to look for critical points of the latter type.
Finding a solution for \eqref{eq:intersection-eq} assuming $F(x_4) = x_4$, see Sec.~\ref{sec:pole-fields}, is straightforward.
On the other hand, integrating $\ddot x_3$, see \eqref{eq:x3dot}, we obtain a second important equation
\begin{align}
F(x_4) = \epsilon G(x_4)
\label{eq:intersection-eq2}
\end{align}
which will determine the true critical Keldysh parameter.
Indeed, it corresponds to the condition in the WKB (Wentzel-Kramers-Brillouin) approach.
However, for weak poleless fields the latter equation is in general transcendental and cannot be solved directly.
We will argue and demonstrate later
that perturbing around $x_4^\mathrm{i}$ that is much easier to obtain proves very powerful in order to solve \eqref{eq:intersection-eq2} analytically.
This will allow us to analytically predict the critical point where both the static strong field and the weak rapid field start to contribute equally.
Furthermore, applying $x_4^\mathrm{i}$ for this purpose is additionally motivated by recent studies. Namely, as we will see, the critical threshold can be estimated to a remarkable order just by applying such intersection points which lead to the same predictions as in \cite{Linder:2015vta,Torgrimsson:2017pzs}.
\begin{figure}[h!]
  \centering
\includegraphics[width=.32\textwidth]{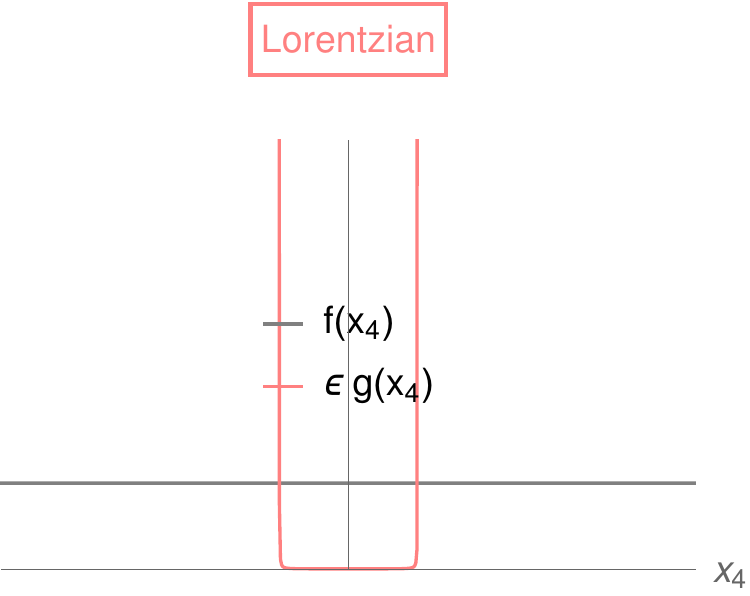}
\hfill%
\includegraphics[width=.32\textwidth]{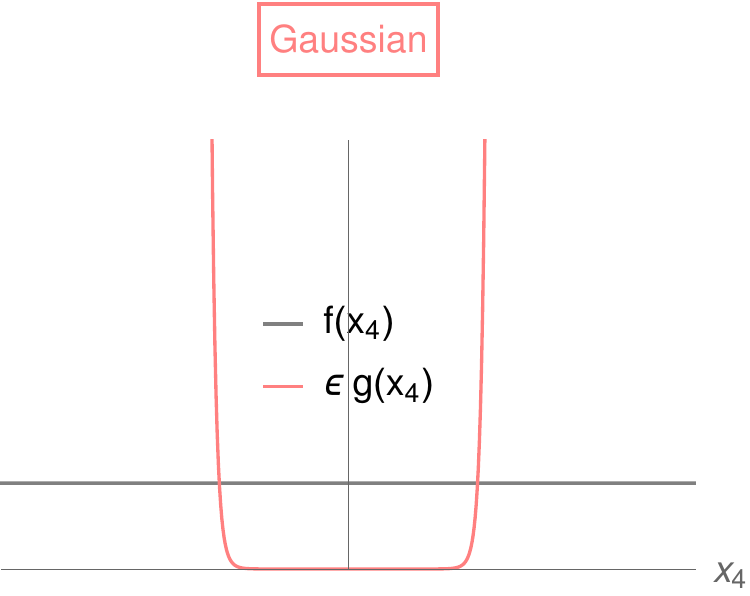}
\hfill%
\includegraphics[width=.32\textwidth]{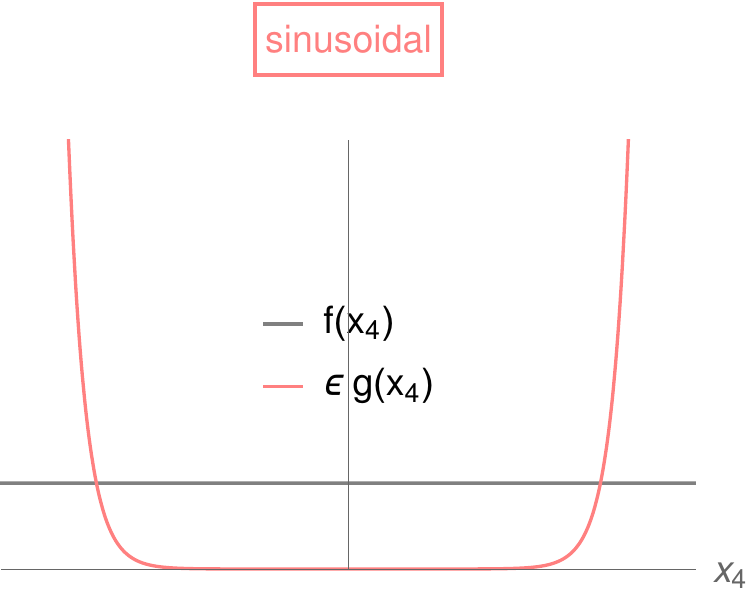}
\hfill%
\caption{Similar plot as in Fig.~\ref{fig:intersection1}. The strength of the weak field is decreased by 5 orders of magnitude compared to the latter plots. The poles for the Lorentzian field (left) do not change. However, the position of the intersection points for the Gaussian (center) and sinusoidal (right) field depend on the parameter $\epsilon$.
}
\label{fig:intersection2}
\end{figure}
From \eqref{eq:intersection-eq} it becomes evident that the effective reflector will depend on the strength parameter $\epsilon$, another common oberservation recently discussed in \cite{Linder:2015vta}.
Such an $\epsilon$ dependence will also apply for the quantum fluctuation prefactor as shown in numerical investigations \cite{Schneider:2016vrl}. Further studies regarding the fluctuation prefactor will be addressed elsewhere \cite{Akal-pre:2017}.
The $\epsilon$ dependence is schematically demonstrated in Fig.~\ref{fig:intersection2}. The location of true poles is fixed, i.e. independent of $\epsilon$. However, for poleless fields one can observe a huge difference where the strength of the weak field is decreased by five orders of magnitude relatively to Fig.~\ref{fig:intersection1}.
One may expect that the prescribed procedure will be more accurate as soon as $\epsilon \rightarrow 0$, since this would lead to a very fast increase of the weak field curve in the intersection points,
similar as one would find in the vicinity of a, let us say, true pole, cf. left plot in Fig.~\ref{fig:intersection2}.
For poleless weak fields we expect that the critical point, in which the former contributes as much as the strong field, will drift more towards the intersection point for $\epsilon \rightarrow 0$.
\begin{figure}[h!]
  \centering
\includegraphics[width=.6\textwidth]{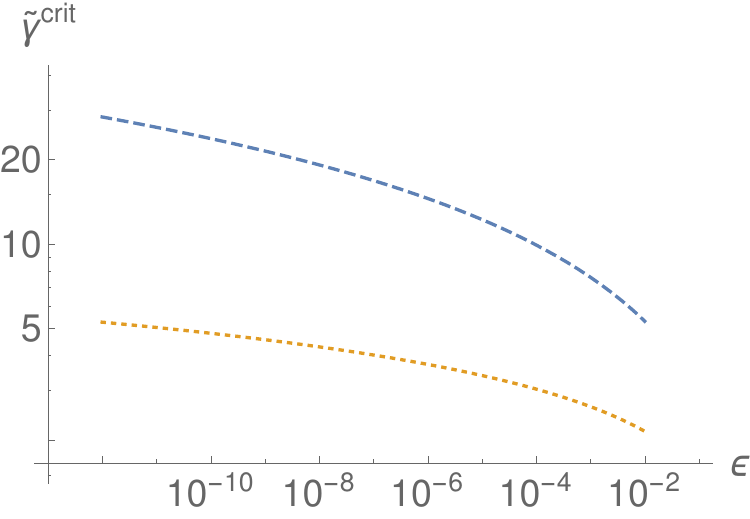}
\caption{Critical combined Keldysh parameter ${\tilde \gamma}^\mathrm{crit}$ for the weak sinusoidal and Gaussian field, superimposed with a strong static field, is plotted versus $\epsilon$, where ${\tilde \gamma}^\mathrm{crit}$ is evaluated assuming the critical point as $x_4^\mathrm{i}$.}
\label{fig:crit-Keldysh-plot}
\end{figure}
\subsection{Intersection points as reflectors}
As a first attempt, in the following two sections \ref{subsec:sinusoid-noimprov} and \ref{subsec:gaussian-noimprov} we will use the previously introduced intersection points as effective reflectors in order to obtain a prediction for the tunneling rate for $\tilde \gamma$ above the critical threshold. Again, improvements for the region around the critical threshold will be derived and discussed in \ref{subsec:real-Keldysh-crit}.
\subsubsection{Weak sinusoid}
\label{subsec:sinusoid-noimprov}
Assume a weak sinusoidal field that is described by
\begin{align}
  g(t) = \cos(\tilde \omega t),\qquad
  G(x_4) = \frac{\sinh(\tilde \omega x_4)}{\tilde \omega}.
  \label{eq:sinusoidal-field}
\end{align}
Inserting the derivatives $F^\prime,G^\prime$ into \eqref{eq:intersection-eq} leads to the intersection point
\begin{align}
  x_4^\mathrm{i} = \frac{\mathrm{arccosh}(1/\epsilon)}{\tilde \omega}.
  \label{eq:intersection-pt-sinusoidal}
\end{align}
Using the equality condition in \eqref{eq:intersection-cond}, we can directly obtain the modified kinematic invariant applying \eqref{eq:mod-a} and \eqref{eq:Keldyshs},
\begin{align}
  a \approx 4 \frac{\gamma}{\omega} \mathrm{arcsin} \left( \frac{1}{ \tilde \gamma} \mathrm{arccosh}(1/\epsilon) \right).
  \label{eq:computed-a-sinusoidal}
\end{align}
For the critical combined Keldysh parameter we find from the latter expression, or alternatively by solving \eqref{eq:crit-relation},
\begin{align}
  {\tilde \gamma}^\mathrm{crit} = \mathrm{arccosh}(1/\epsilon) \approx \ln(2/\epsilon) \approx |\ln(\epsilon)|,
  \label{eq:crit-sinuosid}
\end{align}
since $\epsilon \ll 1$.
It is remarkable that this rough estimation already agrees with the WKB prediction found in \cite{Linder:2015vta}.
The above result is depicted in Fig.~\ref{fig:crit-Keldysh-plot}.

\begin{figure}[!ht]
  \centering
  \hspace*{1.8cm}\includegraphics[width=.8\textwidth]{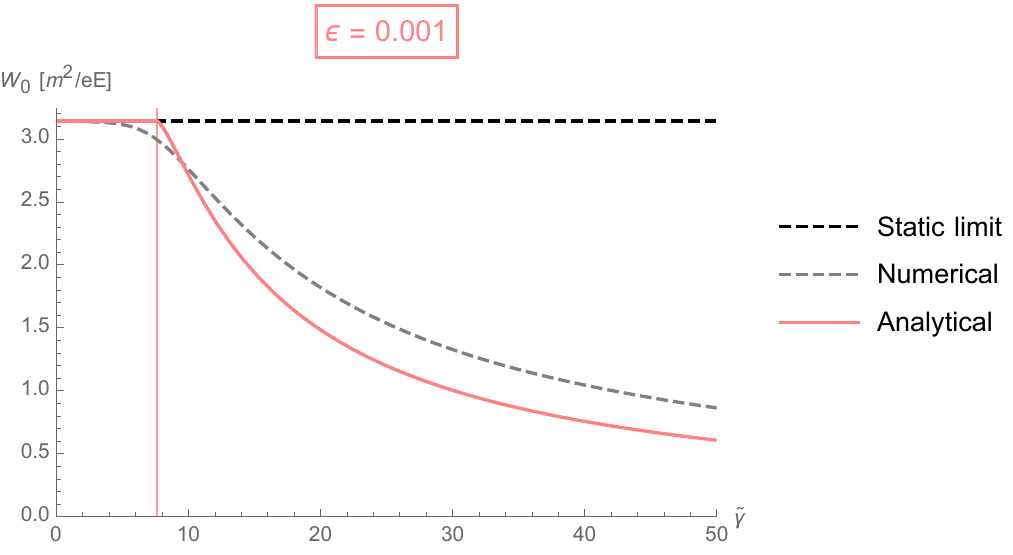}\\
  \hspace*{-1.2cm}\includegraphics[width=.6\textwidth]{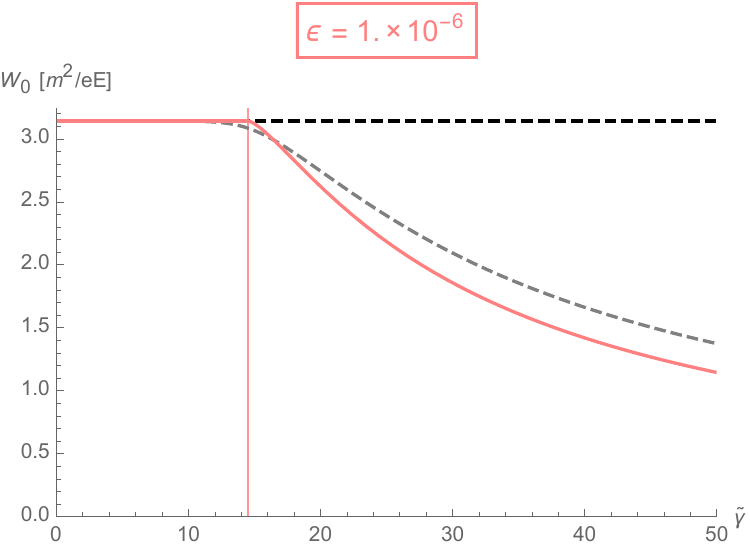}
\caption{$\mathcal{W}_0$ for the weak sinusoidal field \eqref{eq:sinusoidal-field}. The analytical prediction \eqref{eq:computed-W0-sinusoidal} is compared with the exact numerical result for $\epsilon = \{10^{-3},10^{-6}\}$ (top,bottom) with $x_4^\mathrm{p} = x_4^\mathrm{i}$.
}
\label{fig:W0-plot-sinusoid}
\end{figure}

Subsequently, applying \eqref{eq:computed-instanton}, we obtain the approximate instanton path described by
\begin{align}
  \begin{split}
  x_4(u) &\approx \frac{m}{e E}  \sin \left(4 u \mathrm{arcsin}\left(\frac{\mathrm{arccosh}(1/\epsilon)}{\tilde \gamma }\right)\right),\\
  x_3(u) &\approx \frac{m}{e E}  \cos \left(4 u \mathrm{arcsin}\left(\frac{\mathrm{arccosh}(1/\epsilon)}{\tilde \gamma }\right)\right) - \mathcal{C}
  \label{eq:computed-instanton-sinusoidal}
  \end{split}
\end{align}
with $\mathcal{C} = x_3(u=\pm 1/4)$.
Inserting \eqref{eq:intersection-pt-sinusoidal} into \eqref{eq:computed-W}, we get
\begin{align}
  \mathcal{W}_0 \approx \frac{m^2}{e E} \left( \frac{2 \mathrm{arccosh}(1/\epsilon)}{{\tilde \gamma}^2} \sqrt{{\tilde \gamma}^2 - \mathrm{arccosh}^2(1/\epsilon)} + 2 \mathrm{arcsin} \left( \frac{\mathrm{arccosh}(1/\epsilon)}{\tilde \gamma} \right) \right).
  \label{eq:computed-W0-sinusoidal}
\end{align}
The resulting plots for the stationary worldline action \eqref{eq:computed-W0-sinusoidal} are depicted in Fig.~\ref{fig:W0-plot-sinusoid}.
Although the analytical prediction follows the trend of the exact numerical curve, they considerably differ from each other.
There will be some region right after the intersection point which will surely have a non-negligible contribution to the pair production rate. This is completely neglected when we take $x_4^\mathrm{p} = x_4^\mathrm{i}$. An improvement of the effective reflection point is therefore needed. Before we proceed in that direction, let us first introduce a second example which shares similar features.
\subsubsection{Weak Gaussian}
\label{subsec:gaussian-noimprov}
\begin{figure}[!ht]
  \centering
  \hspace*{1.8cm}\includegraphics[width=.8\textwidth]{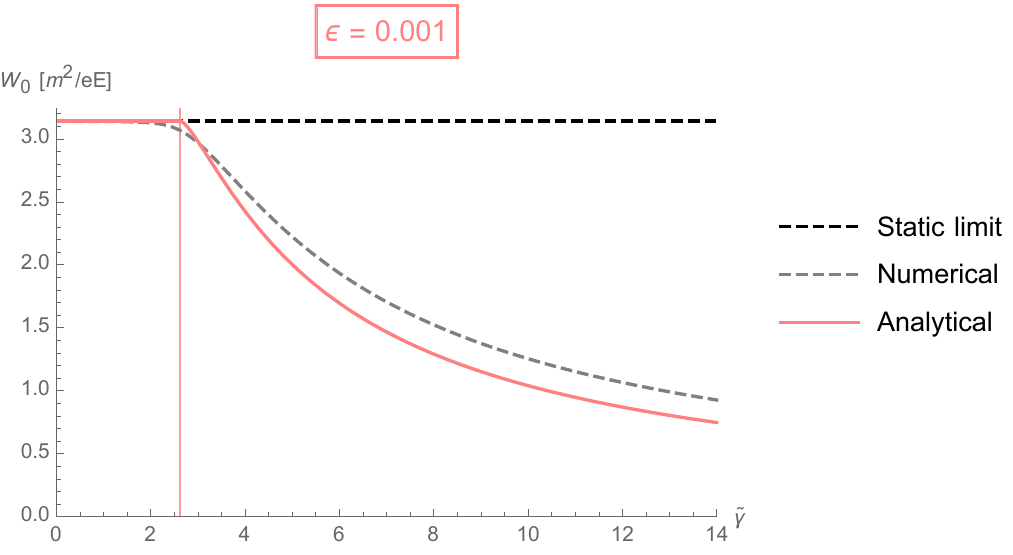}\\
  \hspace*{-1.2cm}\includegraphics[width=.6\textwidth]{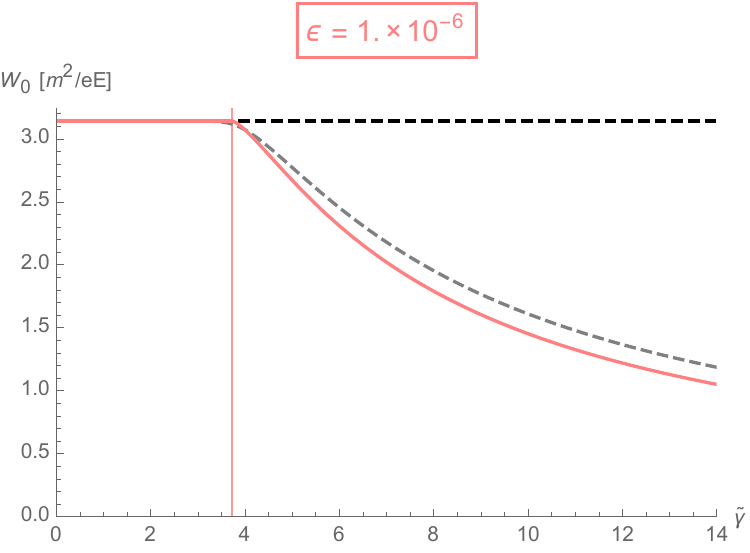}
\caption{$\mathcal{W}_0$ for the weak Gaussian field \eqref{eq:gaussian2-field}. The analytical prediction \eqref{eq:computed-W0-gaussian} is compared with the numerical result for $\epsilon = \{10^{-3},10^{-6}\}$ (top,bottom) with $x_4^\mathrm{p} = x_4^\mathrm{i}$.
}
\label{fig:W0-plot-gaussian}
\end{figure}

We consider a Gaussian field described by
\begin{align}
  g(t) = \exp(-(\tilde \omega t)^2),\qquad
  G(x_4) = \frac{\sqrt{\pi} \mathrm{erfi}(\tilde \omega x_4)}{2 \tilde \omega}.
  \label{eq:gaussian2-field}
\end{align}
The intersection point is
\begin{align}
  x_4^\mathrm{i} = \frac{\sqrt{ \mathrm{ln}(1/\epsilon) }}{\tilde \omega}.
  \label{eq:intersection-pt-gaussian}
\end{align}
Proceeding similarly, we get the approximate invariant from \eqref{eq:mod-a}
\begin{align}
  a \approx 4 \frac{\gamma}{\omega} \mathrm{arcsin} \left( \frac{1}{ \tilde \gamma} \sqrt{\mathrm{ln}(1/\epsilon)} \right)
  \label{eq:computed-a-gaussian}
\end{align}
which leads to the critical Keldysh parameter
\begin{align}
  {\tilde \gamma}^\mathrm{crit} = \sqrt{\mathrm{ln}(1/\epsilon)} = \sqrt{|\ln(\epsilon)|}.
  \label{eq:crit-gaussian}
\end{align}
As in the previous case, this result gives again the prediction obtained within the WKB approach \cite{Linder:2015vta}. The critical value is plotted in Fig.~\ref{fig:crit-Keldysh-plot}.
Inserting the intersection point in \eqref{eq:computed-instanton}, we continue with
\begin{align}
  \begin{split}
  x_4(u) &\approx \frac{m}{e E}  \sin \left(4 u \mathrm{arcsin}\left(\frac{\sqrt{\mathrm{ln}(1/\epsilon)}}{\tilde \gamma }\right)\right),\\
  x_3(u) &\approx \frac{m}{e E}  \cos \left(4 u \mathrm{arcsin}\left(\frac{\sqrt{\mathrm{ln}(1/\epsilon)}}{\tilde \gamma }\right)\right) - \mathcal{C},
  \label{eq:computed-instanton-gaussian}
  \end{split}
\end{align}
where the constant $\mathcal{C} = x_3(u = \pm 1/4)$ plays the same role as before.
We can again plug \eqref{eq:intersection-pt-gaussian} into \eqref{eq:computed-W}, which leads to
\begin{align}
  \mathcal{W}_0 \approx \frac{m^2}{e E} \left( \frac{2 \sqrt{\mathrm{ln}(1/\epsilon)}}{{\tilde \gamma}^2} \sqrt{{\tilde \gamma}^2 - \mathrm{ln}(1/\epsilon)} + 2 \mathrm{arcsin} \left( \frac{\sqrt{\mathrm{ln}(1/\epsilon)}}{\tilde \gamma} \right) \right).
  \label{eq:computed-W0-gaussian}
\end{align}
The plots for \eqref{eq:computed-W0-gaussian} are shown in Fig.~\ref{fig:W0-plot-gaussian}. 
From those, we still find a clear deviation. However, this distinction is significantly smaller than in the case before. This behaviour agrees with the curve trends in Fig.~\ref{fig:crit-Keldysh-plot}.
\subsection{Improved reflection points}
\label{subsec:improve-mirrors}
In the previous examples we have seen that reflecting the instanton at $x_4^\mathrm{i}$ can mimic the reduction of $\mathcal{W}_0$ for large $\tilde \gamma$. However, the predictions still considerably differ from the exact numerical results.
In the following,
we will improve the effective reflection points.
This will confirm the generalization of the reflection picture even for poleless fields.

Before doing so, let us emphasize that such an improvement will only affect the region with a $\tilde \gamma$ larger than the critical threshold.
Strictly speaking, it will in general not allow a prediction around the critical point which we define to be the one where both fields start to contribute equally.
But, as we will discuss in Sec.~\ref{subsec:real-Keldysh-crit}, an appropriate perturbative treatment based on the introduced intersection points
will allow a very accurate analytical prediction for the critical point as well.
Indeed, this is a powerful way to find out when the weak field 
contribution starts to dominate. Actually, this is a highly challenging problem which has been not tackled before analytically, since for most field combinations,
as for instance Gaussian fields, the corresponding equation is of transcendental type. More details will be discussed in Sec.~\ref{subsec:real-Keldysh-crit}.

Let us first elaborate the first improvement mentioned above in order to predict the rate for sufficiently large $\tilde \gamma$.
We argued before that some region after the intersection points will be necessary.
Hence, we first write a correction of the form
\begin{align}
  x_4^\mathrm{p} \rightarrow x_4^\mathrm{i} + \delta/\tilde\omega,
  \label{eq:x4-eff-reflection}
\end{align}
where $\delta/\tilde \omega$ denotes some displacement parameter $\delta$ we have to specify.
Once we have computed $\delta$, the relevant parameter $\tilde \gamma$ in the final expressions has to be modified as well, since ${\tilde \gamma}^\mathrm{crit}$ will still be determined by the intersection point\footnote{This is in general characteristic for fields without poles. There is always a difference between the effective reflection point and the critical point in which the weak field begins to contribute. For fields with a distinct pole structure such a disagreeance is not present, at least in the highly weak limit $\epsilon \ll 1$.}.
Here, we are interested in the behaviour for $\tilde \gamma \gg {\tilde \gamma}^\mathrm{crit}$. Therefore, we keep $x_4^\mathrm{i}$ as some approximate critical point. Then, the relevant modifications can be written as
\begin{align}
  \begin{split}
  x_4^\mathrm{p} \rightarrow x_4^\mathrm{i} + \delta/\tilde\omega,\qquad
  \tilde \gamma \rightarrow \tilde \gamma + \delta.
\end{split}
  \label{eq:improved-steps}
\end{align}
Such a shifted Keldysh parameter signals already that the impact of the weak rapid field on the rate $\mathcal{R}$ substantially differs for fields with and without poles.
In order to compute the correction $\delta$ we
have seen that
weak fields with poles result in almost vertical reflectors intersecting with the strong field curve. 
Despite that this is just a geometric observation, it already provides an explanation why the tunneling rate should not alter with $\epsilon$.
Thus, in case of weak poleless fields we need to increase the accuracy of the effective reflection points.

In order to accomplish this, let us recall some basics in the equivalent WKB approach.
Here, the plan is to obtain conditions that we can combine with our previous analysis in the (worldline) instanton approach to improve our analytical predictions for $\tilde \gamma$ much larger than the critical threshold. This will also illustrate the equivalence between both methods with respect to the (semiclassical) tunneling exponential for which both approaches lead in general to the same result if the momentum spectrum is peaked around zero (canonical) momentum \cite{Strobel:2013vza} as it is the case for fields with one spacetime coordinate as considered in this paper, see e.g. \cite{Dumlu:2011rr}.

Starting with the Dirac equation in an electromagnetic background, one can first identify the evolution of the corresponding Bogoliubov coefficients.
Afterwards the resulting system, which is described by the Riccati equation, can approximately be integrated and one obtains for $p=0$ the condition
\begin{align}
  i e \left( E F(x_4^*) + \tilde E G(x_4^*) \right) = i m
\end{align}
for the singularities $x_4^*$, see e.g. \cite{Dumlu:2011rr,Linder:2015vta}. This relation, for instance, can alternatively be obtained from the RHS of the integrated expression \eqref{eq:x3dot}. The singularities $x_4^*$ determine the pair production probability, at least the correct exponent\footnote{The quantum fluctuation prefactor cannot be correctly determined via WKB, except the (semiclassical) exponential dependence.}.
The solution of this equation gives then the poles of the fields. If $\epsilon \ll 1$ and $G$ is sufficiently small, one gets the usual strong field pole $x_4^* = \frac{m}{e E}$ if $F(x_4) = x_4$. For the assisted mechanism we have to consider the situation where the smallness of $\epsilon$ is counterbalanced by the reflection point. This happens in case of setting
\begin{align}
  \epsilon G(x_4^\mathrm{p}) \overset{!}{=} \frac{\tilde \gamma}{\tilde \omega},
  \label{app-eq:improved-condition}
\end{align}
where $F$ is assumed to be negligible small in the reflection point, i.e.
\begin{align}
  F(x_4^\mathrm{p}) \ll \epsilon G(x_4^\mathrm{p}).
\end{align}
Note that from the important condition
\begin{align}
  F(x_4)\overset{!}{=}\epsilon G(x_4),
  \label{app-eq:equal-condition}
\end{align}
one determines the point in which both fields contribute equally.
Note that equation \eqref{app-eq:equal-condition} can be transcendental\footnote{as for (super) Gaussian fields, cf. Sec.~\ref{sec:fields-without-poles}} which cannot be solved algebraically. In this case, we will follow an alternative approach leading to a drastic simplification of the problem.
Deriving the latter equation after $x_4$ on both hand sides, leads to the intersection condition we have already introduced in \eqref{eq:intersection-eq}.
If the inverse of $G$ does exist, the equation \eqref{app-eq:improved-condition} can be solved directly leading to the solution
\begin{align}
  x_4^\mathrm{p} = G^{-1}\left( \frac{\tilde \gamma}{\epsilon \tilde \omega} \right).
\end{align}
This gives us the improved effective reflection point.
Reminding that we have started from the assumption $x_4^\mathrm{p} \approx x_4^\mathrm{i}$ we try to improve the reflection point via the ansatz \eqref{eq:x4-eff-reflection}.
The correction $\delta > 0$ is then determined by
\begin{align}
  \delta = \tilde \omega G^{-1}\left( \frac{\tilde \gamma}{\epsilon \tilde \omega} \right) - \tilde \omega x_4^\mathrm{i}.
  \label{app-eq:delta}
\end{align}
It is important to note that, according to the present approach, $x_4^\mathrm{i}$ still has to determine the critical Keldysh parameter ${\tilde \gamma}^\mathrm{crit}$ (the true value is for sure different \ref{subsec:real-Keldysh-crit}), i.e.
\begin{align}
  \gamma \overset{!}{=} \omega F(x_4^\mathrm{i}) \rightsquigarrow {\tilde \gamma}^\mathrm{crit}.
  \label{app-eq:crit-condition}
\end{align}
Contributions from the weak field have been neglected before the instanton is reflected, since the drastic enhancement is caused by reflections.
However, the improved effective reflection point \eqref{eq:x4-eff-reflection} will for sure modify ${\tilde \gamma}^\mathrm{crit}$ from above. This value we denote as ${\tilde \gamma}^\mathrm{p,crit}$ which follows from
\begin{align}
  \gamma \overset{!}{=} \omega F(x_4^\mathrm{p}) \rightsquigarrow  {\tilde \gamma}^\mathrm{p,crit}.
  \label{app-eq:critPole-condition}
\end{align}
In order to keep \eqref{app-eq:crit-condition} as the critical threshold, one has to shift $\tilde \gamma$ in the final expressions via \eqref{app-eq:delta},
\begin{align}
  {\tilde \gamma} \rightarrow {\tilde \gamma} + \delta,
\end{align}
where $\delta$ can now be written as
\begin{align}
  \delta = {\tilde \gamma}^\mathrm{p,crit} - {\tilde \gamma}^\mathrm{crit}.
\end{align}
Because of $\delta > 0$ and $\epsilon \ll 1$, we assume
\begin{align}
  2{\tilde \gamma}^\mathrm{crit} > {\tilde \gamma}^\mathrm{p,crit} > {\tilde \gamma}^\mathrm{crit}.
  \label{app-eq:relation-crits0}
\end{align}
Note that the last two steps are justified only if the weak field raises sufficiently fast in the vicinity of the intersection points which usually applies when $\tilde\gamma \gg 1$. Compared to \eqref{app-eq:delta}, we will therefore neglect the explicit $\tilde \gamma$ dependence and rewrite ${\tilde \gamma}^\mathrm{p,crit}$ as
\begin{align}
  {\tilde \gamma}^\mathrm{p,crit} = (1 + \xi){\tilde \gamma}^\mathrm{crit}
  \label{app-eq:relation-crits}
\end{align}
with $0 < \xi < 1$.
With this, we obtain
\begin{align}
  \delta  =  \xi {\tilde \gamma}^\mathrm{crit}.
  \label{app-eq:delta2}
\end{align}
Combining \eqref{eq:x4-eff-reflection} and \eqref{app-eq:delta}, we get from \eqref{app-eq:critPole-condition}
\begin{align}
  {\tilde \gamma}^\mathrm{p,crit} = \tilde \omega G^{-1} \left( \frac{{\tilde \gamma}^\mathrm{p,crit}}{\epsilon \tilde \omega} \right),
\end{align}
where $F(x_4) = x_4$ has been assumed for the strong field.
Subsequently, we apply \eqref{app-eq:relation-crits} to obtain
\begin{align}
  (\xi + 1){\tilde \gamma}^\mathrm{crit} = \tilde \omega G^{-1} \left( \frac{(\xi + 1){\tilde \gamma}^\mathrm{crit}}{\epsilon \tilde \omega} \right).
  \label{app-eq:xi}
\end{align}
Note that ${\tilde \gamma}^\mathrm{crit}$ is known from previous analysis, see condition \eqref{app-eq:crit-condition}. The latter equation is difficult to solve in general, because of the nonlinear $\xi$ dependence on the RHS. However, since we have $\xi < 1$, one may Taylor expand the nonlinearity in the lowest relevant order and compute $\xi$. This solution can be used to obtain the displacement parameter from the expression \eqref{app-eq:delta2}. This is a powerful way to compute $\delta$, specifically, in situations where the inverse function of $G$ is difficult to find or does not exist at all, respectively. Finally, all relevant modifications we need for the improvement are the one given in \eqref{eq:improved-steps}.

\subsection{Reflecting at improved points}
Let us apply the correction $\delta$ to the previously discussed two examples
in order to improve the prediction for the tunneling rate for $\tilde \gamma$ larger than the critical threshold.
\subsubsection{Weak sinusoid}
\begin{figure}[h!]
  \centering
\hspace*{1.8cm}\includegraphics[width=.8\textwidth]{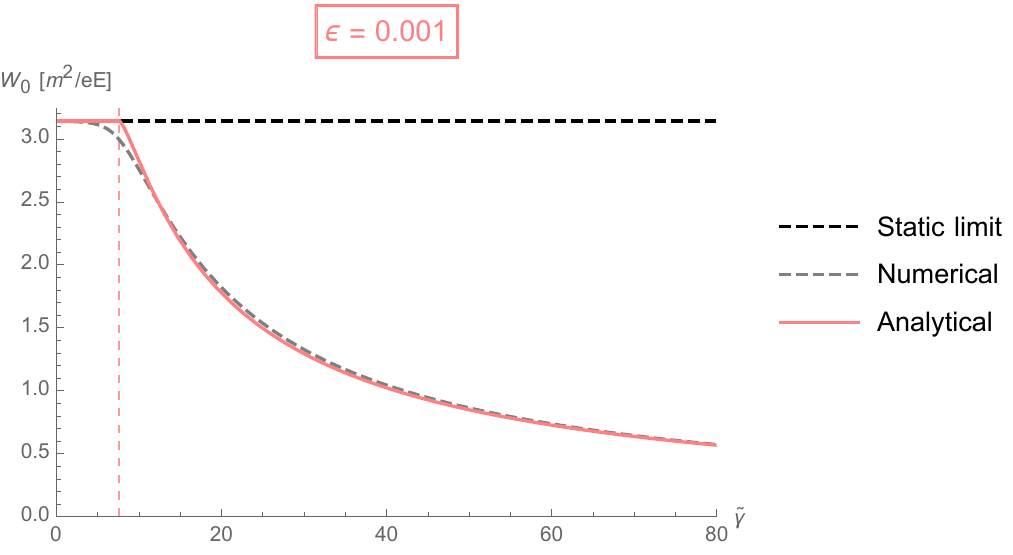}\\
\hspace*{-1.2cm}\includegraphics[width=.6\textwidth]{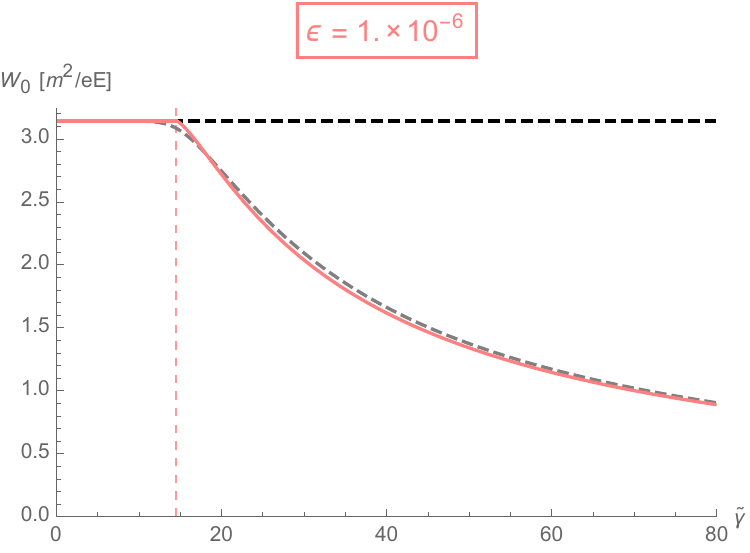}
\caption{$\mathcal{W}_0$ plotted for the sinusoidal weak field \eqref{eq:sinusoidal-field}. The analytical prediction \eqref{eq:computed-W0-sinusoidal-improved} is compared with the exact numerical result. The ratio between the strong and weak field strengths is set to $\epsilon = \{10^{-3},10^{-6}\}$ (top,bottom). The vertical dashed red lines are placed at $\tilde \gamma = {\tilde \gamma}^\mathrm{crit}$ with
$x_4^\mathrm{p} = x_4^\mathrm{i} + \delta/\tilde \omega$.}
\label{fig:W0-plot-sinusoid-improved}
\end{figure}

Again, we begin with the sinusoidal field. According to the modifications in \eqref{eq:improved-steps},
we first need to compute $\delta$. Applying the inverse function
\begin{align}
  G^{-1}(x_4) = \frac{\mathrm{arcsinh}(\tilde \omega x_4)}{\tilde \omega}
\end{align}
to \eqref{app-eq:delta}, we obtain
\begin{align}
  \delta &= \mathrm{arcsinh}\left( \frac{\tilde \gamma}{\epsilon} \right) - \mathrm{arccosh}\left(\frac{1}{\epsilon}\right).
  \label{eq:delta-sinusoid}
\end{align}
Inserting the corresponding replacements afterwards, i.e.
\begin{align}
  \mathrm{arccosh}(1/\epsilon) &\rightarrow \mathrm{arccosh}(1/\epsilon) + \delta,\\
  \tilde \gamma &\rightarrow \tilde \gamma + \delta,
\end{align}
into \eqref{eq:computed-a-sinusoidal}, \eqref{eq:computed-instanton-sinusoidal} and
\eqref{eq:computed-W0-sinusoidal},
we obtain the improved invariant $a$, instanton path and stationary worldline action $\mathcal{W}_0$, respectively.
The latter written out explicitly reads
\begin{align}
  \begin{split}
  \mathcal{W}_0 &\approx \frac{m^2}{e E} \bigg( \frac{2 \left( \mathrm{arccosh}(1/\epsilon) + \delta \right)}{\left({\tilde \gamma} + \delta \right)^2} \sqrt{\left({\tilde \gamma} + \delta \right)^2 - \left( \mathrm{arccosh}(1/\epsilon) + \delta \right)^2}\\
  &+ 2 \mathrm{arcsin} \left( \frac{\mathrm{arccosh}(1/\epsilon) + \delta}{\tilde \gamma + \delta} \right) \bigg).
  \label{eq:computed-W0-sinusoidal-improved}
  \end{split}
\end{align}
The action \eqref{eq:computed-W0-sinusoidal-improved} plotted in Fig.~\ref{fig:W0-plot-sinusoid-improved} for $\epsilon = \{10^{-3},10^{-6}\}$ clearly shows a substantial improvement of the approximate analytical result, cf. Fig.~\ref{fig:W0-plot-sinusoid}. As we expected, despite the region around $\tilde\gamma = \tilde\gamma^\mathrm{crit}$ (vertical dashed red line), the analytical curve is in good agreement with the exact numerical curve.
\subsubsection{Weak Gaussian}
Now, we can follow the same procedure for the Gaussian field \eqref{eq:gaussian2-field}. The displacement parameter $\delta$ we expect to be smaller compared to the sinusoidal field \eqref{eq:sinusoidal-field}. This is simply due to a stronger slope of $g(x_4)$ in the vicinity of the intersection point, cf. Fig.~\ref{fig:intersection2}.
Apart from this expectation, there is another difference. The function $G(x_4)$ is the imaginary error function for which the inverse is difficult to express algebraically.
However, as discussed in the previous section, we can first apply \eqref{app-eq:xi} and Taylor expand the nonlinearity in $\xi$ introduced via the condition \eqref{app-eq:relation-crits}, since $\xi < 1$. Proceeding in that way, we obtain the following result in leading order
\begin{align}
  \xi \approx \frac{1}{\sqrt{2}} \sqrt{\frac{2}{\mathrm{ln} \left(\frac{1}{\epsilon }\right)}-\frac{\sqrt{\pi } \epsilon  \mathrm{erfi}\left(\sqrt{\mathrm{ln} \left(\frac{1}{\epsilon }\right)}\right)}{\mathrm{ln}^{\frac{3}{2}}\left(\frac{1}{\epsilon }\right)}},
\end{align}
depending only on $\epsilon$ which is implicitly required via \eqref{app-eq:relation-crits0} and \eqref{app-eq:relation-crits}.
Finally, using \eqref{eq:crit-gaussian} we get
\begin{align}
  \delta = \xi \sqrt{\mathrm{ln}(1/\epsilon)}.
  \label{eq:delta-gaussian}
\end{align}
Inserting the replacements
\begin{align}
  \sqrt{\mathrm{ln}(1/\epsilon)} &\rightarrow \sqrt{\mathrm{ln}(1/\epsilon)} + \delta,\\
  \tilde \gamma &\rightarrow \tilde \gamma + \delta
\end{align}
into \eqref{eq:computed-a-gaussian}, \eqref{eq:computed-instanton-gaussian} and
\eqref{eq:computed-W0-gaussian},
we obtain again the improved invariant $a$, instanton path and stationary worldline action
\begin{align}
  \begin{split}
  \mathcal{W}_0 &\approx \frac{m^2}{e E} \Bigg( \frac{2 \left( \sqrt{\mathrm{ln}(1/\epsilon)} + \delta \right)}{\left({\tilde \gamma} + \delta \right)^2} \sqrt{\left({\tilde \gamma} + \delta \right)^2 - \left( \sqrt{\mathrm{ln}(1/\epsilon)} + \delta \right)^2}\\
  &+ 2 \mathrm{arcsin} \left( \frac{\sqrt{\mathrm{ln}(1/\epsilon)} + \delta}{\tilde \gamma + \delta} \right) \Bigg),
  \label{eq:computed-W0-gaussian-improved}
  \end{split}
\end{align}
respectively.
\begin{figure}[h!]
  \centering
\hspace*{1.8cm}\includegraphics[width=.8\textwidth]{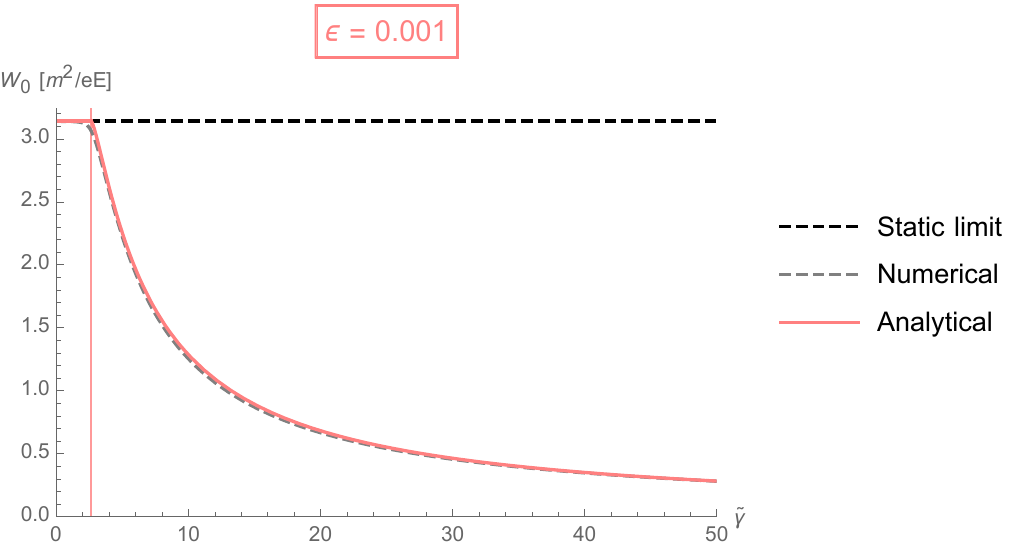}\\
\hspace*{-1.2cm}\includegraphics[width=.6\textwidth]{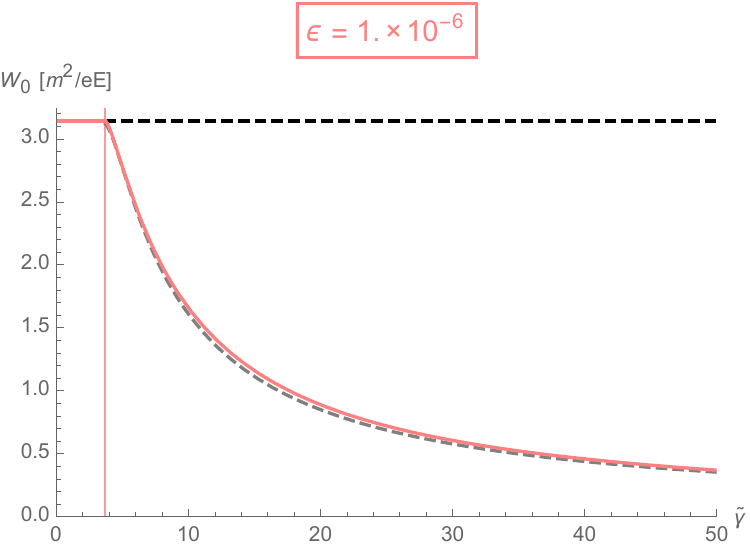}
\caption{$\mathcal{W}_0$ for the weak Gaussian field \eqref{eq:gaussian2-field}. The analytical prediction \eqref{eq:computed-W0-gaussian-improved} is compared with the numerical result. The ratio between the strong and weak field strengths is set to $\epsilon = \{10^{-3},10^{-6}\}$ (top,bottom).
The vertical red lines are placed at $\tilde \gamma = {\tilde \gamma}^\mathrm{crit}$.
The improved effective reflection point $x_4^\mathrm{p} = x_4^\mathrm{i} + \delta/\tilde \omega$ has been applied.
}
\label{fig:W0-plot-gaussian-improved}
\end{figure}
The comparison between \eqref{eq:computed-W0-gaussian-improved} and its exact numerical computation is depicted in Fig.~\ref{fig:W0-plot-gaussian-improved}. As in the previous example, the analytical results for sufficiently large $\tilde \gamma$ are clearly improved, cf. Fig.~\ref{fig:W0-plot-gaussian}. The predicted curve is in good agreement with the exact numerical computation. This observation confirms again the validity of the reflection picture in the case with a poleless weak field.

\subsection{Weak super Gaussian}
\begin{figure}[h!]
  \centering
\includegraphics[width=.58\textwidth]{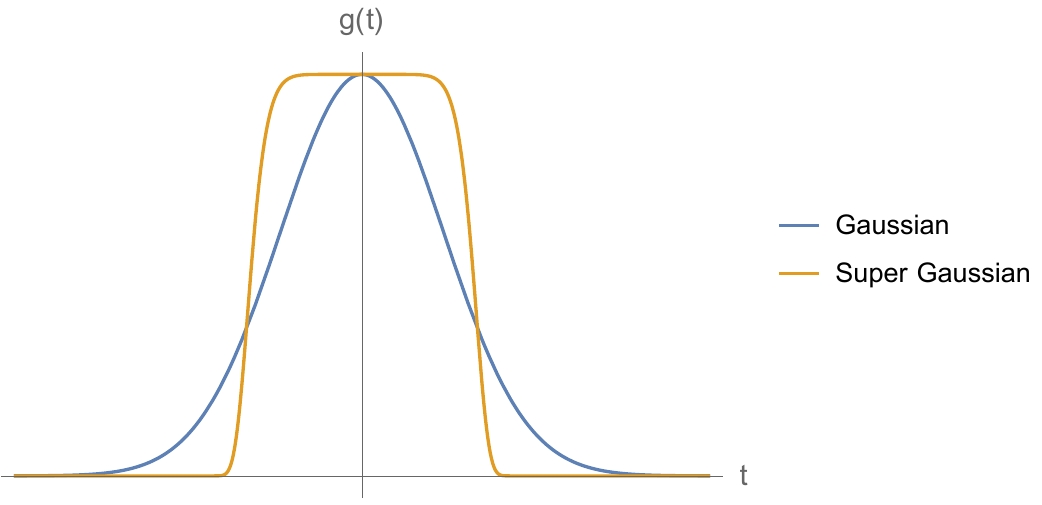}
\hfill%
\includegraphics[width=.41\textwidth]{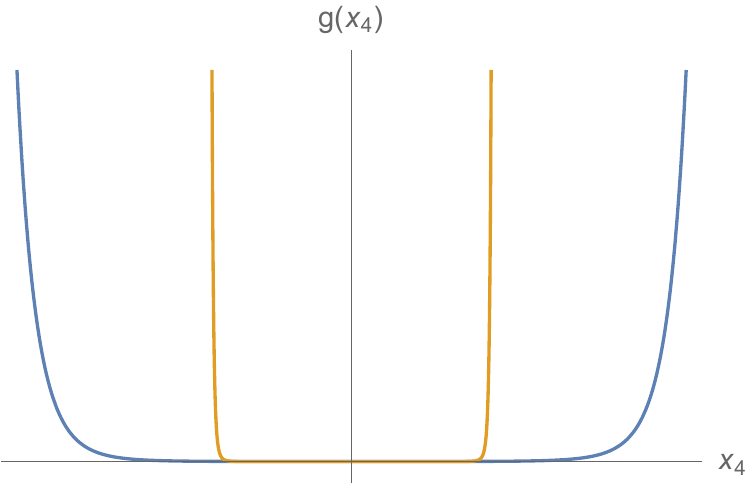}
\caption{
Comparison between the standard \eqref{eq:gaussian2-field} and super Gaussian \eqref{eq:gaussian10-field} profiles in blue and yellow, respectively. In the left panel the function $g(t)$ in Lorentzian time is plotted. The right panel shows the analytic continued function $g(x_4)$.}
\label{fig:gaussians}
\end{figure}
\begin{figure}[h!]
  \centering
\includegraphics[width=.26\textwidth]{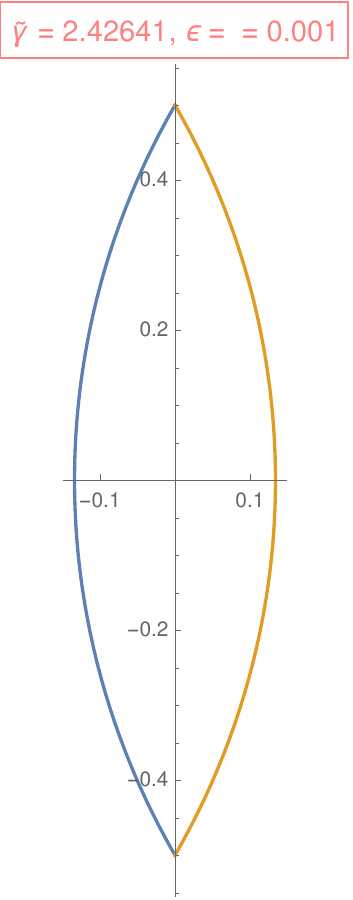}
\qquad\qquad
\includegraphics[width=.284\textwidth]{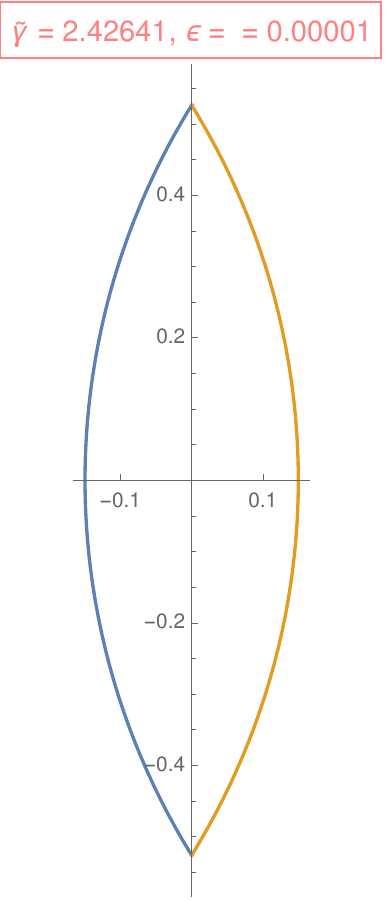}
\caption{Instanton paths for an electric background as superposition of a strong static and weak super Gaussian field from \eqref{eq:gaussian10-field} are plotted for $\epsilon = \{10^{-3},10^{-5}\}$ (left,right). The combined Keldysh parameter is set to $\tilde \gamma = 2 {\tilde \gamma}^\mathrm{crit}(\epsilon = 0.001)$. The instanton is reflected at the intersection point, i.e. $x_4^\mathrm{p} = x_4^\mathrm{i}$.}
\label{fig:instantons-gaussian10}
\end{figure}

The standard Gaussian field \eqref{eq:gaussian2-field} leads in general to more accurate results if we just apply the intersection point as an effective reflector without any further correction $\delta$. It is expected that a field with a stronger slope in the vicinity of such intersection points will lead to more accurate analytical predictions.
For that reason, let us discuss a third example with a weak super Gaussian described by
\begin{align}
  g(t) = \exp(-(\tilde \omega t)^{10}),\qquad
  G(x_4) = -\frac{(\tilde \omega x_4) \mathbf{E}_{\frac{9}{10}}\left((-i \tilde \omega x_4)^{10}\right)}{10 \tilde\omega}.
  \label{eq:gaussian10-field}
\end{align}
The comparison with the previously studied Gaussian field \eqref{eq:gaussian2-field} is depicted in Fig.~\ref{fig:gaussians}. The field profile resembles a rectangular potential-wall with a flat top (left panel). Rotating the function $g$ in the complex plane shows a dramatical increase of the curve slope, similar to the situation with a Sauter/Lorentzian field.
Here, $\mathbf{E}_n(z)$ denotes the exponential integral function.
The intersection point is
\begin{align}
  x_4^\mathrm{i} = \frac{\mathrm{ln}(1/\epsilon)^{1/10}}{\tilde \omega}.
  \label{eq:intersection-pt-gaussian10}
\end{align}
Setting $x_4^\mathrm{p} = x_4^\mathrm{i}$, leads to the modified invariant
\begin{align}
  a \approx 4 \frac{\gamma}{\omega} \mathrm{arcsin} \left( \frac{1}{ \tilde \gamma} \mathrm{ln}(1/\epsilon)^{1/10} \right)
  \label{eq:computed-a-gaussian10}
\end{align}
and critical Keldysh parameter
\begin{align}
  {\tilde \gamma}^\mathrm{crit} = \mathrm{ln}(1/\epsilon)^{1/10} \approx |\ln(\epsilon)|^{1/10}.
\end{align}
Note that the LMA \eqref{eq:lma} condition becomes
\begin{align}
  m a \approx \frac{E_\mathrm{S}}{E} 4 \mathrm{arcsin} \left( \frac{1}{ \tilde \gamma} \left(\mathrm{ln}(1/\epsilon)\right)^{1/10} \right) \gg 1.
\end{align}
For the instanton path we obtain accordingly
\begin{align}
  \begin{split}
  x_4(u) &\approx \frac{m}{e E}  \sin \left(4 u \mathrm{arcsin}\left(\frac{\left(\mathrm{ln}(1/\epsilon)\right)^{1/10}}{\tilde \gamma }\right)\right),\\
  x_3(u) &\approx \frac{m}{e E}  \cos \left(4 u \mathrm{arcsin}\left(\frac{\left(\mathrm{ln}(1/\epsilon)\right)^{1/10}}{\tilde \gamma }\right)\right) - \mathcal{C},
  \label{eq:computed-instanton-gaussian10}
  \end{split}
\end{align}
where the constant $\mathcal{C}$ plays the same role as before, i.e. $\mathcal{C} = x_3(u = \pm 1/4)$.
The predicted instantons are plotted in
Fig.~\ref{fig:instantons-gaussian10} for $\epsilon = \{10^{-3},10^{-6}\}$.
The paths do not differ much from each other, i.e. the $\epsilon$ dependence has become weaker. This is basically in line with the situation for fields which have true poles. Because of the strong curve slope, the position of the intersection points will almost be fixed and do not change with varying $\epsilon$.
Inserting \eqref{eq:intersection-pt-gaussian10} into \eqref{eq:computed-W}, we get
\begin{align}
  \mathcal{W}_0 \approx \frac{m^2}{e E} \left( \frac{2 \left(\mathrm{ln}(1/\epsilon)\right)^{1/10}}{{\tilde \gamma}^2} \sqrt{{\tilde \gamma}^2 - \left(\mathrm{ln}(1/\epsilon)\right)^{1/5}} + 2 \mathrm{arcsin} \left( \frac{\left(\mathrm{ln}(1/\epsilon)\right)^{1/10}}{\tilde \gamma} \right) \right)
  \label{eq:computed-W0-gaussian10}
\end{align}
\begin{figure}[h!]
  \centering
\hspace*{1.8cm}\includegraphics[width=.8\textwidth]{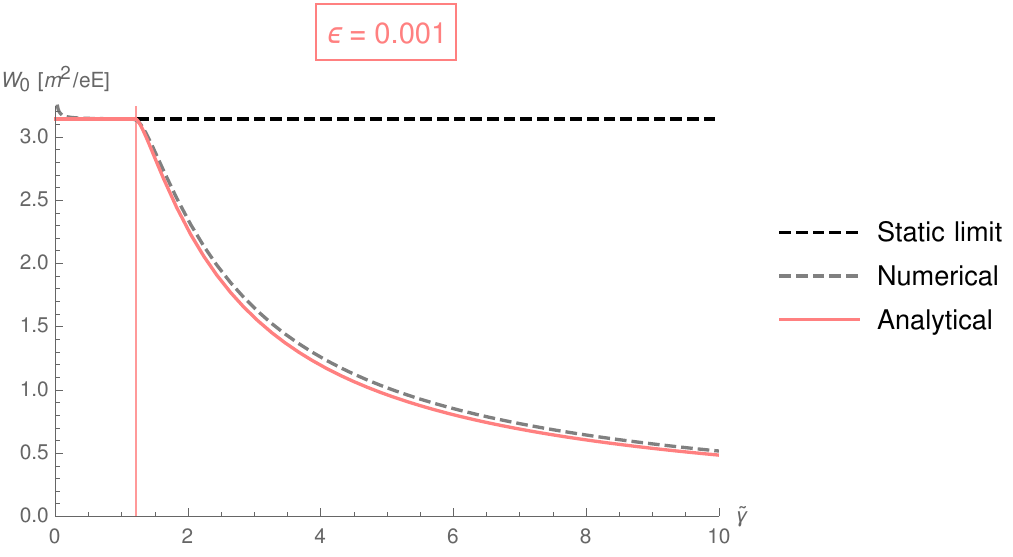}\\
\hspace*{-1.2cm}\includegraphics[width=.6\textwidth]{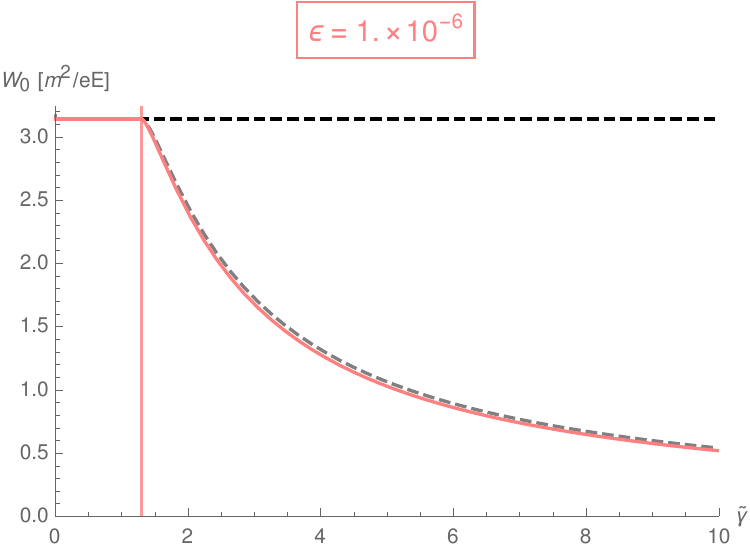}
\caption{$\mathcal{W}_0$ for the weak super Gaussian field \eqref{eq:gaussian10-field}. The analytical prediction \eqref{eq:computed-W0-gaussian10} is compared with the numerical result. Here, the instanton is assumed to be reflected in the intersection point, i.e. $x_4^\mathrm{p} = x_4^\mathrm{i}$.
The vertical red lines are placed at $\tilde \gamma = {\tilde \gamma}^\mathrm{crit}$.
}
\label{fig:W0-plot-gaussian10}
\end{figure}
depicted in Fig.~\ref{fig:W0-plot-gaussian10}. The result agrees well with the numerical curve. Hence, the discussed features from above lead indeed to a  substantial improvement of the analytical estimation, already with $x_4^\mathrm{p} = x_4^\mathrm{i}$.
The results can be generalised for an arbitrary super Gaussian field ($N \geq 1$)
\begin{align}
  g(t) = \exp(-(\tilde \omega t)^{(4N+2)}),\quad N \in \mathbb{N}.
  \label{eq:gaussianN-field}
\end{align}
The corresponding intersection point is
\begin{align}
  x_4^\mathrm{i} = \frac{\mathrm{ln}(1/\epsilon)^\frac{1}{4N+2}}{\tilde \omega}.
  \label{eq:intersection-pt-gaussianN}
\end{align}
Thus, for $N > 2$ we may expect the prediction with $x_4^\mathsf{p} = x_4^\mathrm{i}$ to be even more accurate compared to the latter case with $N=2$, means no additional need for corrections $\delta,\Delta$.
Therefore higher order super Gaussians as in \eqref{eq:gaussianN-field} will behave almost as fields with poles. The $\epsilon$ dependence will be suppressed with increasing $N$. The vacuum pair production rate $\mathcal{R}$ will be enhanced even more, simply due to
\begin{align}
\tilde\omega x_4^\mathrm{i} \longrightarrow 1\quad (N \rightarrow \infty).
\end{align}
Note that the latter limit coincides with the reflection point for a weak Lorentzian field.
\subsection{Comparison of $\mathcal{W}_0$}
\label{subsec:comparison}
In this part we compare the previous predictions for $\mathcal{W}_0$. We consider both types of weak fields, with and without poles.
\begin{figure}[h!]
  \centering
\hspace*{1cm}\includegraphics[width=.9\textwidth]{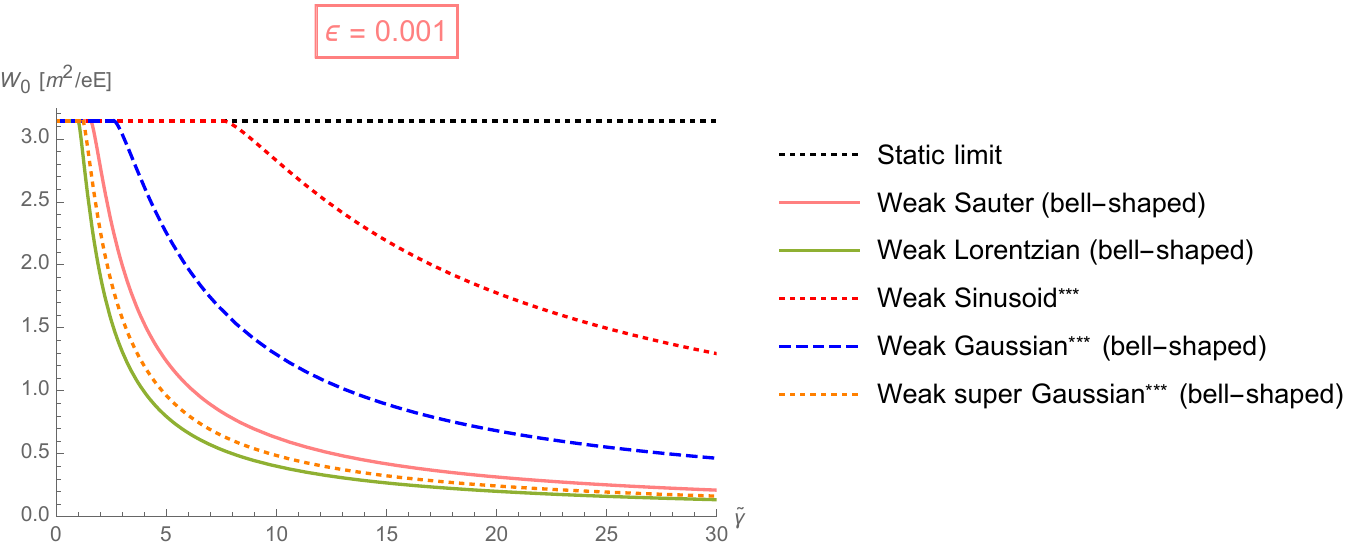}\\
\includegraphics[width=.55\textwidth]{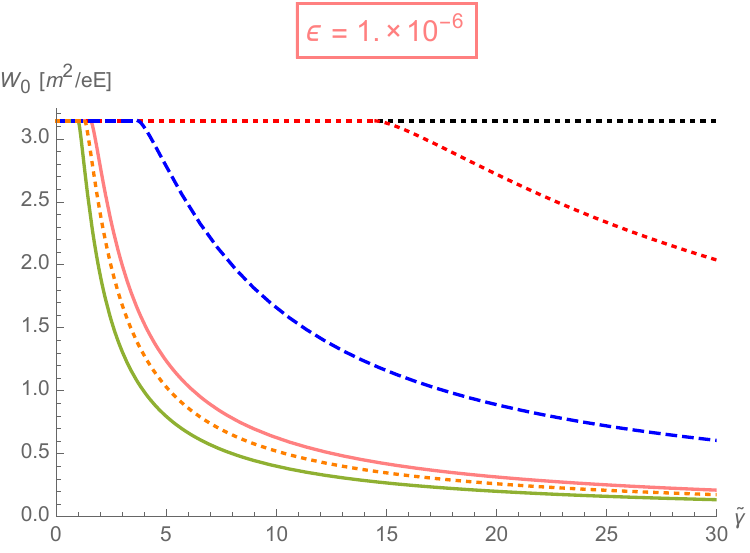}
\caption{Comparison of the stationary worldline action $\mathcal{W}_0$ for different weak field profiles. The ratio between the strong and weak field strengths is set to $\epsilon = \{10^{-3},10^{-6}\}$ (top,bottom). The profiles are listed in the legend. Fields without poles are marked with asterisks in the legend. Note that the applied critical Keldysh parameter for those fields is approximated by ${\tilde \gamma}^\mathrm{crit}$. For more on the true critical value we refer to the discussion in the following subsection.
}
\label{fig:W0-plot-comparison}
\end{figure}
The results are plotted in Fig.~\ref{fig:W0-plot-comparison} for $\epsilon = \{10^{-3},10^{-6}\}$.
The fields without poles are marked with asterisks in the plot legend.
Note that the critical Keldysh parameter for the latter type is determined by the approximate value ${\tilde \gamma}^\mathrm{crit}$. More on the true critical parameter will be discussed in \ref{subsec:real-Keldysh-crit}.
In Fig.~\ref{fig:W0-plot-comparison} the bell-shaped fields enhance the rate much more than the pure (infinitely extended) oscillatory sinusoidal field. Those among them with true poles tend to reduce the stationary action even more.
This is shown upon the direct comparison with the Gaussian field which represents a bell-shaped field but has no poles.

Interestingly, the super Gaussian field with $N = 2$, see \eqref{eq:gaussian10-field}, which obviously does not have true poles as well, leads to comparable enhancement signatures. It is even much more enhancing than the Sauter field.
This can be understood if we rotate the field in the complex plane.
Namely, it results in a very strong slope in the vicinity of the relatively small intersection point,
see Sec.~\ref{sec:fields-without-poles}. The latter is almost equal to the (improved) effective reflection point and (true) critical point. We point out that usually those differ considerably from each other if the field is poleless, see Sec.~\ref{sec:fields-without-poles}.

Bell-shaped fields may have important consequences on oscillatory pulsed fields (wave packets). Such profiles can usually be described by multiplying, for instance, an infinitely extended sinusoidal field with a bell-shaped envelope function. According to the presented results, a weak field of this form will predominantly trigger the assistance via instanton reflections in the poles of the envelope. Varying the pulse width via the frequency of the latter is therefore expected to be dominating the enhancement. However, as we already discussed in Sec.~\ref{sec:pole-fields}, there may be interesting quantum interference effects if we resolve the momentum spectrum of the produced pair, which are generally sensitive to the sub-cycle structure of a wave packet.
However, the total pair production rate would be highly sensitive to the (ideally) finite size of the weak pulse.
This may for sure be substantial for laser experiments where pulses have very short duration.
Such observations may be used for further optimisation studies in order to enhance the rate by an appropriate weak field even more.

\subsection{Improved critical point}
\label{subsec:real-Keldysh-crit}

In this section we discuss how to derive the critical point in order to find the true critical Keldysh parameter. As we have seen before in \ref{subsec:improve-mirrors}, we have computed the correction $\delta$ for the effective reflection point. For the case, where the inverse function $G^{-1}$ is complicated to handle, we have used $x_4^\mathrm{i}$ from \eqref{eq:intersection-eq} in order to obtain $\delta$. 
However, the critical point\footnote{Defined to be the point where both strong, slow and weak, rapid field start to contribute equally to the pair production process.} needs to be improved as well like
\begin{align}
  (1-\Delta)x_4^\mathrm{i},
  \label{eq:x4-crit-real}
\end{align}
which would then lead to the critical threshold
\begin{align}
  (1-\Delta){\tilde \gamma}^\mathrm{crit}.
  \label{eq:crit-Keldysh-real}
\end{align}
In the following, we will compute the correction $\Delta$
via a
perturbation around $x_4^\mathrm{i}$, since the relevant domain to look for is
\begin{align}
 x_4 \in (0,x_4^\mathrm{i}],
 \label{eq:x4-domain}
\end{align}
cf. Figs.~\ref{fig:intersection1} and \ref{fig:intersection2}, which
will correct the previous estimations \eqref{eq:crit-sinuosid} and \eqref{eq:crit-gaussian}, respectively.
The relevant equation is \eqref{app-eq:equal-condition}.
Inserting \eqref{eq:x4-crit-real} into the latter yields
\begin{align}
  (1-\Delta)x_4^\mathrm{i} = \epsilon G\left( (1-\Delta)x_4^\mathrm{i} \right)
  \label{app-eq:approx-Delta-eqn}
\end{align}
after setting $F(x_4) = x_4$. In contrast to the initial equation \eqref{app-eq:equal-condition}, the latter can be Taylor expanded on the RHS for which we find the series
\begin{align}
  (1-\Delta)x_4^\mathrm{i} \approx G( x_4^\mathrm{i} ) - G^\prime( x_4^\mathrm{i}) \Delta - \frac{1}{2} G^{\prime\prime}( x_4^\mathrm{i}) \Delta^2 + \mathcal{O}(\Delta^3).
  \label{eq:Delta-Taylor}
\end{align}
Now, this equation can be solved by truncating after a sufficient order in $\Delta$.
This allows to solve \eqref{app-eq:equal-condition} which in general is hard to tackle directly due to its
transcendental form for various types of backgrounds.
In the following, we explicitly compute $\Delta$ for the sinusoidal and Gaussian weak field.
\subsubsection{Weak sinusoid}
The sinusoidal field is known from \eqref{eq:sinusoidal-field}. Plugging into \eqref{app-eq:approx-Delta-eqn} leads to
\begin{align}
  (1-\Delta) x_4^\mathrm{i} = \frac{\epsilon}{\tilde \omega} \sinh((1-\Delta)\tilde \omega x_4^\mathrm{i}).
\end{align}
With the intersection point \eqref{eq:intersection-pt-sinusoidal}, we obtain up to order $\mathcal{O}(\Delta^2)$,
\begin{align}
  {\tilde\gamma}^\mathrm{crit}  (1-\Delta )\approx \epsilon  \left(\frac{1}{2} {\tilde\gamma}^{\mathrm{crit}^2} \Delta^2 \sinh ({\tilde\gamma}^\mathrm{crit} )-{\tilde\gamma}^\mathrm{crit}  \Delta  \cosh ({\tilde\gamma}^\mathrm{crit} )+\sinh ({\tilde\gamma}^\mathrm{crit} )\right).
\end{align}
Here, we have used the relation $\tilde \omega x_4^\mathrm{i} = {\tilde \gamma}^\mathrm{crit}$.
The parameter $\Delta$ is then determined by
\begin{align}
  \begin{split}
  \Delta &\approx
  -\frac{\text{csch}({\tilde\gamma}^\mathrm{crit} )}{{\tilde\gamma}^\mathrm{crit}  \epsilon }
  +\frac{\coth ({\tilde\gamma}^\mathrm{crit} )}{{\tilde\gamma}^\mathrm{crit} }\\
  &+ \frac{\text{csch}({\tilde\gamma}^\mathrm{crit} ) \sqrt{(2 {\tilde\gamma}^\mathrm{crit}
  -2 {\tilde\gamma}^\mathrm{crit}  \epsilon  \cosh ({\tilde\gamma}^\mathrm{crit} ))^2
  -4 {\tilde\gamma}^{\mathrm{crit}^2} \epsilon  \sinh ({\tilde\gamma}^\mathrm{crit} ) (2 \epsilon  \sinh ({\tilde\gamma}^\mathrm{crit} )
  -2 {\tilde\gamma}^\mathrm{crit} )}}{2 {\tilde\gamma}^{\mathrm{crit}^2} \epsilon }.
\end{split}
\label{eq:Delta-sinusoid}
\end{align}
\subsubsection{Weak Gaussian}
The field has been introduced in \eqref{eq:gaussian2-field}.
The intersection point is given in \eqref{eq:intersection-pt-gaussian}. As before, we plug the corresponding quantities into \eqref{app-eq:approx-Delta-eqn} and obtain
\begin{align}
  (1-\Delta) \tilde \omega x_4^\mathrm{i} = \epsilon \frac{\sqrt{\pi}}{2} \mathrm{erfi}((1-\Delta)\tilde \omega x_4^\mathrm{i}).
\end{align}
This equation can be written up to order $\mathcal{O}(\Delta^2)$ as
\begin{align}
  (1 - \Delta) {\tilde \gamma}^\mathrm{crit} \approx
  \epsilon  \left(-\frac{2 \left(e^{{\tilde \gamma}^{\mathrm{crit}^2} } {\tilde \gamma}^\mathrm{crit} \right) \Delta }{\sqrt{\pi }}+\frac{2 e^{{\tilde \gamma}^{\mathrm{crit}^2} } {\tilde \gamma}^{\mathrm{crit}^3} \Delta^2}{\sqrt{\pi }}+\text{erfi}({\tilde \gamma}^\mathrm{crit} )\right)
\end{align}
and we analogously find
\begin{align}
  \begin{split}
  \Delta &\approx
  \frac{e^{-{\tilde\gamma}^{\mathrm{crit}^2}} \sqrt{\left(\sqrt{\pi } {\tilde\gamma}^\mathrm{crit}
  -2 e^{{\tilde\gamma}^{\mathrm{crit}^2}} {\tilde\gamma}^\mathrm{crit}  \epsilon \right)^2-8 e^{{\tilde\gamma}^{\mathrm{crit}^2}} {\tilde\gamma}^{\mathrm{crit}^3} \epsilon  \left(\sqrt{\pi } \epsilon  \text{erfi}({\tilde\gamma}^\mathrm{crit} )
  -\sqrt{\pi } {\tilde\gamma}^\mathrm{crit} \right)}}{4 {\tilde\gamma}^{\mathrm{crit}^3} \epsilon }\\
  &-\frac{\sqrt{\pi } e^{-{\tilde\gamma}^{\mathrm{crit}^2} }}{4 {\tilde\gamma}^{\mathrm{crit}^2} \epsilon }
  +\frac{1}{2 {\tilde\gamma}^{\mathrm{crit}^2} }.
\end{split}
\label{eq:Delta-gaussian}
\end{align}

\subsubsection{Comparisons}
In case of a distinct pole structure, we can compute $\tilde\gamma^\mathrm{crit}$ by solving the equation
\begin{align}
  x_4^\mathrm{p} = \frac{\gamma}{\omega},
\end{align}
cf. \eqref{eq:crit-relation}. If true poles are not present we have benefited from $x_4^\mathrm{p} = x_4^\mathrm{i}$.
In order to consider ${\tilde \gamma}^\mathrm{crit}$ as the critical threshold, we have assumed that the background after rotation in the complex plane is described by $f(x_4)$ for $\tilde \gamma \leq \tilde\gamma^\mathrm{crit}$, see Figs.~\ref{fig:intersection1} and \ref{fig:intersection2}, respectively. This is for sure not the realistic situation, since usually there is some contribution from the weak field, $g(x_4)$, which leads to an increase of the total effective field strength.
However, doing so allows to obtain analytical predictions for $\mathcal{W}_0$, since the main effect for sufficiently large $\tilde \gamma$ comes from the instanton reflection. One should bear in mind, that further improvements with respect to the effective reflection points were needed, see \eqref{eq:x4-eff-reflection}.
In the absence of true poles simple geometric arguments are not very clear.
Therefore, such agreements have helped to sort previous observations into a more general picture.

The correction to \eqref{eq:crit-sinuosid} and \eqref{eq:crit-gaussian}, respectively, can be obtained according to \eqref{eq:crit-Keldysh-real}. From our previous analysis we expect that the difference between \eqref{eq:x4-crit-real} and the intersection point will decrease for $\epsilon \rightarrow 0$, see
Figs.~\ref{fig:W0-plot-sinusoid-improved} and \ref{fig:W0-plot-gaussian-improved}.
On the other hand, note that if poles exist, intersection, critical and reflection point are just given by the pole itself, $\delta,\Delta \rightarrow 0$.
This is the reason why poleless weak fields assist less strong for $\tilde \gamma$ around the critical Keldysh parameter \eqref{eq:crit-Keldysh-real}. In contrast, for fields with true poles the curve for $\mathcal{W}_0$ decreases very rapidly as soon as $\tilde \gamma$ reaches the threshold, cf. Fig.~\ref{fig:W0-sauter-lorentzian}.
\begin{figure}[h!]
  \centering
\includegraphics[width=.47\textwidth]{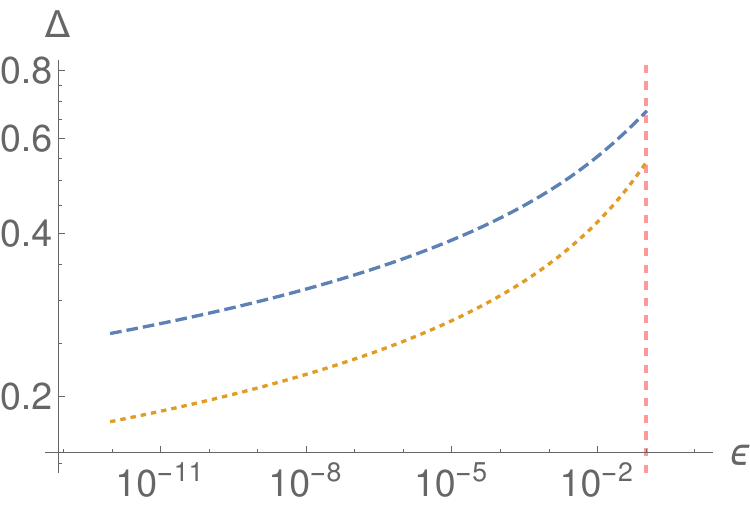}
\includegraphics[width=.52\textwidth]{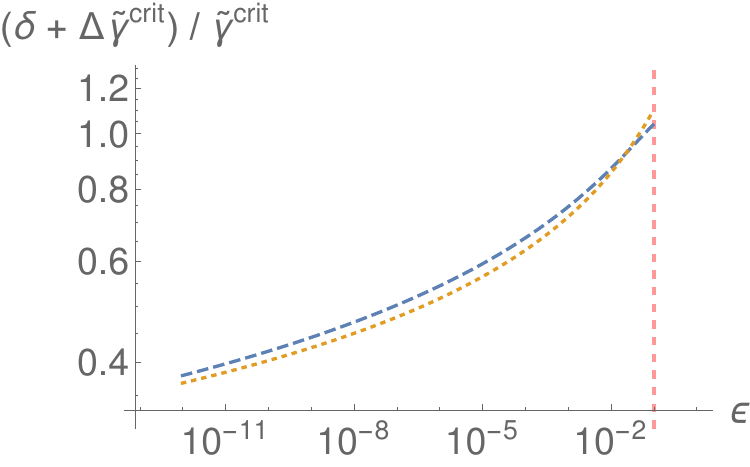}\\
\hspace*{2.9cm}\includegraphics[width=.8\textwidth]{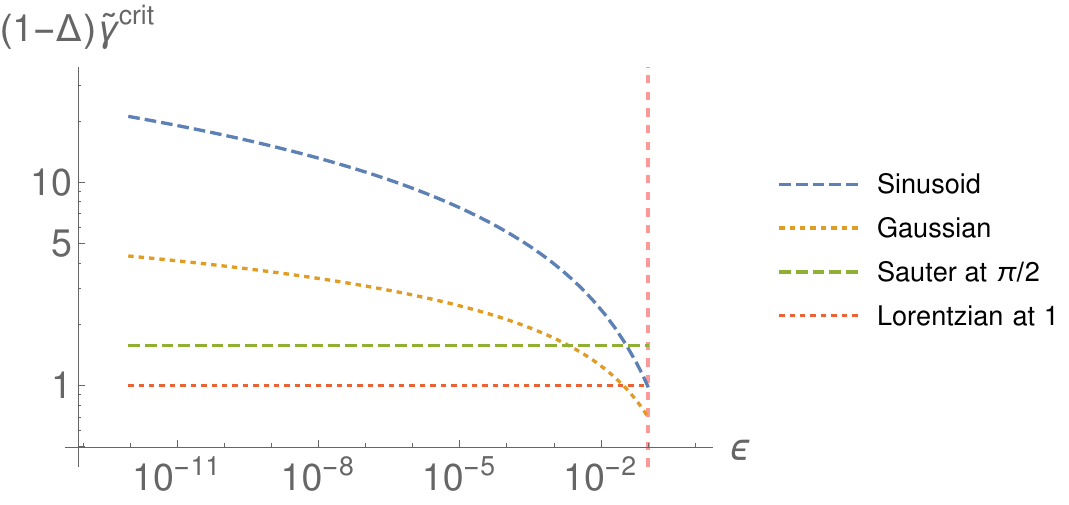}
\caption{Comparison of $\Delta$ for the sinusoidal and Gaussian field (top-left). The difference (with $\tilde \gamma = {\tilde \gamma}^\mathrm{crit}$) between the improved effective reflection point \eqref{eq:x4-eff-reflection} and the true critical point \eqref{eq:x4-crit-real} is shown in units of $\frac{m}{eE}$ (top-right). Note that for fields with a distinct pole structure we find $\delta,\Delta \rightarrow 0$ in the relevant regime $\epsilon \ll 1$. The true critical combined Keldysh parameter, i.e. $(1-\Delta){\tilde \gamma}^\mathrm{crit}$, for fields without poles, introduced in \eqref{eq:crit-Keldysh-real}, is plotted in the bottom panel. Note the difference compared to the previous estimations in
Fig.~\ref{fig:crit-Keldysh-plot}. The correction $\Delta$ is computed up to order $\mathcal{O}(\Delta^2)$. For improvements, in particular, for the sinusoidal field due to a in general larger $\Delta$, cf. top-left panel, we can simply truncate the Taylor series in \eqref{eq:Delta-Taylor} after higher orders in $\Delta$.
The vertical, pink, dashed line is placed at $\epsilon = 0.1$.
}
\label{fig:Delta-sinusoid-gaussian}
\end{figure}
The difference between the intersection point and \eqref{eq:x4-crit-real} is
\begin{align}
  x_4^\mathrm{i} - \eqref{eq:x4-crit-real} = x_4^\mathrm{i} - (1-\Delta)x_4^\mathrm{i} = \Delta x_4^\mathrm{i} = \Delta \frac{{\tilde \gamma}^\mathrm{crit}}{\tilde \gamma} \frac{m}{eE}
  \label{eq:diff-x4intersection-x4critreal}
\end{align}
with $\Delta$ as derived in \eqref{eq:Delta-sinusoid} and \eqref{eq:Delta-gaussian}, respectively. The corrections $\Delta$ are depicted in Fig.~\ref{fig:Delta-sinusoid-gaussian}.
As soon as $\epsilon \rightarrow 0$ we find $\Delta \rightarrow 0$. This is consistent with our expectation and observations in Figs.~\ref{fig:W0-plot-sinusoid-improved} and \ref{fig:W0-plot-gaussian-improved}.
The critical Keldysh parameter \eqref{eq:crit-Keldysh-real} is plotted in the right panel of Fig.~\ref{fig:Delta-sinusoid-gaussian}. The obtained values coincide very well with the critical behaviour in Figs.~\ref{fig:W0-plot-sinusoid-improved} and \ref{fig:W0-plot-gaussian-improved}.
The improved values are of high accuracy, although
$\Delta$ has been computed only up to order $\mathcal{O}(\Delta^2)$, cf. \eqref{eq:Delta-Taylor}. A small deviation, however, occurs in case of the sinusoidal field for $\epsilon = 10^{-3}$ if we compare the analytically predicted value $(1-\Delta){\tilde \gamma}^\mathrm{crit} \approx 4$, see blue, dashed curve,
with the observed value located around $\tilde \gamma \approx 3$, see left panel in Fig.~\ref{fig:W0-plot-sinusoid-improved} (dashed, gray curve).
Such a difference originates due to $\Delta \lesssim 1$, cf. Fig.~\ref{fig:Delta-sinusoid-gaussian}.
To improve the analytical prediction, one can truncate the Taylor series in \eqref{eq:Delta-Taylor} after an appropriate higher order in $\Delta$.
The very well coincidence in the remaining other cases confirm the validity of \eqref{eq:Delta-sinusoid} and \eqref{eq:Delta-gaussian} as solutions for \eqref{app-eq:equal-condition} that is in general very challenging to solve analytically because of its non-trivial transcendental structure, see \cite{Linder:2015vta,Schneider:2016vrl} for numerical studies.

In addition we have plotted the constant lines in Fig.~\ref{fig:Delta-sinusoid-gaussian} (right) at $\frac{\pi}{2}$ and $1$ for the Sauter and the Lorentzian field, respectively. For the remaining two poleless fields we find the non-constant dependence in $\epsilon$.
For $\epsilon \nearrow$, \eqref{eq:crit-Keldysh-real} turns out to be smaller compared to the first estimations \eqref{eq:crit-sinuosid} and \eqref{eq:crit-gaussian}, cf. Fig.~\ref{fig:crit-Keldysh-plot} and \ref{fig:Delta-sinusoid-gaussian}. This is also in line with the presented plots in Figs.~\ref{fig:W0-plot-sinusoid-improved} and \ref{fig:W0-plot-gaussian-improved}.
For sufficiently large $\epsilon$, say $\sim 10^{-2}$, we even achieve values below $\frac{\pi}{2}$ (green, dashed). However, although in this case the Gaussian field will start to assist before the Sauter field, it will reduce the stationary action $\mathcal{W}_0$ much slower with increasing $\tilde \gamma$. This is a direct consequence of
\begin{align}
  \eqref{eq:x4-eff-reflection} - \eqref{eq:x4-crit-real} = \frac{\delta + \Delta {\tilde \gamma}^\mathrm{crit}}{\tilde \gamma} > 0
  \label{eq:diff-x4reflection-x4critreal}
\end{align}
given in units of $\frac{m}{eE}$.
Using \eqref{eq:delta-sinusoid} and \eqref{eq:Delta-sinusoid} (sinusoidal), \eqref{eq:delta-gaussian} and \eqref{eq:Delta-gaussian} (Gaussian), we have plotted the corresponding curves for $\tilde \gamma = {\tilde \gamma}^\mathrm{crit}$
in Fig.~\ref{fig:Delta-sinusoid-gaussian}. Note if the underlying weak field has a pole, we get $\eqref{eq:diff-x4reflection-x4critreal} \rightarrow 0$, namely, already for $\epsilon \sim 10^{-2}$.

We conclude that for $\epsilon \ll 1$ the critical combined Keldysh parameter for fields with poles determines exactly the point where the weak field contribution becomes essential and where the reflection sets in.
However, if poles are not present, the true critical point \eqref{eq:x4-crit-real} does not in general correspond to the effective reflection point \eqref{eq:x4-eff-reflection}. The latter is usually much larger valued as in the former case.
Thus, we find a larger range below the critical Keldysh parameter determining the point where assistance sets in or, say, where the weak field contribution starts to exceed the strong field contribution, see Fig.~\ref{fig:W0-plot-sinusoid-improved}. Consequently, the reduction of $\mathcal{W}_0$ progresses very slow.
There a minimal reduction is mainly triggered by the effectively increased total field strength due to the superposition of the strong and weak field, see left and right panels in Fig.~\ref{fig:intersection1}.
This is somewhat similar to the situation with a single-mode inhomogeneous electric field, see Fig.~\ref{fig:both-frequent} (only gray solid line) in App.~\ref{subsec:effects-strong-mode}.
The large valued effective reflection point is too far away from the true critical point.
However, as soon as both points merge together, i.e. $\eqref{eq:diff-x4reflection-x4critreal} \rightarrow 0$, which happens for $\tilde \gamma$ larger than ${\tilde \gamma}^\mathrm{crit}$, the enhancement becomes stronger due to smaller effective reflection points. Thus, the decrease of $\mathcal{W}_0$ evolves more quickly, similar to the case with true poles acting as reflectors with relatively small values.
One should note that $\tilde \gamma \gg {\tilde \gamma}^\mathrm{crit}$ is exactly the regime where the instanton can be taken as reflected in the effective points, see e.g. Figs.~\ref{fig:W0-plot-sinusoid-improved} and \ref{fig:W0-plot-gaussian-improved}.
Basically, this effect turns out to be the main attribute that distinguishes these two types of fields. 

\section{Assisted dynamical mechanism}
As brought up in the introduction part, we now come to the second part; the assisted dynamical mechanism. The difference is that the strong field will be assumed as dynamical as well. We will study such a background utilising again the reflection approach.

\label{sec:critical-inhomogeneity}
\subsection{Impact on critical threshold}
\label{subsec:criticality-strong-mode}
Let us first discuss how the critical Keldysh parameter depends on the strong field shape.
From \eqref{eq:x4star} we could directly read off the critical point
\begin{align}
  \begin{split}
  x_4^* = F^{-1}\left(\frac{\gamma}{\omega}\right)
  \label{eq:critical-pts}
\end{split}
\end{align}
corresponding to $\dot x_3 = 0$. Closing the instanton in $x_4^*$ results in the usual process.
Here, we need to satisfy \eqref{eq:ref-cond}.
However, from the equality
\begin{align}
  x_4^* = x_4^\mathrm{p},
  \label{eq:crit-cond}
\end{align}
we can obtain immediately a critical frequency
\begin{align}
  {\tilde \omega}^\mathrm{crit} =  \frac{e E}{m} {\tilde \gamma}^\mathrm{crit}.
\end{align}
Below this there will be no assistance.
In case a weak Sauter field is superimposed with a strong static field, the critical value was
\eqref{eq:sauter-gamma-crit}.
For the weak Lorentzian field
see \eqref{eq:lorentzian-gamma-crit}.
On the other hand,
\eqref{eq:crit-cond} clearly shows that the critical Keldysh parameter depends on the strong field, which determines the LHS of the condition \eqref{eq:crit-cond}, and on the weak field itself, which is responsible for the RHS.
This is expected, since with increasing $\gamma$ the closing point $x_4^\mathrm{c}$ will drift towards the origin along the complex time axis as sketched in Fig.~\ref{fig:criticals} left.
Accordingly, the pole $x_4^\mathrm{p}$ has to become smaller as well, see condition \eqref{eq:ref-cond}.
This is why the threshold depends on $\gamma$ and on the strong field profile.
Such a dependence has been analytically obtained for the case of a strong spatial Sauter field combined with a temporal Sauter field, treating both fields nonperturbatively \cite{Schneider:2014mla}.
Following analogous geometric arguments as discussed here, the present approach has been extended to spatiotemporal electric backgrounds $\pmb{E}(t,x)$ with temporal sinusoidal or (super) Gaussian dependence \cite{Akal:2017sbs}
resulting in highly accurate predictions for the critical Keldysh parameter.

For too large $\gamma$ the strong field will drive the vacuum decay alone, since the threshold ${\tilde \gamma}^\mathrm{crit}$ will be large.
This corresponds to the usual anti-adiabatic, perturbative multi-photon process.
For not too large $\gamma$, the weak field assistance is expected for moderate $\tilde \gamma$.
However, as soon as $\gamma$ becomes much smaller than unity, the (locally) static strong field will be again a good approximation.
An explicit example for which the strong field is assumed to be nonstatic
is studied below.
The general question may be, whether the present reflection picture is valid or not if one allows for $\gamma$ values of order unity or larger.
Note, that this situation may in general be not relevant and realistic for upcoming experimental designs.

To resolve the latter question, let us bear in mind that the basic starting point for the reflection picture was based on the negligibility of the weak field contribution away from the reflection point in the original instanton equations \eqref{eq:x4-x3ddot}.
Therefore, even if we allow $\gamma$ to be large, there will be always a reflector from the much more rapid weak field that will dominate above the critical threshold.
Also for $\epsilon \ll 1$, those poles will be much closer to the origin than, if present, the strong field poles, because of $\tilde \omega \gg \omega$.
In order to study the strong field profile dependence of the critical threshold, we use the relation in \eqref{eq:critical-pts} giving the criticality condition
\begin{align}
  \gamma = \omega F\left( x_4^\mathrm{p} \right).
  \label{eq:crit-relation}
\end{align}
For illustrative reasons, let us assume the weak field to be of Lorentzian type described by \eqref{eq:lorentzian}
with the pole \eqref{eq:lorentzian-pole}.
Using this setup, we can compute ${\tilde \gamma}^\mathrm{crit}$ for several strong field profiles starting from \eqref{eq:crit-relation} and using the relation $\omega = \frac{m \gamma E}{E_\mathrm{S}}$:
\begin{figure}[h!]
  \centering
  \includegraphics[width=.4\textwidth]{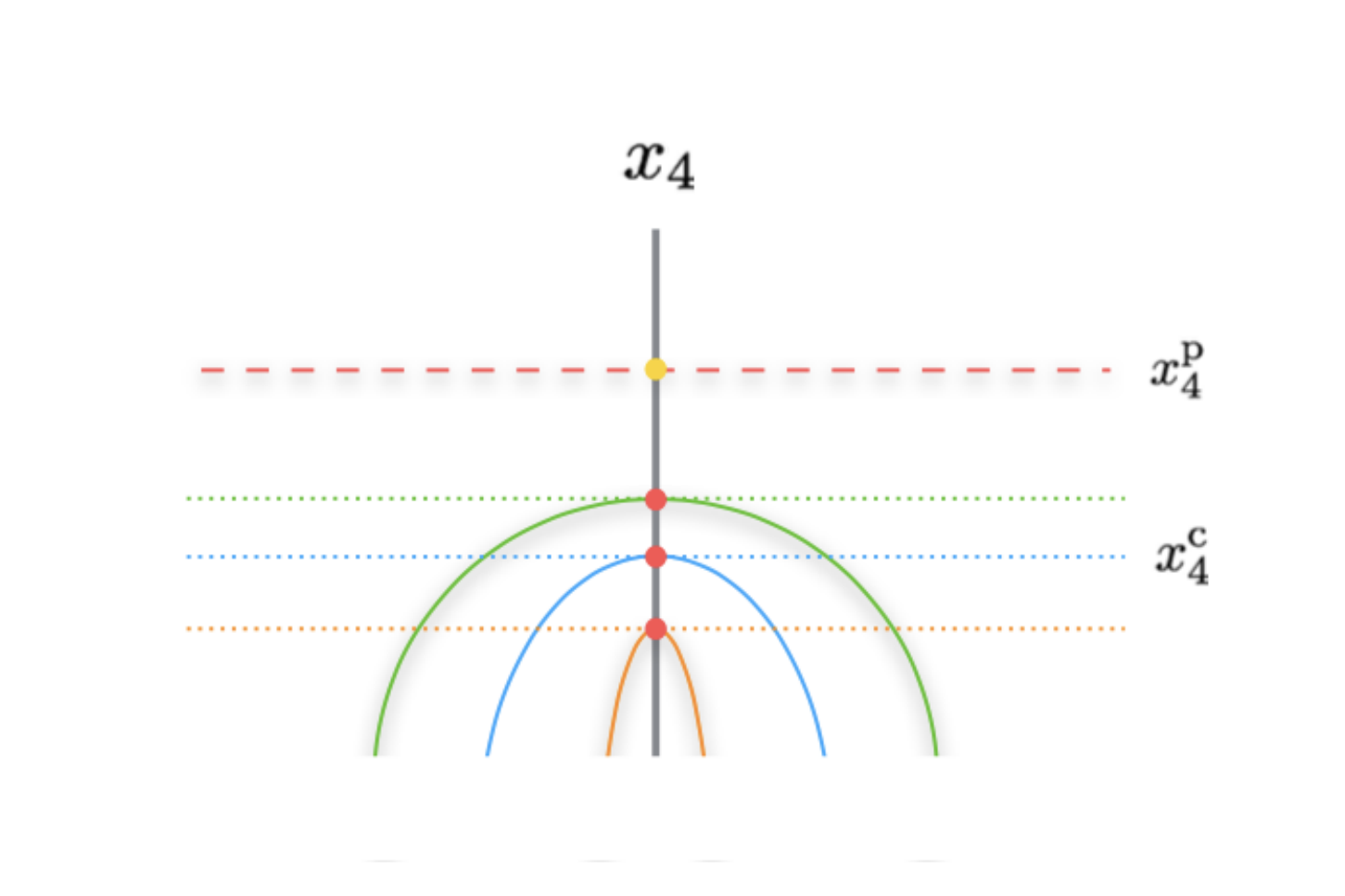}
\includegraphics[width=.58\textwidth]{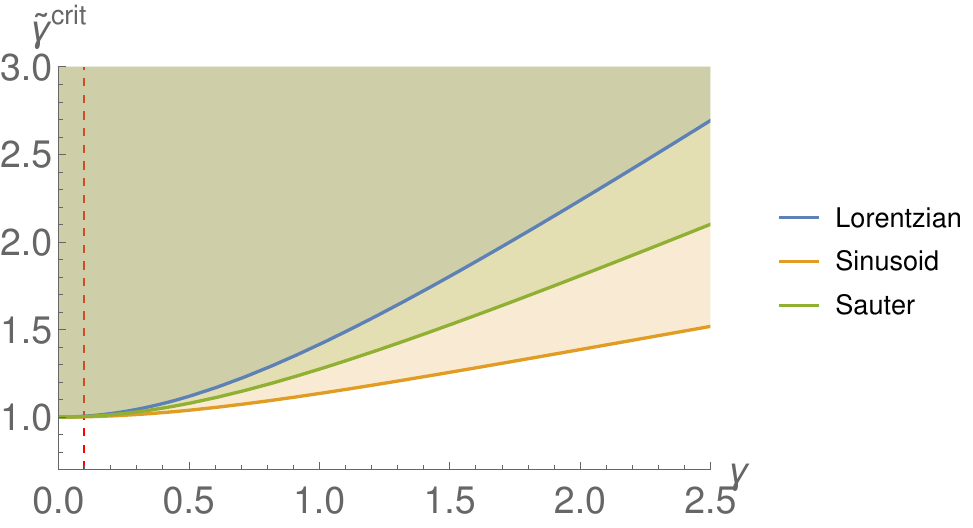}
\caption{Left: condition for the assisted mechansim: in case of reflection both critical points, i.e. closing point of the instanton (red dots) and the weak field pole (yellow dot), have to coincide, cf. \eqref{eq:crit-cond}. For increasing $\gamma$ the closing points drift towards the origin (red dots) and the instanton (solid lines) curves much stronger around its closing point. Right: critical combined Keldysh parameter ${\tilde \gamma}^\mathrm{crit}$ for the case of a weak Lorentzian field superimposed with a strong field of different profiles. The profiles are listed in the legend. The relevant regime for the assisted mechanism ($\tilde \gamma > {\tilde \gamma}^\mathrm{crit}$)
is depicted by the coloured region.
The adiabatic nonperturbative regime, i.e. $\gamma \ll 1$, lies to the left of the vertical dashed red line where ${\tilde \gamma}^\mathrm{crit} \approx 1$.}
\label{fig:criticals}
\end{figure}

\begin{flalign}
  \textbf{Strong static field}\qquad
 f(t) &= 1,\qquad
 F(x_4) = x_4,&&\\
  {\tilde \omega}^\mathrm{crit} &= m \frac{E}{E_\mathrm{S}},\qquad
  {\tilde \gamma}^\mathrm{crit} = 1,
\label{eq:static-criticals}
\end{flalign}

\begin{flalign}
  \textbf{Strong Lorentzian}\qquad
  f(t) &= \frac{1}{\left( 1 + (\omega t)^{2} \right)^{3/2}},\qquad
  F(x_4) = \frac{x_4}{\sqrt{1 - (\omega x_4)^2}},&&\\
  {\tilde \omega}^\mathrm{crit} &= m \frac{E}{E_\mathrm{S}} \sqrt{1 + \gamma^2},\qquad
  {\tilde \gamma}^\mathrm{crit} = \sqrt{1 + \gamma^2},
\label{eq:lorentzian-criticals}
\end{flalign}

\begin{flalign}
  \textbf{Strong sinusoid}\qquad
  f(t) &= \cos(\omega t),\qquad
  F(x_4) = \frac{\sinh(\omega x_4)}{\omega},&&\\
  {\tilde \omega}^\mathrm{crit} &= m \frac{E}{E_\mathrm{S}} \frac{\gamma}{\mathrm{arcsinh}(\gamma)},\qquad
  {\tilde \gamma}^\mathrm{crit} = \frac{\gamma}{\mathrm{arcsinh}(\gamma)},
\label{eq:sinusoid-criticals}
\end{flalign}

\begin{flalign}
  \textbf{Strong Sauter}\qquad
  f(t) &= \mathrm{sech}^2(\omega t),\qquad
  F(x_4) = \frac{\tan(\omega x_4)}{\omega},&&\\
  {\tilde \omega}^\mathrm{crit} &= m \frac{E}{E_\mathrm{S}} \frac{\gamma}{\mathrm{arctan}(\gamma)},\qquad
  {\tilde \gamma}^\mathrm{crit} = \frac{\gamma}{\mathrm{arctan}(\gamma)}.
\label{eq:sauter-criticals}
\end{flalign}

For $\gamma \rightarrow 0$ we approach for all cases the static limit \eqref{eq:static-criticals}, the adiabatic nonperturbative tunneling regime.
However, for larger $\gamma$ the critical value ${\tilde \gamma}^\mathrm{crit}$ increases first parabolic then linear in $\gamma$, cf. right panel in Fig.~\ref{fig:criticals}. The relevant regime for the assisted mechanism, i.e.
$\tilde \gamma > {\tilde \gamma}^\mathrm{crit}$, is indicated by the coloured patterns.
For values $\gamma > 0.1$, i.e. right to the vertical dashed red line, we leave the region with almost constant dependence in $\gamma$, i.e. ${\tilde \gamma}^\mathrm{crit} \approx 1$.
The latter is the nonperturbative regime for the strong field. The slope of the plotted curves turns out to be much stronger for fields that lead in general to smaller reflection points. This explains why for those the weak field inhomogeneity has to be much larger.
Such studies exhibit the two types of mechanisms which lead to a substantial enhancement of the tunneling rate in time-dependent, inhomogeneous electric backgrounds:

1. The enhancement is driven by a single-mode, time-dependent, inhomogeneous electric field (anti-adiabatic, perturbative, multi-photon regime).
This is also known as the standard dynamical mechanism.
The role of a second weak field becomes negligible with increasing $\gamma$. A characteristic threshold in this case does not exist.

2. The electric background is composed of a strong, slow field (adiabatic, nonperturbative, tunneling regime) superimposed with a weak, rapid field. This situation corresponds to the assisted mechanism. The contribution of the weak field is essential for the enhancement. It sets in for $\tilde \gamma$ above the characteristic threshold, the critical combined Keldysh parameter. In the present paper, we have distinguished between the standard assisted mechanism and the assisted dynamical mechanism. The latter is characterized by an additional inhomogeneity of the strong field in addition to the weak but more rapid field.

\subsection{Strong sinusoid and weak Lorentzian}
\label{subsec:effects-strong-mode}
\begin{figure}[h!]
  \centering
\includegraphics[width=.55\textwidth]{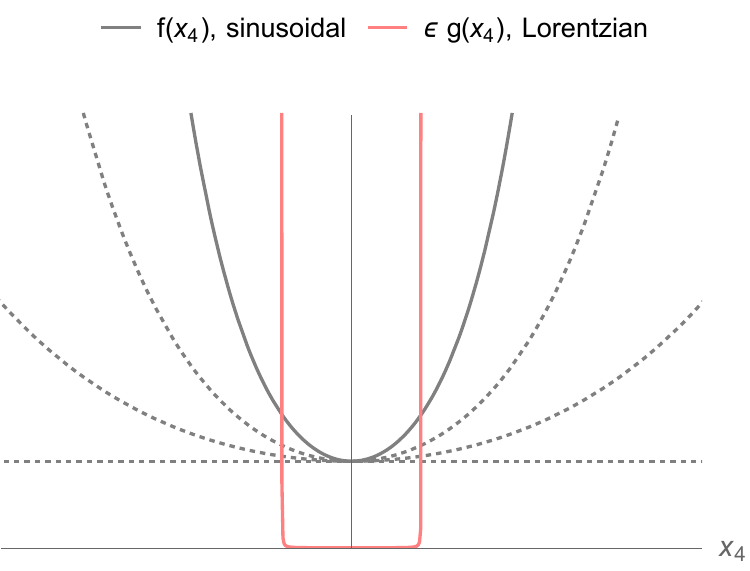}
\caption{Combined electric background after rotation in the complex plane: the strong inhomogeneous field (without poles, grey solid curve) and weak inhomogeneous field (with poles, pink curve) for $\gamma / \tilde \gamma <1$ are plotted separately. Dotted gray curves indicate the increase of $\gamma$ starting at $\gamma = 0$ (horizontal dotted line).}
\label{fig:both-frequent}
\end{figure}
Let us illustrate the effects from Sec.~\ref{subsec:criticality-strong-mode} with an example.
The strong field we assume to be of sinusoidal type and the weak field with Lorentzian profile, i.e.
\begin{align}
  \begin{split}
  f(t) &= \cos(\omega t),\qquad
  F(x_4) = \frac{\sinh(\omega x_4)}{\omega},\\
  g(t) &= \frac{1}{\left( 1 + (\omega t)^{2} \right)^{3/2}},\qquad
  G(x_4) = \frac{x_4}{\sqrt{1 - (\omega x_4)^2}}.
\end{split}
\end{align}
The corresponding modification in comparison to the previously studied cases with a static strong field is depicted schematically in Fig.~\ref{fig:both-frequent}. With increasing $\gamma$ we left the static limit by bending up the initial horizontal line (dotted gray curves), representing the function $f(x_4)$. For $\gamma \neq 0$ there will be a substantial structure (solid gray curve) between the poles of the Lorentzian field (pink curve). The interplay between this parabolic strong field curve and the reflecting weak field poles will be computed.
The pole for the weak Lorentzian field is $x_4^\mathrm{p} = 1 /\tilde \omega$. Using the expression in \eqref{eq:gen-a} we get
\begin{align}
  a = - i \frac{4}{\omega} \mathbf{F}\left(i \frac{\gamma}{\tilde \gamma} \bigg| \frac{-1}{\gamma^2}\right),
  \label{eq:a-sin-lor}
\end{align}
where $\mathbf{F}(\cdot|\cdot)$ is the incomplete elliptic integral of the first kind.
\begin{figure}[h!]
  \centering
\includegraphics[width=.9\textwidth]{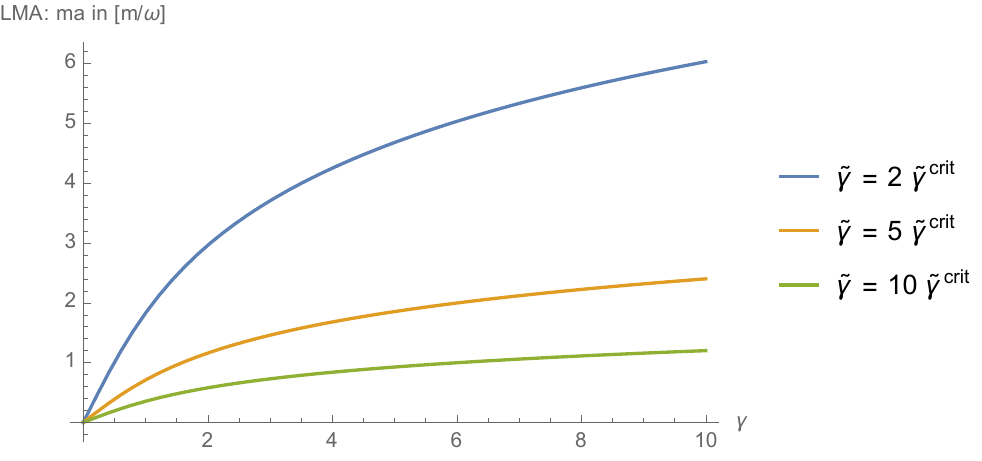}
\caption{Relevant quantity $ma$ expressed in units of $\frac{m}{\omega}$. The LMA requires $ma \gg 1$. Curves are plotted for different $\tilde \gamma$. The values are depicted in the plot legend. The critical combined Keldysh parameter is given in \eqref{eq:sinusoid-criticals}.}
\label{fig:both-frequent-lma}
\end{figure}
From the latter invariant \eqref{eq:a-sin-lor} we can read off the critical Keldysh parameter
${\tilde \gamma}^\mathrm{crit} = \gamma / \mathrm{arcsinh}(\gamma)$  computed in \eqref{eq:sinusoid-criticals}. Using the modified invariant \eqref{eq:a-sin-lor}, the LMA condition \eqref{eq:lma} becomes
\begin{align}
  \frac{m}{\omega} \left( -i4\mathbf{F}\left(i \frac{\gamma}{\tilde \gamma} \bigg| \frac{-1}{\gamma^2}\right) \right) \gg 1.
\end{align}
This quantity we have plotted in Fig.~\ref{fig:both-frequent-lma} versus $\gamma$ and different fixed combined Keldysh parameter $\tilde \gamma$. Very large $\tilde \gamma$ are excluded because of the LMA condition.
Using the equation \eqref{eq:gen-W}, we obtain the stationary worldline action
\begin{align}
  \mathcal{W}_0 \approx \frac{m^2}{e E} \frac{4}{\gamma} \left( - i \mathbf{E} \left( i \frac{\gamma}{\tilde \gamma} \bigg| \frac{-1}{\gamma^2} \right) \right).
\end{align}
The function $\mathbf{E}(\cdot|\cdot)$ denotes the incomplete elliptic integral of the second kind. The instanton solutions for the sinusoidal field are known \cite{Dunne:2005sx}. Based on those, we can write the present modified solution in the right half plane, i.e. $u \in [-1/4,1/4]$, as
\begin{align}
  x_4 &= \frac{m}{e E} \frac{1}{\gamma} \mathrm{arcsinh} \left( \frac{\gamma}{\sqrt{1+\gamma^2}} \mathbf{sd}\left( - i \mathbf{F}\left(i \frac{\gamma}{\tilde \gamma} \bigg| \frac{-1}{\gamma^2}\right) \frac{\sqrt{1+\gamma^2}}{\gamma} u \bigg| \frac{\gamma^2}{1 + \gamma^2} \right)  \right),\\
  x_3 &= \frac{m}{e E} \frac{1}{\gamma} \mathrm{arcsin} \left( \frac{\gamma}{\sqrt{1+\gamma^2}} \mathbf{cd}\left( - i \mathbf{F}\left(i \frac{\gamma}{\tilde \gamma} \bigg| \frac{-1}{\gamma^2}\right) \frac{\sqrt{1+\gamma^2}}{\gamma} u \bigg| \frac{\gamma^2}{1 + \gamma^2} \right)  \right) - \mathcal{C}.
\end{align}
The functions $\mathbf{sd}(\cdot|\cdot)$ and $\mathbf{cd}(\cdot|\cdot)$ denote Jacobi elliptic functions.
The shifting constant along the $\hat x_3$ axis is again determined by
\begin{align}
  \mathcal{C} = x_3(u = \pm 1/4).
\end{align}
The action above applies only for $\tilde \gamma \geq {\tilde\gamma}^\mathrm{crit}$. Taking into account the case when the contribution of the weak field is absent, we can write the complete stationary worldline action as
\begin{align}
  \mathcal{W}_0 =
  \left\{
  \begin{array}{ll}
      \frac{m^2}{e E} \frac{4}{\gamma} \left(-i \mathbf{E} \left( i \frac{\gamma}{\tilde \gamma} \bigg| \frac{-1}{\gamma^2} \right) \right) &\qquad \tilde \gamma \geq {\tilde \gamma}^\mathrm{crit},\\
      4 \frac{m^2}{e E}\frac{\sqrt{\gamma ^2+1}}{\gamma^2} \left(\mathbf{K}\left(\frac{\gamma ^2}{\gamma ^2+1}\right)-\mathbf{E}\left(\frac{\gamma ^2}{\gamma ^2+1}\right)\right) &\qquad \tilde \gamma < {\tilde \gamma}^\mathrm{crit}.
      \label{eq:computed-W0-sinusoid-lorentzian}
\end{array}
\right.
\end{align}
Here, $\mathbf{K}(\cdot)$ and $\mathbf{E}(\cdot)$ denote the complete elliptic integrals of the first and second kind, respectively.
\begin{figure}[h!]
  \centering
\includegraphics[width=.99\textwidth]{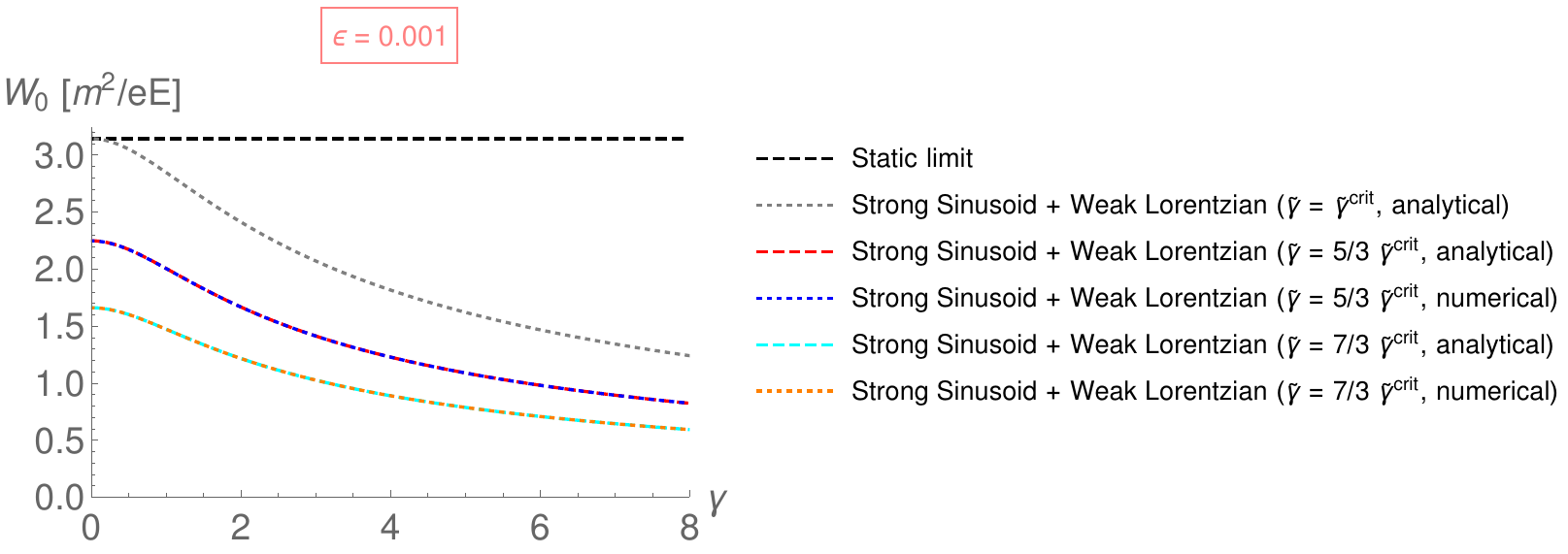}\\
\includegraphics[width=.49\textwidth]{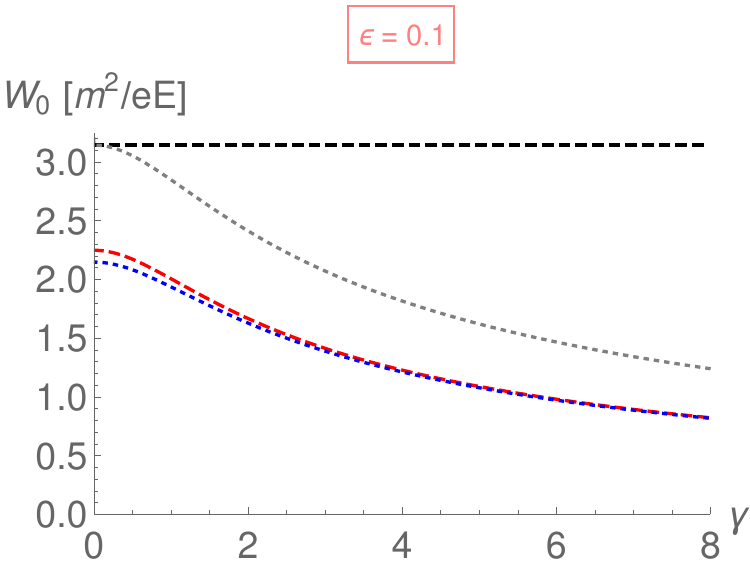}
\includegraphics[width=.49\textwidth]{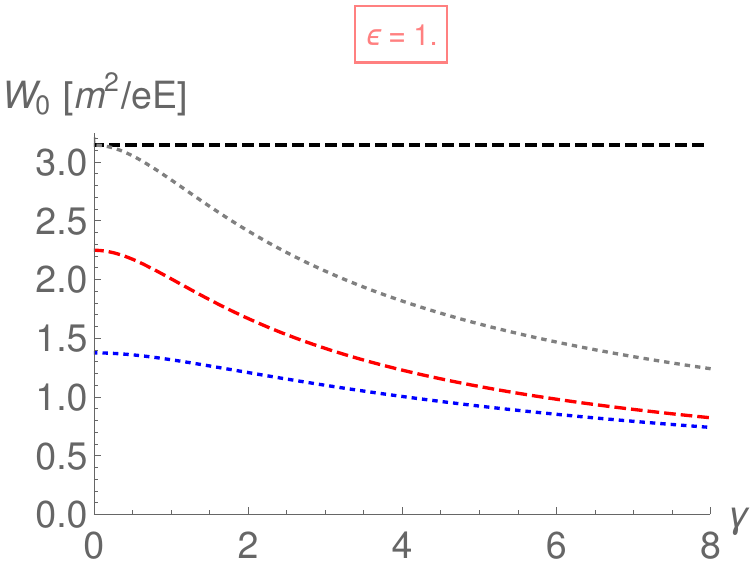}
\caption{Stationary worldline action \eqref{eq:computed-W0-sinusoid-lorentzian} for the strong sinusoidal field superimposed with a weak Lorentzian field \eqref{eq:sinusoidal-field} plotted versus $\gamma$. The analytical prediction is compared with exact numerical computations. The ratio between the strong and weak field strengths is set to $\epsilon = \{10^{-3},10^{-1},10^{0}\}$ (top,bottom-left,bottom-right). The values for the combined Keldysh parameter are given in the plot legend with ${\tilde \gamma}^\mathrm{crit}$ being computed via \eqref{eq:sinusoid-criticals}.
}
\label{fig:W0-plot-sinusoidlorentzian}
\end{figure}

The result \eqref{eq:computed-W0-sinusoid-lorentzian} is plotted in Fig.~\ref{fig:W0-plot-sinusoidlorentzian}. Setting $\tilde \gamma = {\tilde \gamma}^\mathrm{crit}5/3$, we compare between the analytical prediction and the exact numerical computation. Both results do perfectly coincide as long as
$\epsilon \ll 1$ which is the valid regime in the reflection picture. Only if we apply relatively large values $\epsilon = \{0.1,1.0\}$, there appears a notable difference between both curves. The effect of the weak field is well indicated.
A considerable decrease applies in contrast to the situation with $\tilde\gamma={\tilde\gamma}^\mathrm{crit}$,
where the weak field contribution is absent. As one would expect, for $\gamma = 0$ ($\Rightarrow \tilde \gamma = 5/3$) we find again the result from Fig.~\ref{fig:W0-sauter-lorentzian}.
\begin{figure}[h!]
  \centering
\includegraphics[width=.99\textwidth]{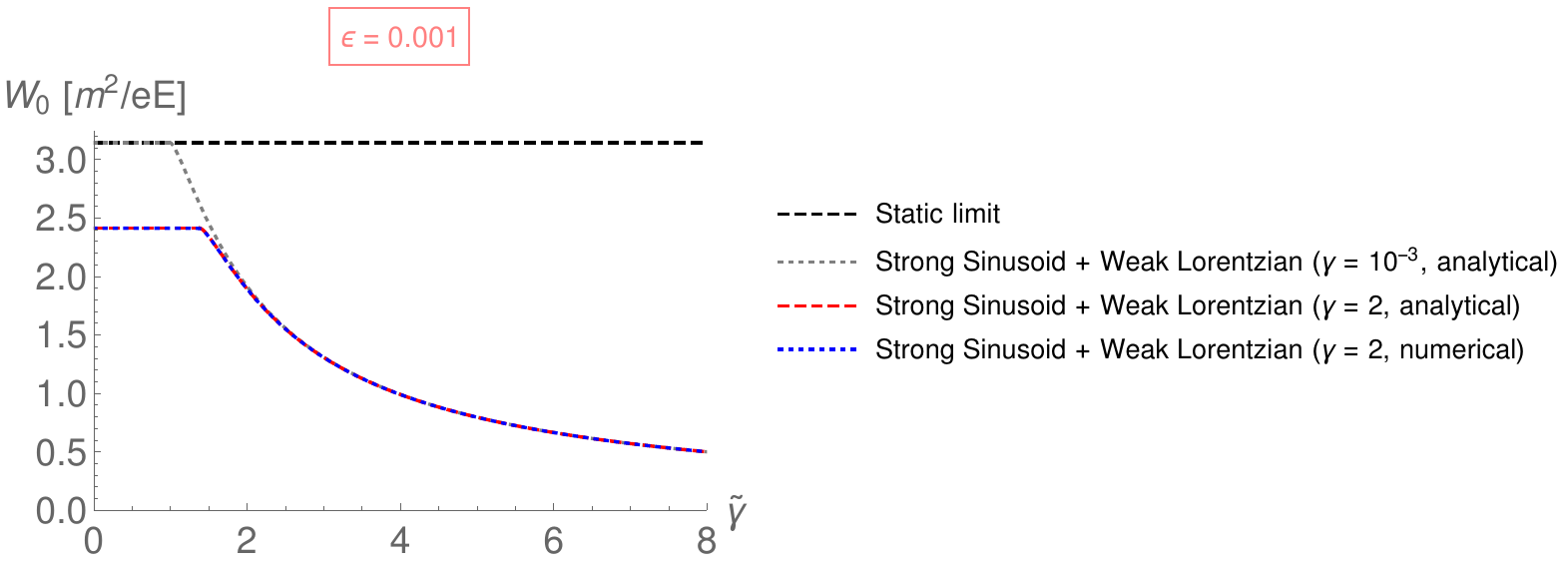}\\
\includegraphics[width=.5\textwidth]{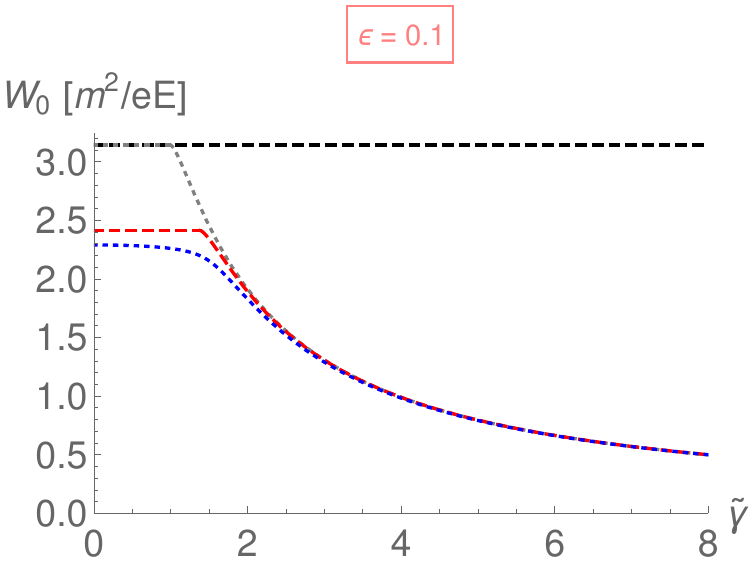}
\caption{Same stationary worldline action as in Fig.~\ref{fig:W0-plot-sinusoidlorentzian} plotted versus the combined Keldysh parameter $\tilde \gamma$. The corresponding strong field inhomogeneities $\gamma$ as well as the ratios $\epsilon$ between the field strengths are given in the plots. The gray dotted curve is the known result in Fig.~\ref{fig:W0-sauter-lorentzian} (i.e. green curve).
}
\label{fig:W0-plot-sinusoidlorentzian-gammac}
\end{figure}

We have seen
that the superposition of a strong field with $\gamma > 0$ and a weak faster field leads to a stronger enhancement. However, this is only operative for $\tilde \gamma$ below the threshold ${\tilde \gamma}^\mathrm{crit}$. For larger values it is again the weak rapid field that mainly drives the enhancement, cf. Fig.~\ref{fig:W0-plot-sinusoidlorentzian-gammac}.

We should note, that in the latter described situation we have set the strong field inhomogeneity parameter as $\gamma = 2$.
Technically, a very strong field (Extreme Light Infrastructure, ELI) with high frequency (European XFEL) is neither realisable with current experimental facilities nor visioned for upcoming setups.
One can alternatively think of the following realistic situation.
Assume we start with a very strong field with strength $E_1$ and frequency $\omega_1 \rightarrow 0$. Superimposing this low-frequent field with a second weak more rapid field, i.e. $E_2/E_1 \ll 1$ and $\omega_2 \gg \omega_1$, resembles the strong field setup depicted in Fig.~\ref{fig:both-frequent}.

\section{Summary}
\label{sec:conclusion}
In this paper we have studied
enhancement effects in
vacuum pair production via two mechanisms, the assisted mechanism (Sec.~\ref{sec:pole-fields} and \ref{sec:fields-without-poles}) and the assisted dynamical mechanism (Sec.~\ref{sec:critical-inhomogeneity}).
Using the worldline formalism, we have obtained from the stationary instanton equations two separate critical points.
While one of them is responsible for the closing of the instanton path, the other serves as a reflecting mirror in Euclidean spacetime.
Employing this reflection picture, we have analysed characteristic features.
Specifically, we have focused on the role of the assisting weak, rapid field. Based on geometric considerations, we have explained the origin for substantial differences due to the analytic structure of the considered backgrounds.
For this, we have distinguished between two types of backgrounds.

The first type is characterised by weak fields which possess a distinct Euclidean pole structure. This is the case where geometrical arguments are very intuitive.
Revisiting previous observations for the assisted mechanism, we have shown that the drastic enhancement is the direct consequence of instanton reflections in such poles (Sec.~\ref{sec:pole-fields}).
This has been illustrated for weak fields of Sauter and Lorentzian type which behave similarly.
We have shown that this common behaviour is caused due to their similar pole structure and not due to their almost indistinguishable bell-shaped profiles.
We have also discussed the impact of a possible sub-cycle structure.
Performing explicit computations, we have illustrated that the assistance is primarily determined by the pole of the bell-shaped envelope function in the instanton plane. Only with sufficiently large $\epsilon$ the encased sub-cycle structure of the considered oscillatory pulse leads to considerable deviations.

In the first main part of this work, concerning the assisted mechanism, we have extended the reflection picture to backgrounds characterised by poleless weak fields (Sec.~\ref{sec:fields-without-poles}).
Based on analogous geometrical arguments,
we have obtained specific conditions from combined analysis based on the instanton equations and the equivalent WKB approach.
By doing so, we have been able to compute analytically the corresponding effective reflection points.
In addition, we have for the first time analytically computed the critical threshold for the combined Keldysh parameter
determined by the critical point where the weak field contribution starts to dominate.
We have shown that the critical point deviates from the relatively large valued effective reflection point, even in the highly weak limit. This feature turns out to be the major difference between fields with and without poles.
In the former case the reflection and the critical point are equal to the pole itself.
We have demonstrated that this discrepancy can be seen as the primary reason why poleless fields, like the sinusoidal one, enhance the vacuum decay less than fields with poles or pole-like behaviour, respectively.
We have shown that the additional $\epsilon$ dependence in this mechanism occurs if the weak rapid field cannot be characterised by a distinct pole structure. However, for super Gaussians this $\epsilon$ dependence becomes increasingly suppressed.

In the second main part, we have studied the assisted dynamical mechanism where
the strong field is assumed to be nonstatic in addition to the weak but more rapid variation (Sec.~\ref{sec:critical-inhomogeneity}). 
Again, applying the reflection approach, we have analytically computed the rate for an explicit example. The additional inhomogeneity has led to a substantial enhancement in comparison to the standard assisted mechanism.
Our analytical predictions in the relevant regime are in perfect agreement with numerical computations.

We can conclude that the dynamical assistance is predominantly determined by instanton reflections, no matter whether poles are present or not.
The location of characteristic critical points for the weak field determines the strength of the assistance. It is notable that reflection points close to the origin lead basically to stronger enhancement signatures. Such insights may allow to pursue further optimisation studies with respect to the weak field in order to maximize these effects.
It is also interesting to work out analogue geometric considerations for the case of an additional spatially inhomogeneous field.
This allows an analytical treatment for electric backgrounds with genuine spatiotemporal dependence.
Furthermore, it facilitates in particular the role of such backgrounds with regard to the nonlocal nature of vacuum pair production.
Recently, results in that direction have been presented in \cite{Akal:2017sbs} demonstrating the validity of the techniques discussed in the present work.
\acknowledgments
We are grateful to Holger Gies for valuable comments and a careful reading of the manuscript. We thank Andreas Ringwald for support and valuable comments during this work. IA thanks Dennis D. Dietrich for interesting discussions.
The authors acknowledge the support of the Colloborative Research Center SFB 676 \textit{Particles, Strings and the Early Universe} of the DFG.

\appendix

\section{Effect of inhomogeneities}
\label{sec:inhomo-effects}
The impact of background inhomogeneities can be very elegantly illustrated with the stationary instanton solutions.
It has been shown that inhomogeneities of latter type tend to shrink the instanton which then leads to a larger pair production rate, see e.g. \cite{Dunne:2005sx} for exact instanton solutions.
The technical reason for such an enhancement can therefore be obtained directly from the instanton equations \eqref{eq:eom}.

However, one should note that for arbitrary inhomogeneous fields, particularly for spatiotemporal type, the situation can be very complicated due to the increasing nonlinear structure of the underlying equations. Hence, it can be quite difficult to get some approximate information directly from the instanton equations. Effects of spatiotemporal fields have been recently studied \cite{Hebenstreit:2010vz,Hebenstreit:2011wk,Schneider:2014mla,Ilderton:2014mla,Ilderton:2015qda,Aleksandrov:2016lxd}.
Here, we focus on time-dependent, inhomogeneous fields as introduced in Sec.~\ref{sec:instantons}. To work out the differences, we will start with a static, electric field and show that the instanton solves the circle equation. This is the most simple case where we can find a closed instanton path with maximal symmetry in the two dimensional plane.
\begin{figure}[h!]
  \centering
\includegraphics[width=.9\textwidth]{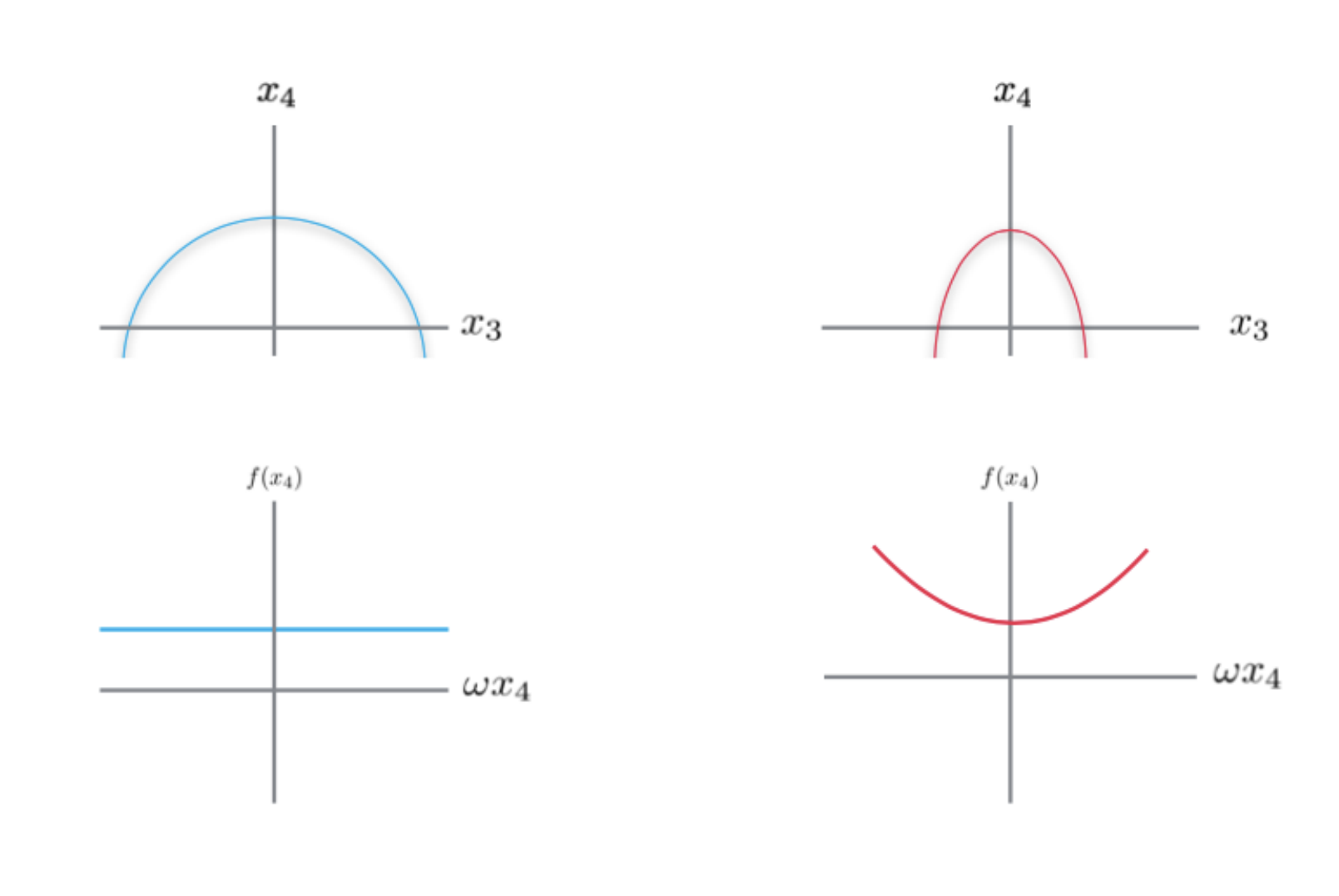}
\caption{Effect of temporal field inhomogeneities: as an example, the comparison between a static and sinusoidal electric field is sketched. For the latter the bounded cosine function becomes after the rotation in the complex plane the (from above) unbounded hyperbolic cosine function (right). The static field remains static (left). Sketched instanton paths around the closing point are depicted on top.}
\label{fig:effect-inhomos}
\end{figure}

We begin with the following relations
\begin{align}
  \begin{split}
  \dot x_4^2 + \dot x_3^2 = a^2,\qquad
  \frac{\ddot x_4}{\dot x_3} = a/R,\qquad
  \frac{\ddot x_3}{\dot x_4} = -a/R.
\end{split}
  \label{eq:static-sys-eqns}
\end{align}
$a$ is the kinematic invariant and $R = \frac{m}{eE}$.
Since $\frac{a}{R}$ is constant, we can integrate the second and third equation in \eqref{eq:static-sys-eqns} to obtain the system
\begin{align}
  \dot x_4 = \frac{a}{R} x_3,\qquad
  \dot x_3 = -\frac{a}{R} x_4
\end{align}
where integration constants vanish due to the periodicity condition $\mathfrak{p}$ in \eqref{eq:rescaled-boundary} and $a^2 = \mathrm{constant}$.
Squaring the latter equations and applying the first relation in \eqref{eq:static-sys-eqns} afterwards, we obtain
the circle equation
\begin{align}
x_4^2 + x_3^2 = R^2.
\end{align}
Hence, the instanton for a static field must be a circle.
This observation
one may also guess just by looking on the equations in \eqref{eq:static-sys-eqns}.
In particular the RHS of the last two equations is a fixed constant $\frac{a}{R}$. In other words, the ratio between the acceleration $\ddot x_4$ ($\ddot x_3$) in one direction and the velocity $\dot x_3$ ($\dot x_4$) in the remaining direction is constant. Kinematically, this situation is realised along a circle path. Hence, the electric field in Euclidean spacetime acts like a magnetic field leading to a circular instanton path.

If this static electric field is oriented in the $\hat x_3$ direction, i.e.
\begin{align}
\mathcal{A}_3 (x_4 ) = - i E x_4,
\label{eq:constant-vecfield}
\end{align}
the resulting circular path in the $(x_3,x_4)$ plane is described by
\begin{align}
x_3(u) = R \cos \left( 2 \pi n u \right),\
x_4(u) = R \sin \left( 2 \pi n u \right),
\label{eq:constant-instanton}
\end{align}
where $a=2 \pi n R$ and $R=\frac{m}{eE}$ following due to the periodicity condition $\mathfrak{p}$ \cite{Affleck:1981bma}.

Note that the distance along the spatial $\hat x_3$ axis at time $x_4 = 0$ is
\begin{align}
  x = 2 R = \frac{2 m}{e E}.
\end{align}
This result one can already obtain from simple energy conservation. Namely, the energy $x e E$ that is needed during delocalising the virtual pair to make it real along a distance $x$ has to be $2 m$.
This is the relation from above.
Note that if $x = 2 \lambda_c$, where $\lambda_c = 1/m$ denotes the usual Compton wavelength, this simple analysis brings us to the usual Schwinger limit $E_\mathrm{S} = \frac{m^2}{e}$, except the prefactor $\pi$, cf. \eqref{eq:W0-n=1}. Therefore, the width of the effective energy gap between the excited particle states and the Dirac sea is naturally encoded in the spatial width\footnote{Fields with temporal inhomogeneities in spacetime lead in general to a substantial reduction of the tunneling barrier, i.e. $m^* < m$, which corresponds then to a smaller spatial width $x^* < x$ of the instanton trajectory at $x_4 = 0$.} of the instanton at zero (Euclidean) time.

Coming back to the circle solution, we find $n$ given in $a$ as the instanton's winding number that counts for the number of times the Euclidean path is traversed.
The higher order instanton contributions with $n>1$ correspond to the production of $n$ pairs \cite{lebedev1984virial,Cohen:2008wz}. It is not clear whether this argument is justified
for the case of strongly coupled non-Abelian gauge theories.
For the circle instanton the LMA from \eqref{eq:lma} becomes immediately $E \ll E_\mathrm{S}$.
For $n=1$ we find the previous evaluated action $\mathcal{W}_0$ from \eqref{eq:W0-n=1} (as the dominating contribution in the weak field limit).
Taking in addition the fluctuation prefactor \cite{Affleck:1981bma,Dunne:2006st} into account, the complete result for the vacuum decay probability reads
\begin{align}
\mathcal{R} \simeq \frac{(eE)^2}{(2 \pi)^3} \sum_{n=1}^\infty \frac{(-1)^{n+1}}{n^2} e^{  - \pi n \frac{E_\mathrm{S}}{E} },
\label{eq:schwinger-res}
\end{align}
which is precisely the original Schwinger formula. The first term is the vacuum pair production rate from \eqref{eq:static-R}.

The situation for a non-static electric field is far more complicated.
The system, one has to solve in this case, is given by
\begin{align}
  \dot x_4^2 + \dot x_3^2 = a^2,\qquad
  \frac{\ddot x_4}{\dot x_3} = f(x_4),\qquad
  \frac{\ddot x_3}{\dot x_4} = -f(x_4).
  \label{eq:nonstatic-sys-eqns}
\end{align}
The constant RHS of the last two equations is now described by $\pm f(x_4)$. The function $f(x_4)$ is nothing but the analytic continuation (except the imaginary prefactor $-i$) of the physical electric field. Therefore, inhomogeneous electric fields with some oscillatory profile may become unbounded positive monotonic functions in the instanton equations, cf. lower right panel in Fig.~\ref{fig:effect-inhomos}.
Let us make this a bit more concrete. For instance, the sinusoidal cosine becomes after the rotation in the complex plane the hyperbolic cosine function. This in addition brings the imaginary prefactor\footnote{Note that it is this complex prefactor which makes the instanton solution real for the presently considered electric backgrounds.} in front. In other words, the complex exponential of cosine becomes the unbounded real exponential.

In this case one will find points where the acceleration in one direction may become much larger than the velocity in the other. The equations of such a system may have ellipse-like solutions which can curve much stronger than the usual circle path. As a consequence, the size of the instanton can drastically be reduced for appropriate field parameters, e.g. sufficiently large temporal inhomogeneities \cite{Dunne:2005sx}.
From the previous discussion, this reduction would correspond to a smaller instanton extension $x^*$ at $x_4 = 0$ or smaller effective mass $m^*$, respectively, means that the tunneling barrier is reduced.
Consequently, the rate for vacuum pair production in such background fields will be increased compared to the static field. Note that the latter remains static even after continuation to Euclidean spacetime, cf. lower left panel in Fig.~\ref{fig:effect-inhomos}.

However, the huge impact of temporal inhomogeneous electric fields on vacuum pair production is not only initiated by their unbounded shape in the instanton equations.
Another effect results due to the appearance of pole structures in the instanton plane.
It is this reason why we think that, despite the differences regarding quantum interferences\footnote{We do not consider interference effects in the present studies. Basically, those are encoded and manifest in the phase-space of the produced pairs, cf. e.g. \cite{Hebenstreit:2009km,Dumlu:2011rr,Orthaber:2011cm,Abdukerim:2013vsa,Akal:2014eua,Aleksandrov:2017owa}. The pole structure of the field is expected to be essential for their appearance, cf. \cite{Linder:2015vta}. For further sensitivity and optimisation studies those effects should be taken into account.}, a weak Sauter field in addition to a strong locally static field leads to a stronger enhancement than a weak poleless sinusoidal field. We aim to study such effects in the present work.

\newpage
\bibliographystyle{JHEP}
\bibliography{mainbib}

\end{document}